\documentclass[13pt]{article}
\usepackage{epsf}
\usepackage{color}
\usepackage{amsmath,amsfonts,amsbsy,amssymb}
\usepackage{indentfirst}
\usepackage{mathtools}
\usepackage{arydshln} 
\usepackage{framed}

\newcommand{\fsl}[1]{\ensuremath{\mathrlap{\not{\phantom{#1}}}#1}}

\newcommand{\nn}{\nonumber}

\def\be{\begin{equation}}
\def\ee{\end{equation}}
\def\bse{\begin{subequations}}
\def\ese{\end{subequations}}
\def\bal{\begin{align}}
\def\ealn{\end{align}}
\def\tr{\text{tr}}
\def\bs{\boldsymbol}
\def\bp{\begin{pmatrix}}
\def\ep{\end{pmatrix}}

\begin{document}

\begin{titlepage}

\def\slash#1{{\rlap{$#1$} \thinspace/}}

\begin{flushright} 

\end{flushright} 

\vspace{0.1cm}

\begin{Large}
\begin{center}

{\bf  $SO(5)$ Landau Models and Nested Nambu Matrix Geometry }
\end{center}
\end{Large}

\vspace{1cm}

\begin{center}
{\bf Kazuki Hasebe}   \\ 
\vspace{0.5cm} 
\it{
National Institute of Technology, Sendai College,  
Ayashi, Sendai, 989-3128, Japan} \\ 

\vspace{0.5cm} 

{\sf
khasebe@sendai-nct.ac.jp} 

\vspace{0.8cm} 

{\today} 

\end{center}

\vspace{1.0cm}

\begin{abstract}
\noindent

\baselineskip=18pt

The $SO(5)$ Landau model is the mathematical platform of the 4D quantum Hall effect and  provide a rare opportunity for a physical realization of the fuzzy four-sphere. We present an integrated analysis of the $SO(5)$ Landau models and the associated matrix geometries  
through the Landau level projection.  With the $SO(5)$ monopole harmonics, we  explicitly derive  matrix geometry of a four-sphere in any Landau level: In the lowest Landau level  the  matrix coordinates are given by the generalized $SO(5)$ gamma matrices of the fuzzy four-sphere satisfying the quantum Nambu algebra, while in higher Landau levels the  matrix geometry becomes a nested fuzzy structure realizing a  pure quantum geometry with no counterpart in classical geometry.   
The internal fuzzy geometry structure is   discussed  in the view of an $SO(4)$ Pauli-Schr\"odinger model and the $SO(4)$ Landau model, where      
we unveil a hidden singular gauge transformation  between their  background  non-Abelian  field configurations.  
Relativistic versions of the $SO(5)$ Landau model  are also investigated and  relationship to the  Berezin-Toeplitz quantization is clarified. We finally discuss the matrix geometry of  the Landau models in even higher dimensions.   

\end{abstract}

\end{titlepage}

\newpage 

\tableofcontents

\newpage 

\section{Introduction}

More than forty years ago, Yang proposed an $SU(2)$ generalization \cite{Yang-1978-1} of the Dirac's monopole  \cite{Dirac-1931}. The set-up behind the Yang's $SU(2)$ monopole stems from a beautiful mathematical concept of  the 2nd Hopf map associated with the generalization of complex numbers to  quaternions \cite{Trautman-1977, Minami-1980}.   The Yang's monopole field configuration on $S^4$ is conformally equivalent to  the BPST instanton  configuration on $\mathbb{R}^4$ \cite{Belavin-Polyakov-Schwartz-Tyupkin-1975}  and possesses the $SO(5)$   global rotational symmetry. Yang also succeeded to construct  
 generalized monopole harmonics in the $SU(2)$ monopole background  \cite{Yang-1978-2}. 
This set-up  was used in the context of the Zhang and Hu's $SO(5)$ Landau model and 4D quantum Hall effect \cite{Zhang-Hu-2001} that realize  natural higher dimensional counterparts of the Wu and Yang's $SO(3)$ Landau model \cite{Wu-Yang-1976} and the Haldane's 2D quantum Hall effect on a two-sphere \cite{Haldane-1983}.\footnote{For early developments of the higher dimensional Landau models and quantum Hall effects, interested readers may consult review articles \cite{Hasebe-2010, Karabali-Nair-2004} and references therein.}    

\begin{figure}[tbph]
\center
\hspace{-0.8cm}
\includegraphics*[width=130mm]{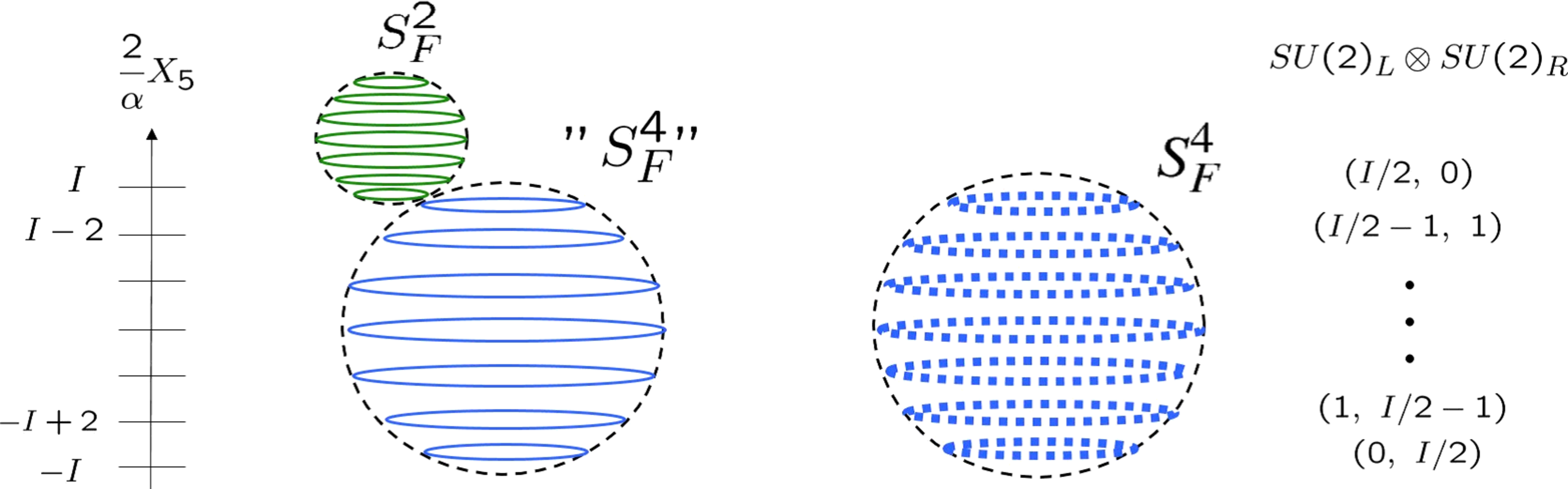}
\caption{Two geometric pictures of the fuzzy four-sphere. In the left figure, $X_{ab}$ (\ref{xxandxab}) span the fuzzy $S^2$-fibre on the original fuzzy manifold ``$S_F^4$''. In the right figure, the internal geometric structure is accounted for by the  degeneracy of the fuzzy three-sphere latitudes on $S_F^4$.  }
\label{fuzzyS4.fig}
\end{figure}

The non-commutative geometry is the emergent geometry of the Landau models and governs the dynamics of the quantum Hall effect \cite{Girvin-Jach-1984,Girvin-MacDonald-Platzman-1986}.  The Landau level projection truncates the whole quantum mechanical Hilbert space to a sub-space and provides a physical set-up where the non-commutative geometry naturally appears.   Along this line, 
the fuzzy four-sphere geometry has been discussed in the context of the $SO(5)$ Landau model \cite{Zhang-Hu-2001, Chen-Hou-Hou-2002, Hasebe-Kimura-2003}.  It is known that  the 
 fuzzy four-sphere exhibits intriguing mathematical structure not observed in the fuzzy two-sphere: While the algebra of the fuzzy two-sphere is given by the $SU(2)$ algebra \cite{Hoppe1982,madore1992}, the five coordinates of  the fuzzy four-sphere \cite{Grosse-Klimcik-Presnajder-1996, Castelino-Lee-Taylor-1997} are not closed by themselves within the Lie algebra but bring extra non-commutative coordinates constituting ``internal'' fuzzy structure \cite{Ho-Ramgoolam-2002,Kimura2002,Kimura2003}. Such a peculiar structure makes the studies of  higher dimensional non-commutative geometry more interesting and attractive.  
There are  two ways to represent the fuzzy four-sphere algebraically: 
$(i)$ Lie algebra \cite{Ho-Ramgoolam-2002,Kimura2002,Kimura2003}: the enlarged algebra of the fuzzy four-sphere is  the $SO(6)\simeq SU(4)$  giving rise to fuzzy fibre space (Fig.\ref{fuzzyS4.fig}):   
\be
[X_a, X_b] =i\alpha X_{ab}. \label{xxandxab}
\ee
$(ii)$ Four-Lie bracket \cite{Jabbari2004,DeBellisSS2010} : With the quantum Nambu bracket \cite{Nambu1973,CurtrightZachos2003}, the fuzzy four-sphere coordinates are closed by themselves without introducing extra fuzzy coordinates. The internal structure is implicit,   and the  internal geometry reflects its existence in the degeneracy of (fuzzy) three-sphere latitudes  (Fig.\ref{fuzzyS4.fig}): 
\be
[X_a, X_b, X_c, X_d]=(I+2)\alpha^3 \epsilon_{abcde}X_e. 
\ee

In the previous studies \cite{Hasebe-2014-1,Hasebe-2014-2, Hasebe-2017}, 
we demonstrated that the quantum Nambu geometry actually appear in  the higher dimensional Landau models, and is elegantly intertwined  with exotic  ideas of   differential topology, quantum anomaly, and string theory.    However, 
the deduction of the non-commutative geometry from the Landau models has been rather heuristic and the obtained results are justified in the thermodynamic limit.\footnote{See \cite{Hatusda-Iso-Umetsu-2003, Zhang-Hu-2001, Hasebe-2010} also.}  A rigorous way to derive the non-commutative geometry is accomplished by  the Landau level projection  not resorting to any approximation, and the results will capture every detail  of the emergent non-commutative geometry.  
The Landau level projection method can also be applied to any Landau level  (not limited to the lowest Landau level) whose non-commutative geometry has  rarely been investigated, in contrast to  the Berezin-Toeplitz quantization focused on  zero-modes.    
The practical procedure of the Landau level projection  is quite straightforward: We just sandwich  coordinates  of interest  by  Landau level basis states   to derive their   matrix-valued counterparts  in a given Landau level.  
Since the total Hilbert space of the Landau model is mathematically well-defined, the truncated subspace of the Landau level necessarily provides  a sound formulation of non-commutative geometry. 
Based on this observation, we derived matrix geometries of the $SO(3)$ Landau models \cite{Hasebe-2015} and the $SO(4)$ Landau models \cite{Hasebe-2018}. 
We extend this project to the $SO(5)$ Landau models. 
Not just rendering the similar analysis, we integrate the  previous results with new $SO(5)$ results to  present a comprehensive view of the emergent fuzzy geometry of the Landau models.  We unveil hidden relations between the background  topological field configurations of the Landau models, and also  discuss the matrix geometry of the Landau models in an arbitrary dimension.

This paper is organized as follows. In Sec.\ref{sec:so5landaumodel}, we review the $SU(2)$ monopole and $SO(5)$ Landau problem in a modern terminology.  Using the $SO(5)$ Landau level eigenstates, we derive the matrix geometry of the $SO(5)$ Landau model in Sec.\ref{sec:fourspherematgeo}.  
Sec.\ref{sec:intfuzzystr} discusses the internal fuzzy three-sphere structure  with emphasis on its relation to the $SO(4)$ Landau model. We also clarify relations among the background topological field configurations in low dimensional Landau models. Relativistic version of the $SO(5)$ Landau model and its associated zero-modes are  analyzed in Sec.\ref{sec:relso5landau}. In Sec.\ref{sec:evenhigherdim}, we extend the  matrix geometry  analysis  to  even higher dimensions.   
Sec.\ref{sec:summary} is devoted to summary and discussions.

\section{Review of the Yang's $SU(2)$ monopole system}\label{sec:so5landaumodel}

 In this section, we  review the Yang's $SU(2)$ monopole system \cite{Yang-1978-1,Yang-1978-2} and the Zhang and Hu's $SO(5)$ Landau model \cite{Zhang-Hu-2001}  adding some more information.

\subsection{$SU(2)$ monopole and $SO(5)$ angular momentum operators}

  With stereographic coordinates of $S^3$-latitude on $S^4$, Yang gave an expression of the $SU(2)$ monopole gauge field \cite{Yang-1978-1}.  
 However, the original expression is  rather cumbersome to handle and we then adopt  the Zhang and Hu's concise  notation of the $SU(2)$ (anti-)monopole gauge field \cite{Zhang-Hu-2001}:   
\be
A_m=-\frac{1}{r(r+x_5)}\bar{\eta}_{mn}^i x_n S_i~~(m ,n=1, 2, 3, 4) ,~~~~A_5=0, \label{zhanghusu2gauge}
\ee
where  
$S_i$ $(i=1,2,3)$ denote the $SU(2)$ matrix of the spin $I/2$ representation: 
\be
S_iS_i =\frac{I}{2}(\frac{I}{2}+1)\bs{1}_{I+1}, 
\ee
and $\bar{\eta}_{mn}^i$ signifies  the  't Hooft symbol: 
\be
\eta_{mn}^i\equiv \epsilon_{mn i4} +\delta_{m i}\delta_{n 4}-\delta_{m 4}\delta_{n i}, ~~\bar{\eta}_{mn}^i\equiv \epsilon_{mn i4} -\delta_{m i}\delta_{n 4}+\delta_{m 4}\delta_{n i}. \label{defthooftsymbols}
\ee
The field strength, $F_{ab}=\partial_a A_b-\partial_b A_a +i[A_a, A_b]$ $(a,b=1,2,3,4,5)$, is given by 
\be
F_{mn}=-\frac{1}{r^2}x_m A_n+\frac{1}{r^2}x_n A_m+\frac{1}{r^2}\bar{\eta}^i_{mn}S_i,~~F_{m 5}=-F_{5m}=\frac{1}{r^2}(r+x_5)A_m. 
\label{constfieldstreng}
\ee
The non-trivial homotopy for the $SU(2)$ monopole field configuration on $S^4$ is guaranteed by  
\be
\pi_3(SU(2)) \simeq \mathbb{Z},  
\ee
and the second Chern number associated with (\ref{zhanghusu2gauge}) is evaluated as 
\be
c_2 =  \frac{1}{8\pi^2}\int_{S^4}\tr ~F^2=-\frac{1}{6}I(I+1)(I+2), \label{chernnumberantimo}
\ee
where $F=\frac{1}{2}F_{ab}dx_a\wedge dx_b$. 
We construct the covariant angular momentum operators,  $\Lambda_{ab}$,  as   
\be
\Lambda_{ab}=-ix_aD_b+ix_bD_a, 
\ee
with 
\be
D_a=\partial_a +iA_a,  \label{covariantderi}
\ee
and the total $SO(5)$ angular momentum operators as  
\be
L_{ab}=\Lambda_{ab}+r^2F_{ab}.  \label{angso5tot}
\ee
In detail, 
\be
L_{mn}={{L}}^{(0)}_{mn} +\bar{\eta}_{mn}^i S_i, ~~~L_{m 5} 
=L_{m 5}^{(0)}-\frac{1}{r+x_5}\bar{\eta}^i_{mn}x_n S_i, 
\label{detailantimonoso5gene}
\ee
where $L^{(0)}_{ab}$ denote the  $SO(5)$ free angular momentum operators: 
\be
L^{(0)}_{ab}=-ix_a\partial_b+ix_b\partial_a. 
\ee

\subsection{The $SO(5)$ Casimir operator and $SO(5)$ monopole harmonics}\label{subsec:so5casmono}

In usual textbook derivation of the spherical harmonics,   the polar coordinates are adopted to represent the $SO(3)$ Casimir. In a similar manner, we decompose the $SO(5)$ Casimir operator to the $SO(4)$  part and the remaining azimuthal angle part. We introduce  the polar coordinates of a four-sphere (with unit radius) as    
\begin{align}
&x_1=\sin\xi\sin\chi\sin\theta\cos\phi,~~x_2=\sin\xi\sin\chi\sin\theta\sin\phi,~~x_3=\sin\xi\sin\chi\cos\theta,\nn\\
&x_4=\sin\xi\cos\chi,~~x_5=\cos\xi, \label{polarx1tox5}
\end{align}
where 
\be
0\le \xi \le \pi,  ~~~0\le \chi \le \pi, ~~~0\le \theta \le \pi, ~~~0\le \phi < 2\pi.
\ee
The $SO(5)$ Casimir is expressed by the sum of the  $SU(2)_L\oplus SU(2)_R$ Casimir parts and  $x_5$-part \cite{Yang-1978-2}
\begin{align}
\sum_{a<b=1}^5{L_{ab}}^2&
= -\frac{1}{1-{x_5}^2}\frac{\partial}{\partial x_5} ((1-{x_5}^2)^2\frac{\partial}{\partial x_5})+ 2\frac{1}{1-x_5} \bs{J}^2+2\frac{1}{1+x_5}\bs{K}^2 +{S_i}^2 
\nn\\&=-\frac{1}{\sin^3\xi}\frac{\partial}{\partial\xi} (\sin^3\xi\frac{\partial}{\partial \xi})+ 2\frac{1}{1-\cos\xi} \bs{J}^2+2\frac{1}{1+\cos\xi}\bs{K}^2 +{S_i}^2 , 
\label{so4andx5partcas} 
\end{align}
where $J_i$ and $K_i$ are the $SU(2)_L$ and $SU(2)_R$ operators given by 
\begin{subequations}
\begin{align}
&J_i=\frac{1}{4}{\eta}_{mn}^i L_{mn}=\frac{1}{4}{\eta}_{mn}^i L^{(0)}_{mn}=J_i^{(0)} ,  \label{detailantimonoso4genej}\\
&K_i=\frac{1}{4}\bar{\eta}_{mn}^i L_{mn}=\frac{1}{4}\bar{\eta}_{mn}^i L_{mn}^{(0)} +S_i=K_i^{(0)}+S_i. \label{detailantimonoso4genek}
\end{align}\label{detailantimonoso4gene}
\end{subequations}
Note that the $SU(2)$ (anti-)monopole gauge field does not act to the $SU(2)_L$ operators but  acts to the  $SU(2)_R$ operators only  (\ref{detailantimonoso4genek}),  as if the right $SU(2)$ angular momentum acquires  additional $SU(2)$ spin degrees of freedom.

\subsubsection{The differential equation for the $S^3$-latitude}\label{subsec:diffons3}

\begin{figure}[tbph]
\center
\includegraphics*[width=70mm]{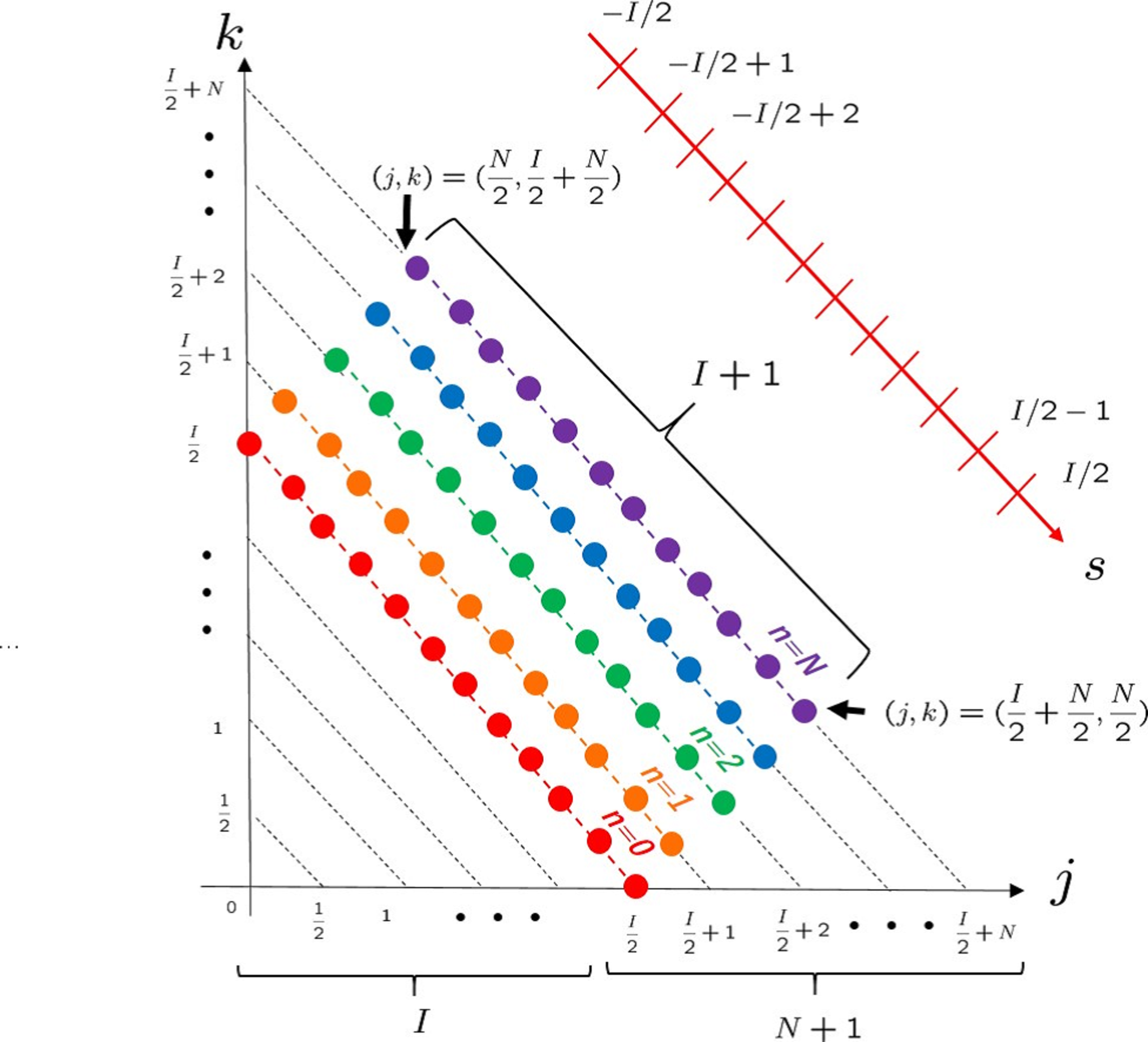}
\caption{Decomposition of the $SO(5)$ irreducible representation $(p,q)_{SO(5)}=(I+N, N )_{SO(5)}$ to the $SO(4)$ irreducible representation $(j,k)_{SO(4)}$.  The oblique lines with fixed  $j+k=n+\frac{I}{2}$ are called the $SO(4)$ lines in the present paper. }
\label{so5diagram.fig}
\end{figure}

Let us first analyze the eigenvalue problem of the $SO(4)$ Casimir operator. 
The $SO(4)$ Casimir eigenstates that satisfy 
\bse
\begin{align}
&\bs{J}^2 Y_{j,m_j; ~k, m_k} (\Omega_3)=j(j+1) Y_{j,m_j; ~k, m_k}(\Omega_3),~~~~J_z Y_{j,m_j; ~k, m_k} (\Omega_3)=m_j Y_{j,m_j; ~k, m_k}(\Omega_3),\\
&\bs{K}^2 Y_{j,m_j; ~k, m_k} (\Omega_3)=k(k+1) Y_{j,m_j; ~k, m_k}(\Omega_3),~~~K_z Y_{j,m_j; ~k, m_k} (\Omega_3)=m_k Y_{j,m_j; ~k, m_k}(\Omega_3),  
\end{align}
\ese
with $\Omega_3\equiv (\chi, \theta,\phi)$  are given by the $SO(4)$ spinor spherical harmonics: 
\be
Y_{j,  m_j;~ k,  m_k}(\Omega_3) =\sum_{m_R=-j}^j\sum_{s_z=-I/2}^{I/2} C_{j, m_R; ~I/2, s_z}^{k, m_k} ~\Phi_{j,m_j;~j, m_R}(\Omega_3)\otimes |I/2, s_z\rangle. 
\label{construcso4eijk}
\ee
Here $C_{j, m_R; ~I/2, s_z}^{k, m_k}$ represent the Clebsch-Gordan coefficients, and $\Phi_{j,~m_j;~j, m_R}(\Omega_3)$ $(j=0, {1}/{2}, 1 , {3}/{2}, \cdots)$ are  
the $SO(4)$ spherical harmonics in the following form \cite{Hasebe-2018}
\be
\Phi_{j,~m_L;~ j,m_R}(\Omega_3)=\sum_{l=0}^p (-i)^l \sum_{m=-l}^l 
C_{\frac{p}{2}, m_L; \frac{p}{2}, m_R}^{l,m}~Y_{plm}(\Omega_3)\biggr|_{p=2j} , 
\label{so4sphericalsu2su2}
\ee
where 
$Y_{plm}(\Omega_3)$ are  the usual $SO(4)$ spherical harmonics \cite{Hochstadt-book}\footnote{The $SO(4)$ spherical harmonics (\ref{so4litspheharmo}) satisfy 
\be
{\bs{J}^{(0)}}^2 Y_{plm}(\Omega_3)={\bs{K}^{(0)}}^2 Y_{plm}(\Omega_3)=\frac{p}{2}(\frac{p}{2}+1)  Y_{plm}(\Omega_3). 
\ee
The dimension of the $SO(4)$ spherical harmonics is given by 
\be
\sum_{l=0}^p (2l+1) =(p+1)^2 = (2j+1)^2|_{j=\frac{p}{2}}. 
\ee
} 
\begin{align}
Y_{p l m}(\Omega_3) &=2^l l!\sqrt{\frac{2(p+1)(p-l)!}{\pi(p+l+1)!}}~\sin^l(\chi)~C_{p-l}^{l+1}(\cos\chi) \cdot Y_{lm}(\theta,\phi). \label{so4litspheharmo}\\
&~~~~~~~~~~~~~(l=0,1,2,\cdots, p~\text{and}~m=-l, -l+1, \cdots, l) \nn 
\end{align}
$C_{p-l}^{l+1}$ denote the Gegenbauer polynomials, and $Y_{lm}(\theta, \phi)$ stand for the $SO(3)$ spherical harmonics:  
\begin{align}
&C_{n}^{\alpha} (x) \equiv 
 \frac{(-2)^n}{n!} \frac{\Gamma(n+\alpha) \Gamma(n+2\alpha)}{\Gamma(\alpha)\Gamma(2n+2\alpha)} (1-x^2)^{-\alpha+\frac{1}{2}}\frac{d^n}{dx^n} [(1-x^2)^{n+\alpha-\frac{1}{2}}], \\
&Y_{l m}(\theta,\phi) = (-1)^m \sqrt{\frac{2l+1}{4\pi}} \frac{(l-|m|)!}{(l+|m|)!} ~
P_l^{|m|}(\cos\theta) e^{im\phi}.  
\end{align}
Since  the (anti-)monopole gauge field only contributes to the $SU(2)_R$ angular momentum operator, in (\ref{so4sphericalsu2su2}) the original $SU(2)_R$ index $j$ of the $SO(4)$ spherical harmonics is contracted with the gauge spin index $I/2$ to form the $SU(2)_R$ composite spin $k$ in (\ref{construcso4eijk}) 
just as  in the case of  
the usual $SU(2)$ angular momentum composition rule. Therefore, $k$ takes 
\be
k= j+\frac{I}{2}, ~j+\frac{I}{2}-1, ~\cdots,~j-\frac{I}{2} \label{rangeofkfromji}
\ee
or 
\be
s \equiv j-k  =\frac{I}{2}, ~\frac{I}{2}-1, \cdots, -\frac{I}{2}. 
\label{jminuskcond}
\ee
$s$ signifies the difference between the left and right $SU(2)$ quantum numbers, and  hence the name the chirality parameter \cite{Hasebe-2014-2, Hasebe-2017}.   
Though $k$ and $j$ are two independent $SU(2)$ group indices,  in the Yang's $SU(2)$ monopole system  the range of $k$ is not arbitrary but restricted as  (\ref{rangeofkfromji}) with a given $j$.   
With $(I+1)$-component vector-like notation of $|I/2, s_z\rangle$ in (\ref{construcso4eijk}), the $SO(4)$ spinor spherical harmonics  can be expressed as  
\be
\bs{Y}_{j,  m_j;~ k,  m_k}(\Omega_3) =\sum_{m_R=-j}^j \begin{pmatrix}
 C_{j, m_R; ~\frac{I}{2}, \frac{I}{2}}^{k, m_k} ~\Phi_{j,m_j;~j, m_R}(\Omega_3) \\
 C_{j, m_R; ~\frac{I}{2}, \frac{I}{2}-1}^{k, m_k} ~\Phi_{j,m_j;~j, m_R}(\Omega_3)\\
\vdots \\
 C_{j, m_R; ~\frac{I}{2}, -\frac{I}{2}}^{k, m_k} ~\Phi_{j,m_j;~j, m_R}(\Omega_3) 
\end{pmatrix}. \label{vectorlikespinsphereharmo}
\ee
From two indices $j$ and $k$, we introduce the $SO(4)$ Landau level index $n$:  
\be
n = j+k-\frac{I}{2}.  \label{defso4landaulevejk}
\ee
$n$  essentially denotes the sum of two $SU(2)$ quantum numbers. With $n$ and $s$, $j$ and $k$ are inversely represented as 
\be
(j , k)_{SO(4)} =(\frac{n}{2}+\frac{I}{4} +\frac{s}{2}, ~\frac{n}{2}+\frac{I}{4} -\frac{s}{2})_{SO(4)}. \label{formulajandk}
\ee
Notice that while 
the (anti-)monopole only acts to the $SU(2)_R$  operator, 
with a given $n$   $j$ and $k$ are totally equivalent in the sense that either of $j$ and $k$ starts from  $n/2$  and ends at $I/2+n/2$, and  the intervals of $n$, $j$ and $k$ are $\Delta n =1$ and $\Delta j=\Delta k =\frac{1}{2}$ (see Fig.\ref{so5diagram.fig}). 

In the view of the $SO(5)$ representation theory,  the $SO(5)$ irreducible representation is specified by  two integers,\footnote{See Appendix \ref{subsec:so5irrepsandso4decomp} for  the $SO(5)$ representation theory and the $SO(4)$ decomposition.}  
\be 
(p,q)_{SO(5)}=(I+N, N)_{SO(5)}.~~~~~(N=0,1,2,\cdots) \label{identpqn}
\ee
The $SO(4)$ decomposition  generally tells  that, 
with a given $N$,  $n$ 
takes 
\be
n=0,1,2,\cdots, N, 
\ee
and the $SO(4)$ decomposition is 
\be
(I+N, N)_{SO(5)} 
=\sum_{n=0}^N \oplus  \biggl(~\sum_{s=-I/2}^{I/2} \oplus ~(j, ~k)_{SO(4)}
\biggr), \label{so5toso4decomp}
\ee
where $j$ and $k$ are given by (\ref{formulajandk}).   
Therefore, the dimension of the $SO(5)$ irreducible representation is obtained as  
\be
D(I, N)=\sum_{n=0}^{N} d(I, n)= \frac{1}{6}(N+1)(I+1)(I+N+2)(I+2N+3), \label{sumso4lldege}
\ee
where 
\be
d(I, n) 
=\sum_{s=-I/2}^{I/2} (2j+1)(2k+1) 
= \frac{1}{6}(I+1)(I^2+(6n+5)I +6(n+1)^2). \label{subso4dimhilb}
\ee 

\subsubsection{Azimuthal part eigenvalue problem}

The $SO(5)$ Casimir operator was decomposed to the azimuthal part $\xi$ and the hyper-latitude $S^3$ part (\ref{so4andx5partcas}). To solve the differential equation of the $SO(5)$ Casimir operator, Yang adopted the  method of separation of variables \cite{Yang-1978-2}: 
\be
\Psi_{j,k}(\xi, \Omega_3)= G(\xi) \cdot Y_{j,k}(\Omega_3).  \label{generalso5monopoharm}
\ee
Here, $Y_{j,k}(\Omega_3)$ denote the $SO(4)$ monopole harmonics (\ref{construcso4eijk}) with the constraint 
\be 
j+k=n+\frac{I}{2}. \label{defnandjk}
\ee 
The $SO(5)$ eigenvalue problem 
\be
\sum_{a<b=1}^5 {L_{ab}}^2 ~\Psi(\xi, \Omega_3) =\lambda ~\Psi(\xi, \Omega_3), 
\ee
is reduced to the eigenvalue equation for $G(\xi)$: 
\be
\biggl[-\frac{1}{\sin^3\xi}\frac{d}{d\xi} (\sin^3\xi\frac{d}{d \xi})+2\frac{1}{1-\cos\xi}j(j+1) + 2\frac{1}{1+\cos\xi} k(k+1)+\frac{I}{2}(\frac{I}{2}+1)\biggr]G(\xi)=\lambda ~G(\xi). 
\label{ordieqg2}
\ee
Yang showed that the difference of two Casimir indices is exactly equal  to the $SU(2)$ monopole index $I$ \cite{Yang-1978-2}: 
\be
p-q=I. 
\ee
Therefore, when we identify  $q$ with the $SO(5)$ Landau level index $N~(=0,1,2,\cdots)$, the $SO(5)$ monopole harmonics\footnote{In \cite{Yang-1978-2},  Yang called the eigenstates (\ref{generalso5monopoharm}) the $SU(2)$ monopole harmonics, but in the present paper  we refer to them as the $SO(5)$ monopole harmonics  emphasizing their $SO(5)$ covariance.} (\ref{generalso5monopoharm}) are specified by the $SO(5)$ indices (\ref{identpqn}). 
The $SO(5)$ Casimir eigenvalues are readily obtained as (see Appendix \ref{subsec:so5irrepsandso4decomp})

\be
\lambda_N=\frac{1}{2}p^2+\frac{1}{2}q^2 +2p+q=N^2+N(I+3)+\frac{1}{2}I(I+4),  
\ee
and  
the corresponding degeneracy  is 
\be
D(I,N)=\frac{1}{6}(p+2)(q+1)(p+q+3)(p-q+1)=\frac{1}{6}(N+1)(I+1)(I+N+2)(I+2N+3), 
\label{so5degelandau}
\ee
which is equal to (\ref{sumso4lldege}). (\ref{so5toso4decomp}) implies that  $N$th $SO(5)$ Landau level consists of  the inner $SO(4)$ Landau levels $(n=0,1,2,\cdots, N)$: 
\be
\mathcal{H}_{SO(5)}^{(N)} =\sum_{n=0}^{N} \oplus ~(\sum_{s=-\frac{I}{2}}^{\frac{I}{2}}\mathcal{H}_{SO(4)}^{(n,s)}).  \label{hilbertsumso45}
\ee
The normalized $SO(5)$ monopole harmonics are derived as  
\be
\bs{\Psi}_{N; j, m_j ;k, m_k}(\Omega_4) 
= G_{N,j,k}(\xi)\cdot \bs{Y}_{j, m_j; k, m_k}(\Omega_3), ~~~~(\Omega_4=(\xi,\chi,\theta,\phi))
 \label{x5so5monopolehamonicsnorm}
\ee
where \cite{Chen-Hou-Hou-2002} 
\begin{align}
G_{N,j,k}(\xi)&=  \sqrt{N+\frac{I}{2}+\frac{3}{2}}~\cdot ~\frac{1}{\sin\xi}~d_{N+\frac{I}{2}+1, j-k, -j-k-1}(\xi)\nn\\
&=\sqrt{N+\frac{I}{2}+\frac{3}{2}}\cdot (-1)^{2j+1}\sqrt{ \frac{(N+\frac{I}{2}+j-k+1)!}{(N+\frac{I}{2}+j+k+2)!}\cdot \frac{(N+\frac{I}{2}-j+k+1)!}{(N+\frac{I}{2}-j-k)!}}~\nn\\
&~~\times \frac{1}{\sin\xi}~(\sin\frac{\xi}{2})^{2j+1}~(\cos\frac{\xi}{2})^{-2k-1} ~P_{N+\frac{I}{2}+1 -j+k}^{2j+1, -(2k+1)}(\cos\xi) . 
\label{exdefgmfunc}
\end{align}
Here $d_{l,m,g}(\xi)$ denotes the Wigner's small $d$-function\footnote{The  small $d$-function can also be expressed as  
\be
d_{l,m,g}(\xi) = (e^{-i\xi S^{(l)}_y})_{m,g}  
\ee 
where $S_y^{(l)}$ denotes  $y$-component of the $SU(2)$ spin matrix with spin magnitude $l$: 
\be
\bs{S}^{(l)} \cdot \bs{S}^{(l)} =l(l+1)~{1}_{2l+1}. 
\ee
} 
and its three indices are identified with  $(l, m ,g)= (N+\frac{I}{2}+1, s, -n-\frac{I}{2}-1)$ as in (\ref{exdefgmfunc}). Note that 
the two magnetic indices, $m$ and $g$, generally take (half-)integer values between $-l$ and $l$, while  in the present case $m=s$ and the  range of $s$ (\ref{jminuskcond}) is restricted to $ |s| \le \frac{I}{2}$ which is smaller than  $l=N+\frac{I}{2}+1$ (except for $N=0$). We thus find that   a subset of  $d$-function is utilized in  (\ref{exdefgmfunc}).     
The orthonormal relation for (\ref{x5so5monopolehamonicsnorm}) is given by 
\begin{align}
&\int d\Omega_4 ~\bs{\Psi}_{N; j, m_j ;k, m_k}(\Omega_4)^{\dagger}~\bs{\Psi}_{N'; j', m'_j ;k', m'_k}(\Omega_4) \nn\\
&=\int_0^{\pi}d\xi \sin^3\xi ~ G_{N, j, k} (\xi)^*~G_{N', j', k'} (\xi) ~\cdot~\int_{S^3}d\Omega_3 ~ \bs{Y}_{j, m_j; k,  m_k}(\Omega_3)^{\dagger}~\bs{Y}_{j', m'_j; k', m'_k}(\Omega_3)  \nn\\
&=\delta_{NN'}\delta_{jj'}\delta_{kk'}\delta_{m_j m_j'}\delta_{m_k m'_k}.
\end{align}
For instance, the $SO(5)$ spinor representation $(N, I)=(0,1)$ is obtained as 
\bse
\begin{align}
&\bs{\Psi}_{0; 1/2, 1/2; 0, 0}(\Omega_4) = -\sqrt{\frac{3}{4\pi^2}}\sin\frac{\xi}{2}\begin{pmatrix}
\cos\chi-i\sin\chi\cos\theta \\
-i\sin\chi\sin\theta e^{i\phi}
\end{pmatrix}  ~\propto~ \bs{\psi}_1\equiv   \frac{1}{\sqrt{2(1+x_5)}}
\begin{pmatrix}
x_4-ix_3 \\
-ix_1+x_2 
\end{pmatrix}
, \nn\\
&\bs{\Psi}_{0; 1/2, -1/2; 0, 0}(\Omega_4)= -\sqrt{\frac{3}{4\pi^2}}\sin\frac{\xi}{2}\begin{pmatrix}
-i\sin\chi\sin\theta e^{-i\phi} \\
\cos\chi+i\sin\chi\cos\theta 
\end{pmatrix} ~\propto~ \bs{\psi}_2\equiv  ~\frac{1}{\sqrt{2(1+x_5)}}
\begin{pmatrix}
-ix_1-x_2 \\
x_4+ix_3 
\end{pmatrix}
, \\
&\bs{\Psi}_{0 ; 0, 0; 1/2, 1/2}(\Omega_4)= -\sqrt{\frac{3}{4\pi^2}}\cos\frac{\xi}{2}
\begin{pmatrix}
1 \\
0
\end{pmatrix} ~\propto~ \bs{\psi}_3\equiv \sqrt{\frac{1+x_5}{2}}
\begin{pmatrix}
1 \\
0
\end{pmatrix}
, \nn\\
&\bs{\Psi}_{0; 0, 0;  1/2, -1/2}(\Omega_4) =-\sqrt{\frac{3}{4\pi^2}}\cos\frac{\xi}{2}
\begin{pmatrix}
0 \\
1
\end{pmatrix} ~\propto~ \bs{\psi}_4\equiv  \sqrt{\frac{1+x_5}{2}}
\begin{pmatrix}
0 \\
1
\end{pmatrix}. 
\end{align}\label{simplestrepso5}
\ese
From  
\be
G_{N; j, k}(x_5) =(1-x_5)^j (1+x_5)^{-k-1} ~\overbrace{P_{N+\frac{I}{2}+1-j+k}^{2j+1, -(2k+1)}(x_5)}^{\sim ~{x_5}^{N+\frac{I}{2}+1-j+k}} ~\sim~ {x_5}^{N+\frac{I}{2}}
\ee
and 
\be
Y_{j,k} (\Omega_3) \sim {y_{\mu}}^{2j} \sim \frac{1}{(1-{x_5}^2)^j} {x_{\mu}}^{2j}, 
\ee
the behavior of  $\Psi_{N; j,k}$ can be read off as  
\be
\Psi_{N; j,k} ~\sim~{x_a}^{N+\frac{I}{2}}.
\ee
At $I=0$, (\ref{x5so5monopolehamonicsnorm}) is reduced to the $SO(5)$ spherical harmonics  as expected (see Appendix \ref{append:so5sphericalharmon}).

\subsection{The $SO(5)$ Landau model}\label{subsec:so5landaumodel}

The $SO(5)$ Landau model \cite{Zhang-Hu-2001} is a Landau model on a four-sphere   in the $SU(2)$ monopole background.  
With the covariant derivatives  $D_a$ (\ref{covariantderi}),  
the $SO(5)$ Landau Hamiltonian is given by 
\be
H=-\frac{1}{2M}\sum_{a=1}^5 {D_a}^2 \biggr|_{r={1}} 
=\frac{1}{2M}\sum_{a<b}{\Lambda_{ab}}^2 , \label{origiso5landauhaml}
\ee
which can be  rewritten as 
\be
H=\frac{1}{2M}\sum_{a<b} ({L_{ab}}^2 -{F_{ab}}^2),   
\ee
where we used $\Lambda_{ab}F_{ab} =F_{ab}\Lambda_{ab}=0$. 
From (\ref{constfieldstreng}),  we can readily derive  
\be 
\sum_{a<b}{F_{ab}}^2=\sum_{m<n}(\bar{\eta}_{mn}^i S_i)^2=2{S_i}^2=\frac{1}{2}I(I+2), 
\ee 
and  the $SO(5)$ Landau Hamiltonian is diagonalized as 
\be
E_N(I) =\frac{1}{2M}( N^2+N(I+3)+I) , 
\ee
with the Landau level degeneracy (\ref{so5degelandau}).  
The Landau level eigenstates are given by the $SO(5)$ monopole harmonics (\ref{x5so5monopolehamonicsnorm}).

The lowest Landau level  degeneracy is given by 
\be
D_{\text{LLL}}(I) \equiv D(I,N=0) =\frac{1}{6}(I+1)(I+2)(I+3), \label{s05llldege}
\ee
which is simply understood as the number of  the fully symmetric representation \cite{Zhang-Hu-2001},  
\be
\frac{1}{\sqrt{m_1! ~m_2 !~ m_3! ~m_4 !}}~{\psi_1}^{m_1} {\psi_2}^{m_2}{\psi_3}^{m_3}{\psi_4}^{m_4} \label{fullsymmrep}
\ee
where  $m_1,m_2, m_3, m_4$ are non-negative integers  subject to 
\be 
m_1+m_2+m_3+m_4=I 
\ee
and $\psi$s denote the components of the 2nd Hopf spinor\footnote{We will discuss the 2nd Hopf map in Sec.\ref{sec:foursphereendhopf}.}  
\be
\begin{pmatrix}
\psi_1 \\
\psi_2 \\
\psi_3 \\
\psi_4 
\end{pmatrix}\equiv  \frac{1}{\sqrt{1+x_5}}
\begin{pmatrix}
x_4 -ix_i{\sigma_i}^*  \\
(1+x_5)1_2  
\end{pmatrix}
\begin{pmatrix}
\phi_1 \\
\phi_2 
\end{pmatrix}. 
\ee
We can see equivalence between  the fully symmetric representation (\ref{fullsymmrep}) and the $N=0$ $SO(5)$  monopole harmonics  as follows.      
With the higher spin basis  
\be
\bs{e}_A^{(I/2)} =\frac{1}{\sqrt{(\frac{I}{2}+A)! (\frac{I}{2}-A)!}} ~{\phi_1}^{\frac{I}{2}+A}{\phi_2}^{\frac{I}{2}-A}, ~~~(A=I/2, I/2-1, \cdots, -I/2)
\ee
(\ref{fullsymmrep}) can be expanded as 
\be
\frac{1}{\sqrt{m_1! m_2 ! m_3! m_4 !}}~{\psi_1}^{m_1} {\psi_2}^{m_2}{\psi_3}^{m_3}{\psi_4}^{m_4} =\sum_{A=-I/2}^{I/2}
{\psi}_{A}^{(m_1,m_2,m_3,m_4)} \bs{e}_A^{(I/2)}.  \label{expfulsym}
\ee
From the expansion coefficients, 
 we can construct a $(I+1)$-component ``vector'' as 
\be 
\bs{\Psi}_{N=0; j, m_j; k, m_k}\equiv  
-\frac{\sqrt{(I+2)(I+3)}}{4\pi}
 \begin{pmatrix}
{\psi}_{I/2}^{(m_1,m_2,m_3,m_4)} \\
{\psi}_{I/2-1}^{(m_1,m_2,m_3,m_4)} \\
\vdots \\
{\psi}_{-I/2}^{(m_1,m_2,m_3,m_4)})
\end{pmatrix}, 
\ee
which is exactly equal to the $N=0$ $SO(5)$ monopole harmonics (\ref{x5so5monopolehamonicsnorm}) under the identification 
\be
j =\frac{1}{2}(m_1+m_2),~~m_j= \frac{1}{2}(m_1-m_2), ~~k=\frac{1}{2}(m_3+m_4),~~m_k=\frac{1}{2}(m_3-m_4). 
\ee

\section{Four-sphere matrix geometry}\label{sec:fourspherematgeo}

In this section, we investigate the matrix geometry of the $SO(5)$ Landau model. First, we discuss a general structure of the matrix geometry deduced from the $SO(5)$ irreducible decomposition rule. Next, we discuss  the lowest Landau level matrix geometry  at the quantum limit $I=1$ and at the classical limit $I>\!>1$. Finally, through the Landau level projection, we explicitly derive the matrix geometry interpolating between these two limits and applicable in  any Landau level.   

\subsection{General form of matrix coordinates}

\begin{figure}[tbph]
\hspace{3cm}
\includegraphics*[width=70mm]{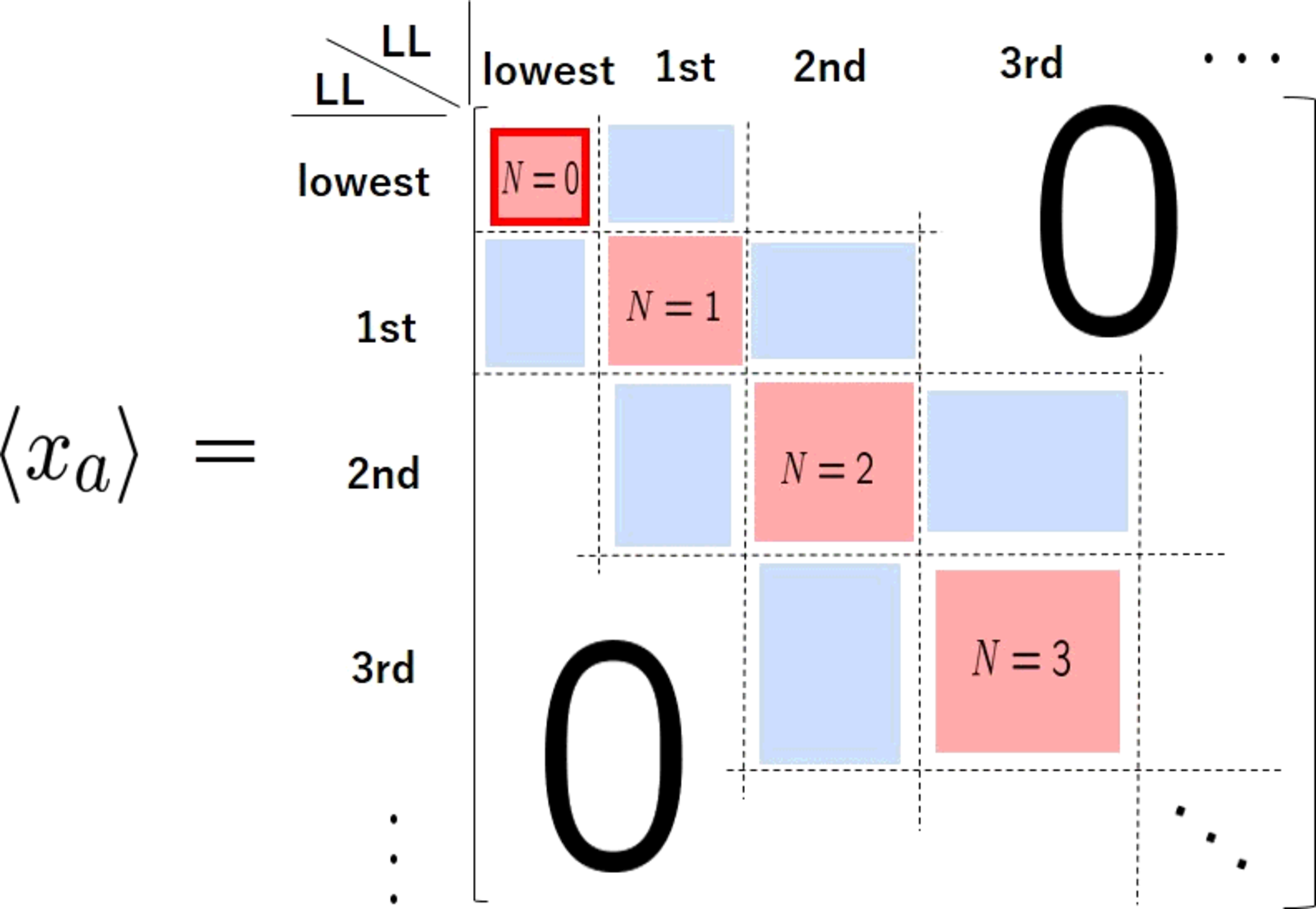}
\caption{The matrix elements of four-sphere coordinates. The shaded regions stand for non-zero blocks: The red shaded squares denote the matrix elements in intra Landau levels, while the blue shaded rectangles represent the matrix elements between inter Landau levels. }
\label{LLsmatrix.fig}
\end{figure}

In this section, we utilize the notation $[[N, I]]$ to specify the $SO(5)$ irreducible representation instead of $(p,q)_{SO(5)} =(N+I, N)$.  
First, let us see a general structure of the matrix elements of the four-sphere coordinates: 
\be
\langle N', \frac{I}{2}|x_a|N, \frac{I}{2}\rangle .  
\ee
Here $|N,\frac{I}{2}\rangle$ is the abbreviation of  the $N$th  Landau level eigenstates (\ref{x5so5monopolehamonicsnorm}), and   the $SO(5)$ vector 
$x_a$ carries the $SO(5)$ index $[[1,0]]$, and hence  
 the $SO(5)$ index of $x_a|N, \frac{I}{2}\rangle$  is  given by 
\be
[[1,0]]\otimes [[N,I]],  \label{tensor10ni}
\ee
which is irreducibly decomposed as \cite{Girardi-Sciarrino-Sorba-1982-1,Vlasii-Rutte-Wiese-2016,Feger-Kephart-2012} 
\be
[[1,0]] \otimes [[N, I]] =  [[N+1, I ]]\oplus [[N-1, I]]\oplus [[N, I]]\oplus[[ N+1, I-2]] \oplus [[ N-1, I+2]],  \label{dcompfavo}
\ee 
where\footnote{(\ref{speequire}) is a special case of more general formula  
\be
[[N+I+1, -I-2]] = -[[N, I]] 
\label{geneformni}
\ee
or 
\be
D_{[[N+I+1, -I-2]]} = -D_{[[N, I]]}. 
\ee
(\ref{geneformni}) is verified by the fact that the $[[N ,I]]$ irreducible representation is specified by the polynomial \cite{Vlasii-Rutte-Wiese-2016} 
\be
\xi(x,y)_{[[N, I]]} =   x^{N+I+2} y^{N+1}-x^{N+1} y^{N+I+2} +\frac{y^{N+I+2}}{x^{N+1}}  -\frac{{y^{N+1}}}{x^{N+I+2}} +\frac{1}{x^{N+I+2} y^{N+1}} -\frac{1}{x^{N+1}y^{N+I+2}}+\frac{x^{N+1}}{y^{N+I+2}}-\frac{x^{N+I+2}}{y^{N+1}}, 
\ee
which has the property 
\be
\xi(x,y)_{([[N+I+1, -I-2]]} = -\xi(x,y)_{[[N, I]]}.
\ee
} 
\bse
\begin{align}
&[[N+1,-2]] \equiv -[[N,0]], \label{speequire}\\
&[[N, -1]] =[[-1, I]] \equiv \phi ~(\text{empty ~set}). 
\end{align}
\ese
See Appendix \ref{directproso5} for  several  examples of (\ref{dcompfavo}). 
The corresponding dimension-counting is given by 
\be
5 \otimes D_{[[N, I]]} = D_{[[N, I]]}\oplus D_{[[N+1, I ]]}\oplus D_{[[N+1, I-2]]} \oplus D_{[[N-1, I+2]]} \oplus D_{[[N-1, I]]}. 
\ee
With  an  $SU(2)$ monopole background fixed $I$,    (\ref{dcompfavo}) implies that the Landau level transition, if occurred, only takes place  between  the adjacent Landau levels:  
\be
 |N, \frac{I}{2}\rangle ~~ \overset{x_a}{\longrightarrow}  ~~ |N+1, \frac{I}{2}\rangle,~~|N-1, \frac{I}{2}\rangle~~|N, \frac{I}{2}\rangle.  \label{transittedstates}
\ee
Consequently, the matrix elements only have finite values between the adjacent inter Landau levels and intra Landau levels:   
\be
\langle N', \frac{I}{2} |x_a | N, \frac{I}{2}\rangle \neq 0 ~~\text{only for}~~ \Delta N \equiv N-N'=0,\pm 1, 
\ee
as depicted in {Fig.\ref{LLsmatrix.fig}}. 

\subsection{The 2nd Hopf map and Bloch four-sphere (quantum limit: $I=1$)}\label{sec:foursphereendhopf}

The Yang's $SU(2)$ monopole is closely related to the 2nd Hopf map \cite{Trautman-1977, Minami-1980, Zhang-Hu-2001}.  Using quaternions $q_m$ $(m=1,2,3,4)$, the 2nd Hopf map, $S^7~\overset{S^3}{\rightarrow}~S^4$, is realized as  
\be
\psi ~~\rightarrow~~\psi^{\dagger}\gamma_a\psi =x_a, ~~~(a=1,2,3,4,5) \label{2ndhopfmap}
\ee
where $\gamma_a$  are  
\be
\gamma_{m}\equiv 
\begin{pmatrix}
0 & \bar{q}_{m} \\
q_{m} & 0 
\end{pmatrix}, ~~~
\gamma_5\equiv  ~\begin{pmatrix}
-1 & 0 \\
0 & 1
\end{pmatrix}, 
\label{fundgammamatrix}
\ee
with $\bar{q}_m=\{-q_i, q_4\}$. 
$\psi$ which we refer to as the 2nd Hopf spinor  is a two-component quaternionic spinor  $\psi=(\psi_1 ~\psi_2)^t$ ($\psi_1$ and $\psi_2$ are quaternions) subject to 
\be
\psi^{\dagger}\psi=1,   
\ee
and signifies the total manifold $S^7$. $x_a$ (\ref{2ndhopfmap}) satisfy the normalization condition $\sum_{a=1}^5{x_a}^2 =(\psi^{\dagger}\psi)^2=1$ and are regarded as the coordinates of the base-manifold $S^4$, while the $S^3$-fibre part of $S^7$ is projected out  in the map (\ref{2ndhopfmap}).   The four-sphere associated with the 2nd Hopf map can be considered as   a 4D version of  the Bloch sphere \cite{Mosseri-Dandoloff-2001}. 
The algebras  of the quaternions implies  that $\gamma_a$ satisfy 
\be
\{\gamma_a, \gamma_b\} =2\delta_{ab} ,  
\ee
and act as the $SO(5)$ gamma matrices. This will be more transparent when we introduce a matrix realization of the quaternions: 
\be
q_{m}=\{-i\sigma_{i=1,2,3}, 1_2\},~~~~~~\bar{q}_{m} =\{i\sigma_{i=1,2,3}, 1_2\}. 
\label{quaternionsmatrix}
\ee
Substituting (\ref{quaternionsmatrix}) to (\ref{fundgammamatrix}), $\gamma_a$ now become the familiar $SO(5)$ $4\times 4$ gamma matrices,   
and the corresponding $SO(5)$ generators are obtained as 
\be
\sigma_{ab} =-i\frac{1}{4}[\gamma_{a}, \gamma_{b}],  
\ee
where 
\be
\sigma_{mn} =\frac{1}{2}
 \begin{pmatrix}
\eta_{mn}^i \sigma_i & 0 \\
0 & \bar{\eta}_{mn}^i \sigma_i
\end{pmatrix},  ~~~~~\sigma_{m 5} = 
i\frac{1}{2}\begin{pmatrix}
0 & -\bar{q}_{m} \\
q_{m} & 0 
\end{pmatrix} \label{so5genespinorrep}
\ee
with the 't Hooft symbols (\ref{defthooftsymbols}). 
The  2nd Hopf spinor $\psi$ is also promoted to  a $4\times 2$ matrix $\Psi$  subject to 
\be
\Psi^{\dagger}\Psi=1_2. 
\ee
The $S^3$-fibre part represents the $SU(2)$ gauge degrees of freedom that acts to $\Psi$ as 
\be
\Psi~\rightarrow~\Psi\cdot g. ~~~(g\in SU(2)) 
\label{gaugetrapsig}
\ee
A possible choice of $\Psi$ is  
\be
\Psi(\Omega_4)=  \frac{1}{\sqrt{2(1+x_5)}}\begin{pmatrix}
x_m \bar{q}_m \\
1+x_5 
\end{pmatrix}=  \frac{1}{\sqrt{2(1+x_5)}}\begin{pmatrix}
x_4 1_2+i x_i \sigma_i \\
(1+x_5)1_2 
\end{pmatrix}.  
\label{2ndhopfmatr}
\ee
Interestingly, (\ref{2ndhopfmatr}) consists of the $N=0$ $SO(5)$ spinor multiplet for $I=1$ (\ref{simplestrepso5}): 
\be
\Psi(\Omega_4)^{\dagger}  =\begin{pmatrix}
\bs{\psi}_1 & 
  \bs{\psi}_2 & 
  \bs{\psi}_3 & 
 \bs{\psi}_4  
\end{pmatrix}. 
\ee
This implies  that the 2nd Hopf map  encodes information of  the  lowest Landau level  of  the  minimum $SU(2)$ monopole index $I=1$, which we   
 call  the quantum limit. For the $SO(5)$ spinors,  
the $SU(2)$ gauge transformation (\ref{gaugetrapsig}) acts  as 
\be
\bs{\psi}_{\alpha}~~\rightarrow~~g^{\dagger} \cdot \bs{\psi}_{\alpha},~~(\alpha=1,2,3,4)
\ee
and the gauge field is given by 
\be
A =-i  \Psi^{\dagger} d \Psi = -i\sum_{\alpha=1}^4 ~\bs{\psi}_{\alpha} ~d\bs{\psi}^{\dagger}_{\alpha}=-\frac{1}{2(1+x_5)}\bar{\eta}_{mn}^ix_n\sigma_i dx_m,  
\ee
which is exactly equal to the Yang's monopole gauge field (\ref{zhanghusu2gauge}) for 
 $I=1$.   
Under the gauge transformation (\ref{gaugetrapsig}), the gauge field is transformed as expected: 
\be
A=-i\sum_{\alpha=1}^4 \bs{\psi}_{\alpha} d\bs{\psi}^{\dagger}_{\alpha}~~\rightarrow~~-i\sum_{\alpha=1}^4 g^{\dagger}\bs{\psi}_{\alpha} (d\bs{\psi}^{\dagger}_{\alpha} \cdot g +\bs{\psi}^{\dagger}_{\alpha}\cdot dg)=g^{\dagger}A g -ig^{\dagger} dg. 
\ee
Including the $SU(2)$ gauge degrees of freedom, 
the 2nd Hopf spinor is generally given by 
\be
\psi(\Omega_4, G)=
\begin{pmatrix}
\psi_{1}(\Omega_4, G) \\
\psi_{2}(\Omega_4, G) \\
\psi_{3}(\Omega_4, G)  \\
\psi_{4}(\Omega_4, G) 
\end{pmatrix}
=\Psi(\Omega_4) ~\begin{pmatrix}
\phi_1 (G)
\\
\phi_2(G)
\end{pmatrix} 
\ee
or 
\be
\psi_{\alpha} (\Omega_4, G)=\sum_{i=1}^2 \Psi_{\alpha i} (\Omega_4)~ \phi_i(G)  . ~~~(\alpha=1,2,3,4)
\ee
Here $\Psi(\Omega_4)$ signifies the base-manifold $S^4$ and $(\phi_1~~\phi_2)^t$ denotes a normalized $SU(2)$ spinor taking its value on the  $S^3$-fibre.  
With some appropriate inner product, we orthonormalize $\phi_{i}$ $(i=1,2)$ as  
\be
\langle \phi_i| \phi_j\rangle =\delta_{ij},  
\ee
and the normalization condition of $\psi$ is restated as 
\be
\Psi(\Omega_4)^{\dagger}\Psi(\Omega_4)=\sum_{\alpha=1}^4 \bs{\psi}_{\alpha} \bs{\psi}_{\alpha}^{\dagger}=1_2. \label{norlargpsi}
\ee

With such a simple set-up,   
we discuss the $SU(2)$ gauge invariance and the $SO(5)$  covariance of the matrix geometry.  
The $SU(2)$ gauge transformation (\ref{gaugetrapsig}) can be reinterpreted  as the transformation of the $\phi$-part:  
\be
\phi_i~~\rightarrow~~g_{ij}\phi_j , 
\ee
while the $SO(5)$ global transformation acts to $\Psi$-part as\footnote{
 $\bs{\psi}_{\alpha=1,2,3,4}$ (\ref{simplestrepso5}) constitute  the lowest Landau level eigenstates 
\be
L_{ab}|_{I=1}\bs{\psi}_{\alpha} =(\sigma_{ab})_{\beta\alpha}\bs{\psi}_{\beta}  , 
\ee
and under the $SO(5)$ transformation generalized by $L_{ab}$, 
they behave as 
\be
\bs{\psi}_{\alpha} ~~\rightarrow~~\bs{\psi}_{\beta}U_{\beta\alpha}, 
\ee
or (\ref{psimattransug}). 
}  
\be
\Psi ~\rightarrow ~U^{\dagger}\Psi, \label{psimattransug}
\ee
where 
\be
U\equiv e^{i\sum_{a<b}\omega_{ab} \sigma_{ab}} 
\ee
with the transformation parameter $\omega_{ab}$. 
We define the  matrix elements of observable $O(\Omega_4)$ as 
\footnote{The factor in front of the integration  is introduced for the normalization  
\be
\frac{2}{A(S^4)}\int_{S^4}d\Omega_4 \Psi\Psi^{\dagger}=1_4.  
\ee
}  
\be
\langle \psi_{\alpha}|O|\psi_{\beta}\rangle \equiv \frac{2}{A(S^4)}\int d\Omega_4 ~\Psi_{\alpha i} ~O ~\Psi_{\beta j}^*~ \langle \phi_j |\phi_i\rangle =\frac{2}{A(S^4)}\sum_{i=1}^2 \int d\Omega_4 ~\Psi_{\alpha i} ~O ~\Psi_{\beta i}^* =\frac{2}{A(S^4)}\int d\Omega_4 ~\bs{\psi}_{\alpha }^{\dagger} ~O ~\bs{\psi}_{\beta }.
\ee
Thus the evaluation of the matrix elements is boiled down to the integration of the operator  sandwiched by the $SO(5)$ spinors.   
In particular,  the matrix elements of $x_{a}$ are given by 
\be
(X_{a})_{\alpha\beta}  =\frac{2}{A(S^4)}
\sum_{i=1}^2 \int d\Omega_4 ~x_a ~\Psi_{\alpha i} ~\Psi_{\beta i}^*  
=\frac{2}{A(S^4)}\int d\Omega_4 ~x_{a} ~\bs{\psi}_{\alpha }^{\dagger} ~\bs{\psi}_{\beta }, 
\label{matemupsppsi}
\ee 
or 
\be
X_{a}=\frac{2}{A(S^4)}\int d\Omega_4 ~x_{a} ~P,  
\label{largexfromx}
\ee
where $P$ denotes a $4\times 4$ projection matrix 
\be
P=\Psi \Psi^{\dagger} =\frac{1}{2}(1+\sum_{a=1}^5 x_a\gamma_a).  
\ee
From (\ref{norlargpsi})  $P^2 
=P$, and $P$ is  invariant under the $SU(2)$ gauge transformation (\ref{gaugetrapsig}). Therefore,  $X_a$ (\ref{largexfromx}) are obviously gauge invariant as they should be. 
From the 2nd Hopf map $x_a =\Psi^{\dagger}\gamma_a\Psi$,    $X_a=\int d\Omega_4 \Psi x_a \Psi^{\dagger}$ can also be represented as 
\be
X_a =\frac{2}{A(S^4)}\int d\Omega_4 ~P~{\gamma_a} ~P .   \label{pgammap}
\ee
Using the formulas 
\be
P~{\gamma_a} ~P =\frac{1}{2}x_a (1+x_b{\gamma_b}), ~~~~
\int_{S^4}d\Omega_4 ~x_a =0, ~~~~\int_{S^4}d\Omega_4 ~x_a x_b =\frac{1}{5}A(S^4)\delta_{ab}, 
\ee 
we can easily evaluate (\ref{pgammap}) as\footnote{
Alternatively, we can  obtain (\ref{normxas}) by  performing the integration  (\ref{matemupsppsi}) with (\ref{simplestrepso5}).    
}  
\be
X_a =\frac{1}{5}{\gamma_a}. \label{normxas}
\ee
Thus in the quantum limit,  the lowest Landau level matrix geometry is  given by the $SO(5)$ gamma matrices (\ref{fundgammamatrix}) up to a proportional factor.  
Under the $SO(5)$ global transformation (\ref{psimattransug}),  
$X_{a}$ are transformed as 
\be
X_{a} =\frac{2}{A(S^4)}\int d\Omega_4 ~x_{a}~\Psi\Psi^{\dagger} ~~\rightarrow~~\frac{2}{A(S^4)}~U^{\dagger}\cdot\int d\Omega_4 ~x_{a}~\Psi\Psi^{\dagger} \cdot U =U^{\dagger} X_{a}U=R_{ab} X_{b}, \label{transxaundu}
\ee
where we used  the $SO(5)$ covariance of the  gamma matrices   
\be
U^{\dagger} {\gamma_{a}} U=R_{ab}\gamma_{b} ~~~~~(R \equiv  e^{i\sum_{a<b}\omega_{ab} \Sigma^{\text{Ad}}_{ab}},~~~\Sigma^{\text{Ad}}_{ab}\equiv -i\delta_{ac}\delta_{bd} +i\delta_{ad}\delta_{bc}).
\ee
(\ref{transxaundu}) indicates that 
the matrix coordinates  transform as an $SO(5)$ vector as expected.

\subsection{Heuristic derivation of the fuzzy geometry (classical limit: $I>\!>1$)}

Next, we consider the opposite limit  $I>\!>1$, which we refer to as the classical limit by the analogy  of quantum spin model $S>\!>1$.  
Refining the heuristic discussions of \cite{Hasebe-2014-1}, 
we will show how the non-commutative geometry takes place in this limit.   

At $I>\!>1$,  the field strength term becomes dominant in  $L_{ab}$ (\ref{angso5tot}): 
\be
L_{ab} ~~\rightarrow~~r^2 F_{ab}. \label{labappfab}
\ee
The coordinates $x_a$ can be extracted  from the $SU(2)$ field strength (\ref{constfieldstreng}) as \cite{Hasebe-2014-1}
\be
\frac{1}{r^{5}} x_a =\frac{2}{4! c_{2}(I)} \epsilon_{a b cde}\tr(F_{b c}F_{de}). 
\label{xafromfff}
\ee
Here $c_2(I)$ denotes the 2nd Chern number (\ref{chernnumberantimo}): 
\be
c_2(I) =-\frac{1}{6}I(I+1)(I+2)=-D(I-1,0).
\ee
Using (\ref{labappfab}), we replace  $F_{ab}$ with $L_{ab}$ in (\ref{xafromfff}) to have  
\be
X_a  ~\sim~\frac{2r}{4! c_{2}(I)} \epsilon_{a bcde} L_{bc}L_{de}~\tr (\bs{1}_{\text{internal~space}}).   
\label{xfromlll}
\ee
Since $L_{ab}$ are the $SO(5)$ operators,  the  coordinates now become  the operators.   
$\tr$ in (\ref{xafromfff}) is taken in the ``internal'' fuzzy space $S_{F}^{2}$  with dimension $I+1$ \cite{Hasebe-2014-1,Hasebe-Kimura-2003}, and so in (\ref{xfromlll})  
\be
\tr (\bs{1}_{\text{internal~space}})= I+1.   
\ee
In the lowest Landau level, we may replace  the $SO(5)$ operators $L_{ab}$ with  the $SO(5)$ matrices $\Sigma_{ab}$ of  the fully symmetric irreducible representation:  
\be
L_{ab}~~\rightarrow~~\Sigma_{ab}, 
\ee
and (\ref{xfromlll}) turns into 
\be
X_a =\frac{2}{4!}\frac{I+1}{ c_{2}(I)} \epsilon_{a b cde} \Sigma_{bc}\Sigma_{de} 
=-\frac{1}{4I(I+2)} \epsilon_{a bcde} \Sigma_{bc}\Sigma_{de} . 
\label{preciseXam}
\ee
Since in the fully symmetric representation $\Sigma_{ab}$ satisfy\footnote{
The gamma matrices in the fully symmetric representation  are constructed as  
\be
\Gamma_a^{(I)} \equiv (\overbrace{\gamma_a \otimes \bs{1} \otimes \bs{1} \cdots \cdots \bs{1}}^{I} + \overbrace{\bs{1}\otimes \gamma_a \otimes \bs{1} \cdots \cdots \bs{1}}^{I}+\cdots +  \overbrace{\bs{1} \otimes \bs{1} \cdots \cdots \bs{1} \otimes \gamma_a }^{I}   )_{sym.}, 
\ee
which  satisfy 
\begin{subequations}
\begin{align}
&\sum_{a=1}^5 {\Gamma_a^{(I)}}\Gamma_a^{(I)} = I(I+4) \cdot \bs{1}_{\frac{1}{6} (I+1) (I+2)(I+3)},  \\
&[\Gamma_{a}^{(I)}, \Gamma_{b}^{(I)}, \Gamma_{c}^{(I)}, \Gamma_{d}^{(I)}]  = 8(l+2) \epsilon_{a b  c d e}\Gamma_{e}^{(I)}. 
\label{propgammas}
\end{align}
\end{subequations}
In this paper, we will drop $(I)$ on the shoulder of $\Gamma_a^{(I)}$ for brevity otherwise stated. }  
\be
\epsilon_{a bcde} \Sigma_{bc}\Sigma_{de} = -2(I+2)\Gamma_a, 
\ee
(\ref{preciseXam})  is greatly simplified as 
\be
X_a =\frac{1}{I}~\Gamma_a. 
\label{xagammaa}
\ee
Therefore in the classical limit,  the lowest Landau level matrix coordinates are  given by the $SO(5)$ gamma matrices in the fully symmetric representation.  

 From (\ref{xagammaa}), we have 
\be
[X_a, X_b] =i(\frac{2}{I})^2 \Sigma_{ab},   \label{xxsigab}
\ee
and around the north-pole $X_5 =\frac{1}{I} \Gamma_5 ~\sim~ -\bs{1}_{I+1}$,\footnote{$\Gamma_{5}$  is   a $D(I,0) \times D(I,0)$ block diagonal matrix whose most upper-left/lower-right block is given by  $\mp I \cdot \bs{1}_{I+1}$.} (\ref{xxsigab}) is reduced to 
\be
[X_{n}, X_{n}] =i(\frac{2}{I})^2 ~\Sigma_{mn} ~~~
{\sim} ~~~i(\frac{2}{I})^2 ~\eta_{mn}^iS_i, 
\ee
which realizes the non-commutative algebra of Zhang and Hu \cite{Zhang-Hu-2001}.

\subsection{Landau level projection and matrix geometry (arbitrary $I$ and $N$) }\label{subsec:matrixgeoarbitrary}

We  have obtained the matrix geometry either at the quantum limit and at the classical limit. Here, we apply the Landau level projection to derive more general results.   
The explicit form of the  $SO(5)$ monopole harmonics is crucial in 
 the analysis.

\subsubsection{Landau level matrix elements}

We perform integrations in the azimuthal part and the $S^3$-latitude part separately. 
The $S^4$-coordinates are decomposed to the azimuthal part and the $S^3$-latitude part:
\be
x_{m} =\sin\xi ~{y}_{m} ,~~x_5=\cos\xi,    
\ee
where $x_m$  are expressed by the product of the radius of $S^3$-latitude and  the (normalized) $S^3$-coordinates: 
\be
y_1=\sin\chi\sin\theta\cos\phi, ~~y_2=\sin\chi\sin\theta\sin\phi, ~~y_3=\sin\chi\cos\theta, ~~y_4=\cos\chi.       
\label{expliys}
\ee 
The area element of $S^4$ is expressed as 
\be
d\Omega_4 =d\xi \sin^3\xi ~d\Omega_3 , 
\ee
with the  $S^3$ area element 
\be
d\Omega_3 =\sin^2\chi \sin\theta ~d\chi ~d\theta ~d\phi. 
\ee
For instance, an integration on $S^4$ is carried out as 
\begin{align}
&\langle \Psi_{N; j', m'_j; k', m'_k} |\Psi_{N; j, m_j; k, m_k}\rangle = \langle G_{N, j',k'} |G_{N, j, k} \rangle \cdot \langle Y_{j', m'_j; k', m'_k}| Y_{j, m_j; k, m_k} \rangle \nn\\
&= \int_0^{\pi} d\xi ~\sin^3\xi~ G_{N,j',k'}(\xi)^* ~G_{N,j,k}(\xi) \cdot \int d\Omega_3 ~ \bs{Y}_{j',m'_j; k', m'_k}(\Omega_3)^{\dagger} ~\bs{Y}_{j,m_j; k, m_k}(\Omega_3). 
\end{align}

\begin{figure}[tbph]
\center
\hspace{-1.0cm}
\includegraphics*[width=100mm]{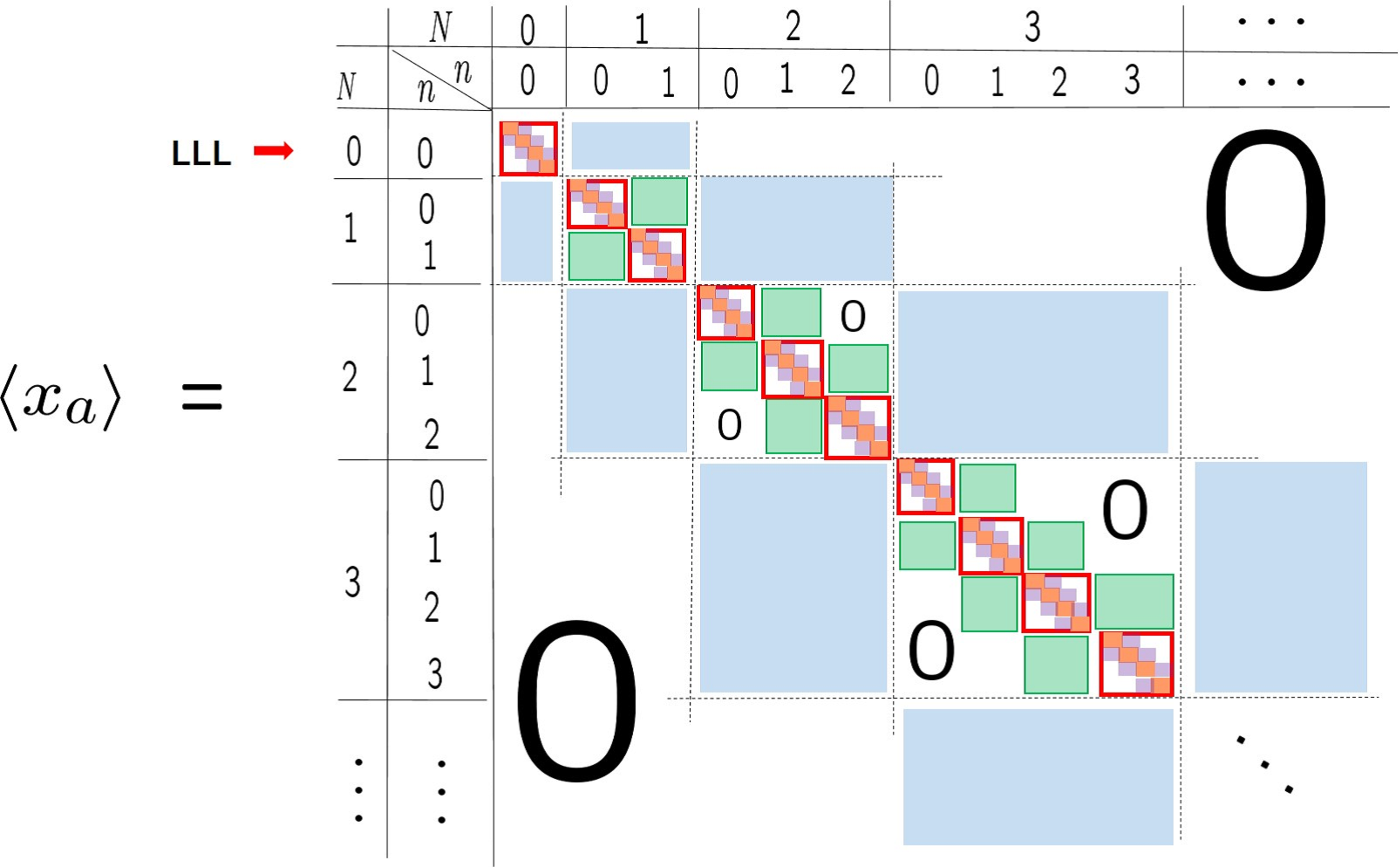}
\caption{Matrix coordinates for $I=3$. There are non-zero matrix elements in the shaded color regions. The blue, green, purple shaded regions are specified by $\Delta N =\pm 1$, $(\Delta n, \Delta N)=(\pm 1, 0)$ and $(\Delta s, \Delta n, \Delta N)= (\pm 1, 0, 0)$, respectively.  The red shaded regions correspond to $\Delta N=\Delta n=\Delta s=0$.  The red-framed squares (with inner red and purple squares) denote the $SO(4)$ Landau level subspaces. Obviously, the matrix geometry  exhibits a nesting structure.   }
\label{nestcoord.fig}
\end{figure}

As discussed in Sec.\ref{subsec:so5casmono}, the $N$th $SO(5)$ Landau level consists of  inner $SO(4)$ Landau levels with  $n=0,1,2,\cdots N$.  
In the $SO(4)$ language, $x_m$ acts as a vector with the  $SO(4)\simeq SU(2)_L\otimes SU(2)_R$ index $(j,k)=(1/2, 1/2)$  and $x_5$ acts as a  scalar with $(j,k)=(0,0)$. For the $SO(4)$ Landau level index $n$ (\ref{defnandjk}) and the chirality parameter $s$ (\ref{jminuskcond}),  the differences are represented as  $\Delta n =\Delta j +\Delta k$ and $\Delta s =\Delta j-\Delta k$, and then the $SO(4)$ selection rule  tells that 
 the matrix coordinates have non-zero values  only for the cases     
\begin{subequations}
\begin{align}
&\langle x_m \rangle ~:~(\Delta n, \Delta s) =(\pm 1, 0), ~~(0, \pm 1), \label{transxm14}\\
&\langle x_5 \rangle ~:~(\Delta n, \Delta s) =(0, 0). 
\end{align}
\end{subequations}
Regions of the non-zero matrix elements are  depicted in Fig.\ref{nestcoord.fig}  that expresses fine internal  structures of Fig.\ref{LLsmatrix.fig}. 
With this in mind, we shall evaluate the matrix elements of $x_5$ and $x_m$.

\begin{itemize}
\item {Matrix coordinates for $x_5$}
\end{itemize}

The matrix elements of $x_5$ are diagonalized as  
\begin{align}
\langle \Psi_{N; j', m'_j; k', m'_k}|x_5 |\Psi_{N; j, m_j; k, m_k}\rangle& = \langle G_{N, j',k'} |x_5|G_{N, j, k} \rangle \cdot \langle Y_{j', m'_j; k', m'_k}| Y_{j, m_j; k, m_k} \rangle\nn\\
&= \langle G_{N, j,k} |x_5|G_{N, j, k} \rangle \cdot  \delta_{j, j'}\delta_{k, k'}\delta_{m_j, m'_j}\delta_{m_k, m'_k}, \label{expx5mat}
\end{align}
with 
\begin{align}
\langle G_{N, j,k} |x_5|G_{N, j, k} \rangle &=
(N+\frac{I}{2}+\frac{3}{2}) \int_0^{\pi} d\xi ~\sin\xi~d_{N+\frac{I}{2}+1, s, -n-\frac{I}{2}-1}(\xi)~\cos\xi~d_{N+\frac{I}{2}+1, s, -n-\frac{I}{2}-1}(\xi)
\nn\\
&=-\frac{2n+I+2}{(2N+I+2)(2N+I+4)} \cdot 2s ,   
\label{impint}
\end{align}
where we used  (\ref{exdefgmfunc}) and  a formula for the small $d$-function.\footnote{
\be
\int_0^{\pi} d\theta ~\sin\theta~d_{l, m, g}(\theta)~\cos\theta~d_{l, m, g}(\theta)
=\frac{2g}{l(l+1)(2l+1)}m. 
\ee
} 
The matrix coordinate (\ref{impint}) takes equally spaced discrete values specified by the chiral parameter $s=I/2, I/2-1, \cdots, -I/2$, which are regarded as latitudes of a fuzzy four-sphere.   Such a structure is very similar to that of  the fuzzy two-sphere \cite{Hasebe-2015}, but  
 while the latitudes of fuzzy two-sphere are not degenerate,  the latitudes of fuzzy four-sphere are  degenerate giving rise to the internal structure.

\begin{itemize}
\item{Matrix coordinates for $x_{m=1,2,3,4}$}
\end{itemize}

As indicated by (\ref{transxm14}),  there are two cases in which  $\langle x_m \rangle$ take  finite values.  The first case is  $(\Delta n, \Delta s) =(\pm 1, 0)$ representing  transition between two adjacent $SO(4)$ Landau levels (two adjacent $SO(4)$ lines in Fig.\ref{so5diagram.fig})  corresponding to the green shaded regions in Fig.\ref{nestcoord.fig}, while the second case 
\be 
(\Delta n, \Delta s) =(0, \pm 1)
\ee
represents transition between the two adjacent sub-bands specified by $s$ inside a $SO(4)$ Landau level  (two adjacent dots on an identical $SO(4)$ line in Fig.\ref{so5diagram.fig})  corresponding to the small purple shaded regions in Fig.\ref{nestcoord.fig}.   
 In the following, we focus on the second case, which  in the language of  
the $SU(2)_L\otimes SU(2)_R$ corresponds to   
\be
j'=j+\frac{\sigma}{2},~~k'=k-\frac{\sigma}{2}. ~~~(\sigma=+,~ -) 
\label{j'-k'=j-k}
\ee
Under the condition (\ref{j'-k'=j-k}), we have 
\begin{align}
&\langle \Psi_{N; j', m'_j; k', m'_k}|x_{m} |\Psi_{N; j, m_j; k, m_k}\rangle \nn\\
&~~~~~~~~~~~~
= \sum_{\sigma=+,-} \langle G_{N, j+\frac{\sigma}{2},k-\frac{\sigma}{2}} |\sin \xi|G_{N, j, k} \rangle~\cdot  ~Y_{m}^{(\sigma, -\sigma)}(j, ~k)_{(m_j',m'_k; ~m_j, m_k)}~\cdot~ \delta_{j',j+\frac{\sigma}{2}}~\delta_{k',k-\frac{\sigma}{2}} , \label{matrixelexmu} 
\end{align}
where 
\be 
Y_{m}^{(\sigma, -\sigma)}(j,k)_{(m_j',m'_k;~m_j, m_k)} \equiv \langle \bs{Y}_{j+\frac{\sigma}{2}, m'_j; k-\frac{\sigma}{2}, m'_k}|{y}_{m}| \bs{Y}_{j, m_j; k, m_k}\rangle. 
\label{evalcmu} 
\ee
$Y_{m}^{(\sigma,-\sigma)}(j,k)$ are regarded as  $(2j+\sigma+1)(2k-\sigma+1)\times (2j+1)(2k+1)$ rectangular matrices with magnetic indices $(m_j', m_k';  m_j, m_k)$, and  $Y_{m}^{(\sigma,-\sigma)}(j,k)$  and $Y_{m}^{(-\sigma,\sigma)}(j+\frac{\sigma}{2}, k-\frac{\sigma}{2})$ are in the relation of Hermitian conjugate. 
 We can evaluate the $S^3$-radius part of  (\ref{matrixelexmu}) as   
\begin{align}
\langle G_{N, j+\frac{\sigma}{2},k-\frac{\sigma}{2}} |\sin \xi|G_{N, j, k} \rangle 
 &=(N+\frac{I}{2}+\frac{3}{2})\int_0^{\pi} d\xi ~\sin^2\xi~d_{N+\frac{I}{2}+1, s', -n-\frac{I}{2}-1}(\xi)~d_{N+\frac{I}{2}+1, s, -n-\frac{I}{2}-1}(\xi)\biggr|_{s'=s+\sigma} \nn\\
&=-\frac{2n+I+2}{(2N+I+2)(2N+I+4)} \cdot  2\sqrt{(N+\frac{I}{2}-\sigma s +1 )(N+\frac{I}{2}+\sigma s +2) }. 
\label{impintt}
\end{align}
In the last equation, we used another formula of the small $d$-function.\footnote{ 
\be
\int_0^{\pi} d\theta ~\sin\theta~d_{l, m', g}(\theta)~\sin\theta~d_{l, m, g}(\theta)|_{ m'= m\pm 1}  
=\frac{2g}{l(l+1)(2l+1)}  \sqrt{(l\mp m)(l\pm m+1)} . 
\ee
} 
Next, we turn to the unit-$S^3$ part (\ref{evalcmu}).  
Notice first that  $y_m$ (\ref{expliys})  can be expressed by  the $SO(4)$ spherical harmonics (\ref{so4sphericalsu2su2}):   
\begin{align}
&{y}_1=-i\frac{\pi}{2}({\Phi}_{\frac{1}{2}, \frac{1}{2}; \frac{1}{2}, \frac{1}{2}}  - {\Phi}_{\frac{1}{2}, -\frac{1}{2}; \frac{1}{2}, -\frac{1}{2}} ), ~~~~{y}_2=-\frac{\pi}{2}({\Phi}_{\frac{1}{2}, \frac{1}{2}; \frac{1}{2}, \frac{1}{2}}  + {\Phi}_{\frac{1}{2}, -\frac{1}{2}; \frac{1}{2}, -\frac{1}{2}}  ), \nn\\
&{y}_3=i\frac{\pi}{2}({\Phi}_{\frac{1}{2}, \frac{1}{2}; \frac{1}{2}, -\frac{1}{2}}  +  {\Phi}_{\frac{1}{2}, -\frac{1}{2}; \frac{1}{2}, \frac{1}{2}}), ~~~~~~~{y}_4=\frac{\pi}{2}({\Phi}_{\frac{1}{2}, \frac{1}{2}; \frac{1}{2}, -\frac{1}{2}}  - {\Phi}_{\frac{1}{2}, -\frac{1}{2}; \frac{1}{2}, \frac{1}{2}}) .  
\label{s3coordinatessphericalrel}
\end{align}
With an integration formula for the $SO(4)$ spherical harmonics, 
 a bit of calculation (see Appendix \ref{appendix:so4appendspin}) shows
\footnote{
Similarly, 
\begin{align}
Y_{m=1,2}^{(-+)}(j,k) &=-(-i)^m \frac{1}{2} (-1)^{n+I}  \begin{Bmatrix}  
j-\frac{1}{2} & k+\frac{1}{2} & \frac{I}{2} \\
k & j & \frac{1}{2}
\end{Bmatrix}\nn\\
&\times  (\delta_{m_j', m_j+\frac{1}{2}}\delta_{m_k', m_k+\frac{1}{2}} \sqrt{(j-m_j)(k+m_k+1)} -(-1)^{m} \delta_{m_j', m_j-\frac{1}{2}}\delta_{m_k', m_k-\frac{1}{2}} \sqrt{(j+m_j)(k-m_k+1)} ), \nn\\
Y_{m=3,4}^{(-+)}(j,k) &=-(-i)^m \frac{1}{2} (-1)^{n+I}  \begin{Bmatrix}  
j-\frac{1}{2} & k+\frac{1}{2} & \frac{I}{2} \\
k & j & \frac{1}{2}
\end{Bmatrix}\nn\\
&\times  (\delta_{m_j', m_j+\frac{1}{2}}\delta_{m_k', m_k-\frac{1}{2}} \sqrt{(j-m_j)(k-m_k+1)} +(-1)^{m} \delta_{m_j', m_j-\frac{1}{2}}\delta_{m_k', m_k+\frac{1}{2}} \sqrt{(j+m_j)(k+m_k+1)} ). 
\label{explfcmu2}
\end{align}
}  
\begin{align}
Y_{m=1,2}^{(+-)}(j,k) &=(-i)^m \frac{1}{2} (-1)^{n+I}  \begin{Bmatrix}  
j+\frac{1}{2} & k-\frac{1}{2} & \frac{I}{2} \\
k & j & \frac{1}{2}
\end{Bmatrix}\nn\\
&\times  (\delta_{m_j', m_j+\frac{1}{2}}\delta_{m_k', m_k+\frac{1}{2}} \sqrt{(j+m_j+1)(k-m_k)} -(-1)^{m} \delta_{m_j', m_j-\frac{1}{2}}\delta_{m_k', m_k-\frac{1}{2}} \sqrt{(j-m_j+1)(k+m_k)} ), \nn\\%
Y_{m=3,4}^{(+-)}(j,k) &=-(-i)^m \frac{1}{2} (-1)^{n+I}  \begin{Bmatrix}  
j+\frac{1}{2} & k-\frac{1}{2} & \frac{I}{2} \\
k & j & \frac{1}{2}
\end{Bmatrix}\nn\\
&\times  (\delta_{m_j', m_j+\frac{1}{2}}\delta_{m_k', m_k-\frac{1}{2}} \sqrt{(j+m_j+1)(k+m_k)} +(-1)^{m} \delta_{m_j', m_j-\frac{1}{2}}\delta_{m_k', m_k+\frac{1}{2}} \sqrt{(j-m_j+1)(k-m_k)} ). 
\label{explfcmu1}
\end{align}
%
$Y_{m}^{(\sigma, -\sigma)}(j,k)$  do not depend on the $SO(5)$ Landau level $N$ and denote the matrix coordinates of three-sphere as we shall discuss in Sec.\ref{sec:intfuzzystr}.
The matrix coordinates (\ref{matrixelexmu}) are  thus completely determined as   
 (\ref{impintt}) and  (\ref{explfcmu1}) in any Landau level. 

\subsubsection{Fuzzy four-sphere in the lowest Landau level }

\begin{figure}[tbph]
\center
\includegraphics*[width=160mm]{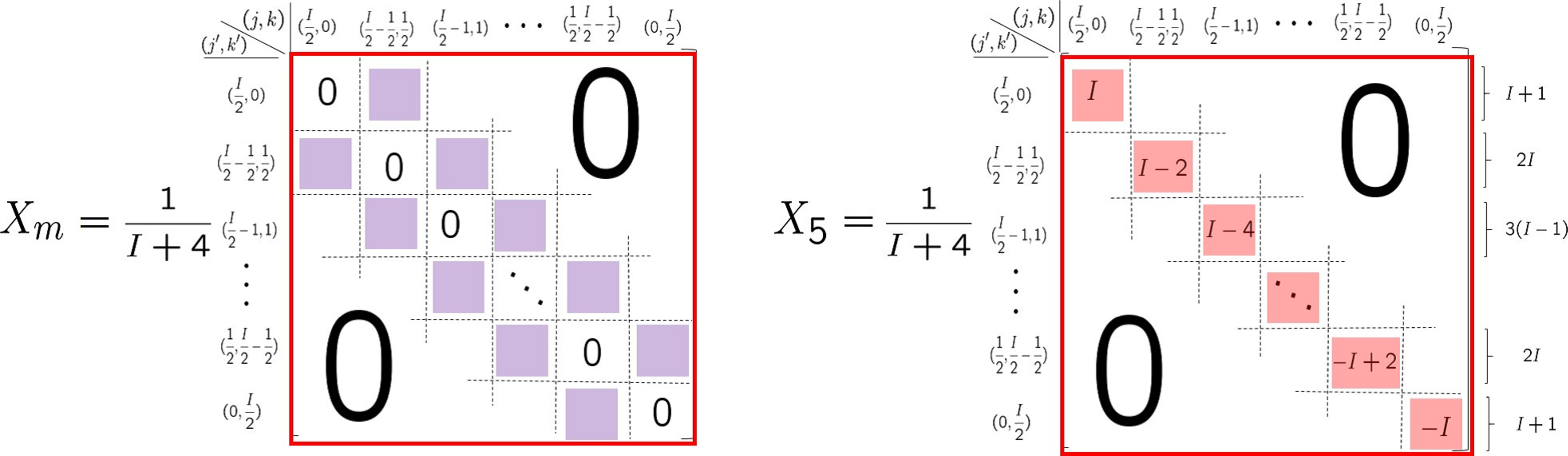}
\caption{The lowest Landau level matrix coordinates: the  red-framed square of the top leftmost in Fig.\ref{nestcoord.fig}. }
\label{matele.fig}
\end{figure}

With the general results above, 
the lowest Landau level $(N=n=0)$   matrix coordinates  are readily obtained as\footnote{
In the special case $s=\frac{1}{2}\sigma$, (\ref{generadoxmu14}) becomes 
\be
\langle \Psi_{N=0; j'=j+\frac{\sigma}{2}, m'_j; k'=k-\frac{\sigma}{2}, m'_k}|x_{m} |\Psi_{N=0; j, m_j; k, m_k}\rangle =-\frac{1}{I+4} ~(I+3)~Y_m^{(\sigma,-\sigma)}(j,k). 
\ee
$Y_m^{(\sigma,-\sigma)}(j,k)$ realizes the matrix for the fuzzy three-sphere \cite{Hasebe-2018}.}   
\begin{subequations}
\begin{align}
&X_{m=1,2,3,4}\equiv \langle \Psi_{N=0; j', m'_j; k', m'_k}|x_{m} |\Psi_{N=0; j, m_j; k, m_k}\rangle = -\frac{2}{I+4}~\times 
\nn\\
&\biggl( \sqrt{(\frac{I}{2}+s+1)( \frac{I}{2}-s+2)}~~ Y_{m}^{(+,-)}(j,k) ~\delta_{j',j+\frac{1}{2}} \delta_{k',k-\frac{1}{2}} 
+\sqrt{(\frac{I}{2}-s+1)( \frac{I}{2}+s+2)}~~ Y_{m}^{(-,+)}(j,k) ~\delta_{j',j-\frac{1}{2}} \delta_{k',k+\frac{1}{2}} \biggr), 
\label{generadoxmu14} \\
&X_5\equiv \langle \Psi_{N=0; j', m'_j; k', m'_k}|x_5 |\Psi_{N; j, m_j; k, m_k}\rangle=-\frac{2}{I+4} ~ s  ~\delta_{j, j'}\delta_{k, k'}\delta_{m_j, m'_j}\delta_{m_k, m'_k}. \label{lllx5matrixelemnts}
\end{align}\label{consmatxa}
\end{subequations}
(\ref{consmatxa}) is concisely expressed as 
\be
X_a 
=\frac{1}{I+4}\Gamma_a.   \label{xalllmat}
\ee
where $\Gamma_a$  denote  the $SO(5)$ gamma matrices in the fully symmetric representation $(p,q)=(I, 0)$. 
See Fig.\ref{matele.fig} also.   
(\ref{xalllmat})  interpolates between the  quantum limit at $I=1$ (\ref{normxas}) and the classical limit at $I >\!>1$ (\ref{xagammaa}).  
 The matrix geometry (\ref{xalllmat})  realizes the quantum Nambu geometry of the  fuzzy four-sphere \cite{Jabbari2004,DeBellisSS2010}: 
\begin{subequations}
\begin{align}
&\sum_{a=1}^5 X_a X_a = {\frac{I}{I+4}} \cdot \bs{1}_{\frac{1}{6}(I+1)(I+2)(I+3) } 
\label{condfuzzs4rad}, \\
&[X_{a}, X_{b}, X_c, X_{d}] = ~(I+2)~(\frac{2}{I+4})^{3}~\epsilon_{a bcde}X_{e},   \label{nambualgebra5}
\end{align}\label{deffuzzy4sph}
\end{subequations}
where $[\cdots]$ of (\ref{nambualgebra5}) signifies the quantum Nambu bracket \cite{Nambu1973,CurtrightZachos2003}, 
\be
[O_1, O_2, \cdots, O_{2k}] \equiv \epsilon_{\mu_1 \mu_2 \cdots \mu_{2k}}O_{\mu_1}O_{\mu_2} \cdots O_{\mu_{2k}}. \label{defofquantumnambubrack}
\ee
In the thermodynamic limit $I\rightarrow \infty$, (\ref{condfuzzs4rad}) is reduced the condition of a four-sphere with unit radius. 

Since $X_5$ is a diagonal matrix with eigenvalues of equal spacing  (\ref{lllx5matrixelemnts}), the fuzzy four-sphere geometry can be regarded as a stacking of the matrix-valued three-spheres along $x_5$-axis with equal spacing  as depicted in  Fig.\ref{fuzzyS4.fig}.  
One may wonder  the stacking along the $x_5$-axis might break the $SO(5)$ symmetry of the four-sphere. However, this is not the case. 
Recall   that we have adopted  $x_5$ as a special axis. 
  If we had chosen $x_1$ as a special axis, we would have had the stack  along the $x_1$-axis.   
Therefore, the picture of the stack   along $x_5$-axis is a kind of ``gauge-artifact'' by choosing  $x_a$ as a special axis in $\mathbb{R}^5$,   and  the fuzzy four-sphere certainly respects the  $SO(5)$ symmetry.

\subsubsection{Nested matrix geometry in higher Landau levels }

\begin{figure}[tbph]
\center
\includegraphics*[width=60mm]{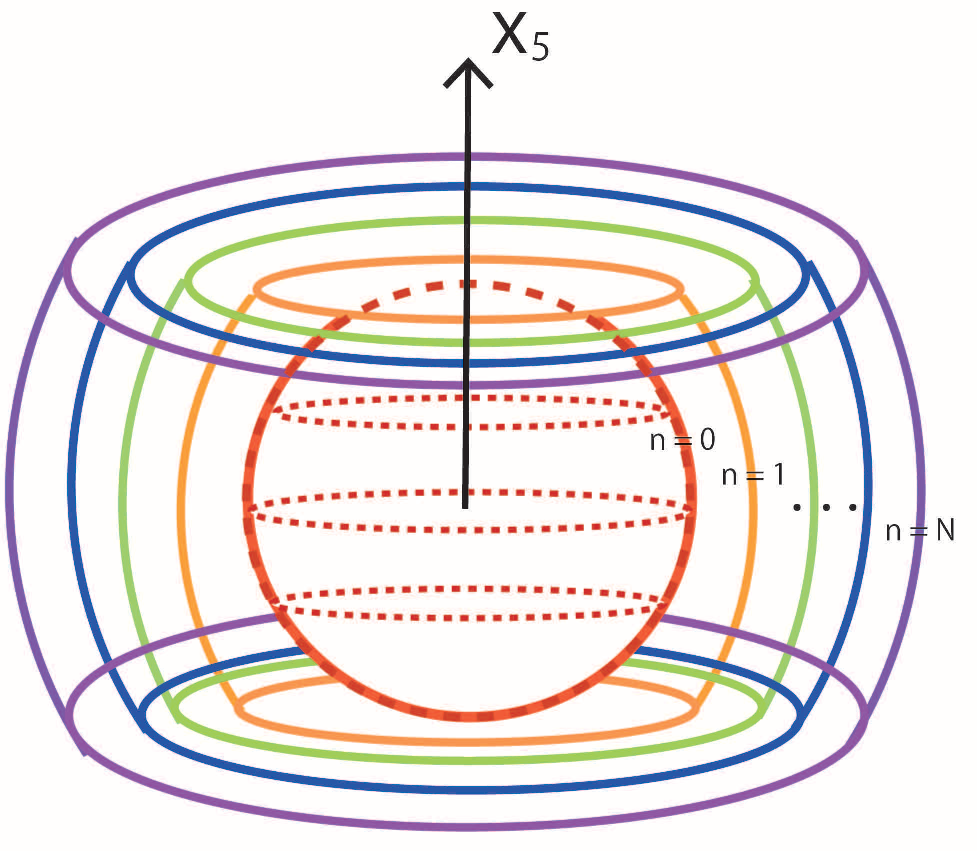}
\caption{ The nested fuzzy structure in the $SO(5)$ Landau level. There are $N+1$ quasi-fuzzy four-spheres indexed by $n$ 
in the $N$th $SO(5)$ Landau level. The colors of the quasi-fuzzy four-spheres correspond to 
the $SO(4)$ lines in Fig.\ref{so5diagram.fig}.    }
\label{nestedspheres.fig}
\end{figure}

Let us consider the matrix geometry in higher $SO(5)$ Landau levels. With a given $SO(5)$  Landau level $N$, there are $N+1$ inner $SO(4)$ Landau levels indexed by $n=0,1,2,\cdots, N$, and further in each of the $SO(4)$ Landau levels there are $I+1$ sub-bands  indexed by the chiral parameter $s$. 
Each sub-band $s$ realizes the matrix-valued  $S^3$-latitude, and a stack of such  $(I+1)$ matrix-valued $S^3$-latitudes  along the $x_5$-axis constitute   
 a fuzzy 4D geometry in each of the $SO(4)$ Landau levels. Therefore inside the $N$th $SO(5)$ Landau level,  there are $N+1$ fuzzy 4D geometries  that form a nested structure as a whole  [Fig.\ref{nestedspheres.fig}].  
Recall that the range of the chiral parameter $s$   is restricted to $|s|=\frac{I}{2}$  and does not cover  the whole range of the matrix size specified by $j+k =n+\frac{I}{2}$ (except for $n=0$). 
This implies that 
the corresponding matrix geometry  is not a complete fuzzy four-sphere but a  fuzzy four-sphere like geometry with  north and south ``caps'' removed  due to   the uncovered parameter regions of $s$.  We referred to such a fuzzy 4D geometry as the quasi-fuzzy four-sphere. 
Each $SO(4)$ Landau level accommodates  a quasi-fuzzy four-sphere geometry,  and so  $N$th $SO(5)$ Landau level realizes  $N+1$ quasi-fuzzy four-spheres with different matrix size depending on the $SO(4)$ index $n$. In this way, $N+1$ quasi-fuzzy four-spheres exhibit a concentric nested structure in the $N$th $SO(5)$ Landau level as depicted in Fig.\ref{nestedspheres.fig}. 
 The lowest Landau level ($N=n=0$) is exceptional, because the nested structure no longer exists and  only a fuzzy four-sphere geometry remains.   

The $SO(4)$  irreducible representations on an $SO(4)$ line 
(Fig.\ref{so5diagram.fig}) constitute each  quasi-fuzzy four-sphere,  which does not possess the $SO(5)$ covariance. However as an aggregation of the quasi-fuzzy four-spheres, the nested fuzzy geometry possesses the $SO(5)$ covariance since its quantum states are realized by an $SO(5)$ irreducible representation (or the $SO(5)$ monopole harmonics).  There exist non-vanishing  off-diagonal matrix elements between the adjacent $SO(4)$ Landau levels (as represented by the green shaded rectangular blocks in Fig.\ref{nestcoord.fig}). Borrowing the  string theory interpretation that the off-diagonal parts signify interactions between the fuzzy objects represented by the diagonal block matrices,  one may say that the quasi-fuzzy four-spheres of the adjacent $SO(4)$ Landau levels interact and  conspire to  bring the $SO(5)$ covariance to the nested fuzzy geometry.   Furthermore,  the nested fuzzy geometry has the $SO(5)$ symmetry also. 
Apparently as a classical geometry the nested structure [Fig.\ref{nestedspheres.fig}]  does  not   have the $SO(5)$ symmetry, but it does have in a quantum mechanical sense. 
The reason is essentially same as of the discussion  below Eq.(\ref{defofquantumnambubrack}).  
We had chosen $x_5$ as a special axis, and  we obtained the truncated nested structure without the north and south caps along the $x_5$-axis  like  Fig.\ref{nestedspheres.fig}, but   
if we had chosen the $x_1$ axis, we would have had a similar structure along  the $x_1$-axis.   
Actually we can adopt any axis in $\mathbb{R}^5$, and then the nested structure has to have the $SO(5)$ symmetry. 
Therefore, the nested fuzzy geometry is considered to be  $SO(5)$ symmetric in a quantum mechanical sense, which does not have   its counterpart in classical geometry.

\section{Internal fuzzy structure and the $SO(4)$ Landau models}\label{sec:intfuzzystr}

We discuss a physical model  that realizes the matrix-valued three-sphere geometry inside the $SO(5)$ Landau model.
We also clarify  relations among Landau models in different dimensions. 

\subsection{$SU(2)$ meron gauge field and $SO(4)$ Pauli-Schr\"odinger Hamiltonian}

 We first construct a  physical model whose eigenstates are given by the $SO(4)$ spinor spherical harmonics (\ref{vectorlikespinsphereharmo}). 
The expression of the $SO(4)$ part of the $SO(5)$ free angular momentum operators are exactly equal to the $SO(4)$ free angular momentum operators (see Appendix \ref{append:so5sphericalharmon}): 
\be 
L_{mn}^{(0)} =-ix_m\frac{\partial}{\partial x_n} +ix_n \frac{\partial}{\partial x_m} 
=-iy_m\frac{\partial}{\partial y_n} +iy_n \frac{\partial}{\partial y_m}. 
\ee
The $SO(4)$ angular momentum $L_{mn}$ (\ref{detailantimonoso5gene}) can also be represented only in terms of  the $S^3$-coordinates (\ref{expliys}): 
\be
L_{mn} 
=-iy_m\frac{\partial}{\partial y_n} +iy_n \frac{\partial}{\partial y_m}+\bar{\eta}_{mn}^iS_i. 
\label{so4angopysrep}
\ee
Therefore, the $SO(4)$ analysis in Sec.\ref{subsec:diffons3} can be restated entirely in the language of $S^3$  without resorting to any information of the original manifold  $S^4$.  
We then explore the $SO(4)$ problem as  an $\it{independent}$ problem defined on $S^3$, and  utilize the $S^3$-coordinates only, 
$y_{m=1,2,3,4}$, in this section.  
Interestingly, (\ref{so4angopysrep}) can be realized as the $SO(4)$ angular momentum operators in the meron gauge field  introduced by 
Alfaro, Fubini and Furlan as a solution of pure Yang-Mills field equation \cite{Alfaro-Fubini-Furlan-1976,Alfaro-Fubini-Furlan-1979}: 
\be
A^{\text{AFF}}=-\frac{1}{2r^2}\bar{\eta}_{mn}^i y_{n}\sigma_i ~dy_{m}  
\label{merongaugefielda}
\ee
where $r=\sqrt{y_my_m}$.  The meron gauge field with arbitrary spin magnitude is simply obtained by the dimensional reduction of the Yang's $SU(2)$ monopole gauge field (\ref{zhanghusu2gauge}): 
\be
A=-\frac{1}{r(r+x_5)}\bar{\eta}_{mn}^ix_n S_i  dx_m ~~\overset{x_5\rightarrow 0}{\longrightarrow}~~ A^{\text{AFF}} =-\frac{1}{r^2}\bar{\eta}_{mn}^i y_n S_i  dy_m.  
\label{meronspinmagarbit}
\ee
Notice that the Yang's monopole has the string-like singularity, while the meron only has the point-like singularity at the origin.  
The corresponding field strength 
is given by\footnote{The associated 2nd Chern number is evaluated 
\be
Q= \frac{1}{32\pi^2}\int_{\mathbb{R}^4} d^4x~\epsilon_{mnpq}\tr(F_{mn}^{\text{AFF}} F_{pq}^{\text{AFF}}) =-\frac{1}{2}, 
\ee
leading to the name  ``meron''. For the meron field configuration with general spin $S_i^{(I/2)}$, the 2nd Chern number is evaluated as 
\be
Q= -\frac{1}{12}I(I+1)(I+2). 
\ee
} 
\be
F_{mn}^{\text{AFF}} =-\frac{1}{r^2}y_{m}A^{\text{AFF}}_{n}+\frac{1}{r^2}y_{n}A^{\text{AFF}}_{m} +\frac{1}{2r^2}\bar{\eta}_{mn}^i \sigma_i, 
\ee
and the total angular momentum operator is  
\begin{align}
L_{mn} &=\Lambda_{mn} +r^2 F^{\text{AFF}}_{mn} =-iy_{m}\partial_{n}+iy_{n}\partial_{m} + y_{m} A^{\text{AFF}}_{n}-y_{n}A^{\text{AFF}}_{m} +r^2 F_{mn}^{\text{AFF}}\nn\\
&=  -iy_{m}\partial_{n}+iy_{n}\partial_{m} +\frac{1}{2}\bar{\eta}_{mn}^i\sigma_i, 
\label{so4angopysrep12}
\end{align}
where 
\be
\Lambda_{mn}=-iy_m (\frac{\partial}{\partial y_n}+iA^{\text{AFF}}_n)+iy_n  (\frac{\partial}{\partial y_m}+iA^{\text{AFF}}_m). 
\ee
With the replacement of $\frac{1}{2}\sigma_i$ with higher $SU(2)$ spin matrix $S_i$, (\ref{so4angopysrep12}) turns to the $SO(4)$ angular momentum (\ref{so4angopysrep}). 
 The $SO(4)$ Casimir is given by 
\be
{L_{mn}}^2 =4(\bs{J}^2 +\bs{K}^2), 
\ee
where $\bs{J}$ and $\bs{K}$ are the   $SU(2)_L$ and $SU(2)_R$ operators  
(\ref{detailantimonoso4gene}) :   
\begin{subequations}
\begin{align}
&J_i
=J_i^{(0)}=-i\frac{1}{2}\eta_{mn}^i y_m\frac{\partial}{\partial y_n}, \\
&K_i
=K_i^{(0)}+S_i =-i\frac{1}{2}\bar{\eta}_{mn}^i y_m\frac{\partial}{\partial y_n}+S_i,  
\end{align} 
\end{subequations}
and  the $SU(2)_L$ and $SU(2)_R$ Casimir eigenvalues are given by 
\be
\bs{J}^2 =j(j+1), ~~~\bs{K}^2=k(k+1),
\ee
with 
\be
j+k=n+\frac{I}{2}~~(n=0,1,2,\cdots),~~ ~~~s\equiv j-k =\frac{I}{2}, \frac{I}{2}-1, \cdots, -\frac{I}{2}, \label{so4vernis}
\ee
or 
\be
j =\frac{n}{2} +\frac{I}{4}+\frac{s}{2}, ~~~~k=\frac{n}{2} +\frac{I}{4}-\frac{s}{2}. \label{identifjkandns}
\ee
Their simultaneous eigenstates are given by the $SO(4)$ spinor spherical harmonics 
 (\ref{vectorlikespinsphereharmo}).

\begin{figure}[tbph]
\center
\hspace{-0.8cm}
\includegraphics*[width=140mm]{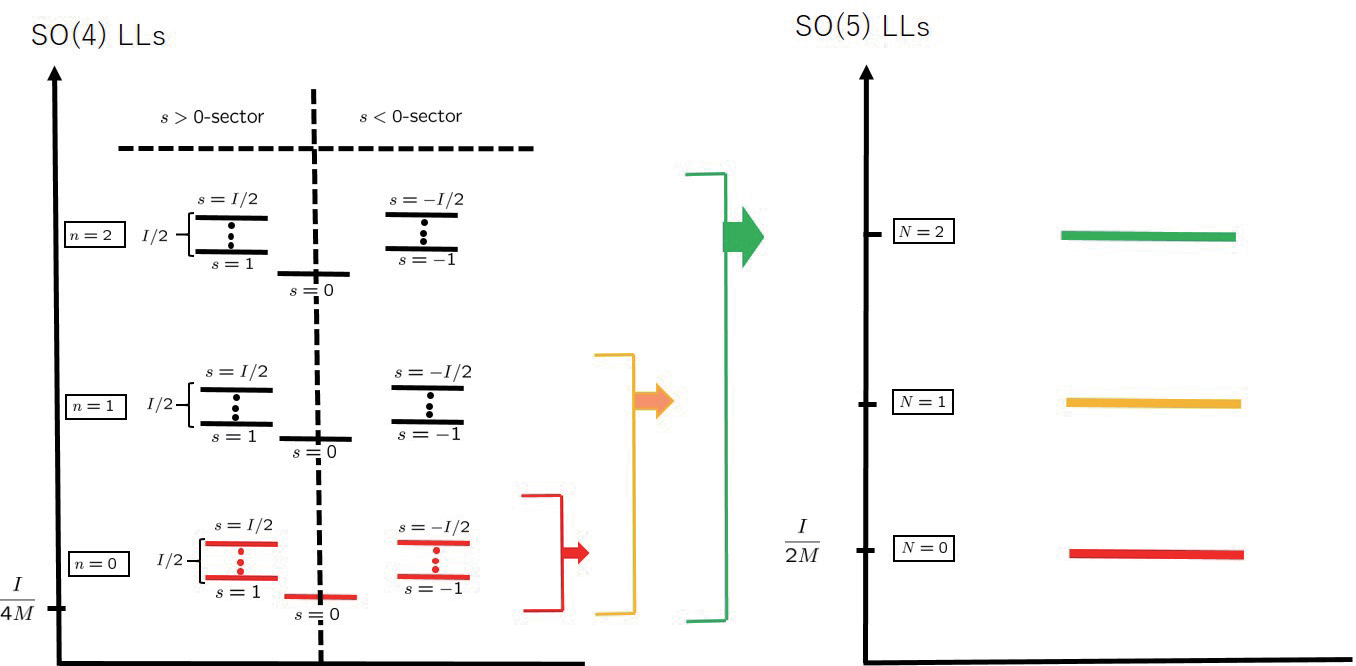}
\caption{The $SO(4)$ Landau levels $E_n(s) = \frac{1}{2M} (n(n+2) +I (n+\frac{1}{2}) +s^2)$ (left figure) and $SO(5)$ Landau levels $E_N =\frac{1}{2M} (N(N+3) +I(N+1))$ (right figure). }
\label{so5so4.fig}
\end{figure}

In the meron field background, we  introduce  an  $SO(4)$ Landau-like Hamiltonian 
\be
H_{\text{PS}}=\frac{1}{2M}\sum_{m<n=1}^4 {\Lambda_{mn}}^2. \label{meronhamillandau}
\ee
As usual, (\ref{meronhamillandau}) can be rewritten as 
\be
H_{\text{PS}}=\frac{1}{2M}\sum_{m<n} ({L_{mn}}^2 -{F_{mn}}^2)=\frac{1}{2M}(2\bs{J}^2 +2\bs{K}^2 -\bs{S}^2), \label{meronspinham}
\ee
where we used  
$\sum_{m<n}\Lambda_{mn}F_{mn} =\sum_{m<n}F_{mn}\Lambda_{mn}=0$ and $\sum_{m<n} {F_{mn}}^2=\bs{S}^2.$  
(\ref{meronhamillandau}) can also  be expressed as 
\be
H_{\text{PS}} 
=\frac{1}{2M} (2{\bs{J}^{(0)}}^2 +2{\bs{K}^{(0)}}^2 + 4\bs{K}^{(0)}\cdot \bs{S}+\frac{I}{2}(\frac{I}{2}+1)),  
\ee 
which realizes an  $SO(4)$ generalization of the original Pauli-Schr\"odinger  Hamiltonian \cite{Pauli-1927} with spin-orbit coupling.  For this reason, we refer to (\ref{meronhamillandau}) as   the $SO(4)$ Pauli-Schr\"odinger  Hamiltonian in this paper. 
From (\ref{identifjkandns}), 
the eigenvalues of the Pauli-Schr\"odinger Hamiltonian are readily obtained as 
\be
E_n(s)=\frac{1}{2M}(n(n+2)+\frac{I}{2}(2n+1)+s^2)  ,  
\label{eigenvaluespaulischso4}
\ee
 where $n$ denotes the $SO(4)$ Landau levels and $s$ denotes the sub-bands in the $SO(4)$ Landau levels [Fig.\ref{so5so4.fig}]. 
The $SO(4)$ Landau level eigenstates are actually  the $SO(4)$ spinor spherical harmonics $\bs{Y}_{j,m_j;~k,m_k}$ with (\ref{so4vernis}), and  so the previous  three-sphere matrix geometry   (\ref{evalcmu}) is considered to be 
 realized in the  $SO(4)$ Landau level.  
In this way, we can reformulate the $SO(4)$ part of the $SO(5)$ Landau model with the $SO(4)$ Pauli-Schr\"odinger  model. In other words,  the $SO(5)$ Landau model  accommodates the $SO(4)$ Pauli-Schr\"odinger  model as its internal model. 

\subsection{Singular gauge transformation and $SO(4)$ matrix geometry}

Curiously, the energy levels (\ref{eigenvaluespaulischso4}) are exactly equal to the Landau levels of the $SO(4)$ Landau Hamiltonian proposed by  Nair and Daemi  \cite{Nair-Daemi-2004}. 
This coincidence suggests a hidden relation between the $SO(4)$ Pauli-Schr\"odinger  model and the $SO(4)$ Landau model. 
In the following, we adopt the notation of \cite{Hasebe-2014-2, Hasebe-2018}. 
The $SO(4)$ Landau Hamiltonian  is given by 
\be
H=\frac{1}{2M}\sum_{m<n=1}^4{\Lambda_{mn}}^2, \label{s04landauhamdr} 
\ee
where 
\be
\Lambda_{mn}=-iy_m(\frac{\partial}{\partial y_n}+iA^{\text{ND}}_n) + iy_n (\frac{\partial}{\partial y_m}+iA^{\text{ND}}_m), 
\ee
with the Nair-Daemi $SU(2)$ gauge field\footnote{The Nari-Daemi $SU(2)$ monopole gauge field is equivalent to the spin connection of  $S^3$.} 
\be 
A^{\text{ND}} =-\frac{1}{r(r+y_4)} \epsilon_{ij k} y_j S_k ~dy_i. \label{nairdaemisu3gaugefield}
\ee 
Obviously, the Nair-Daemi $SU(2)$ gauge field  has a Dirac string-like singularity. 
The corresponding field strength 
is derived as 
\be
F^{\text{ND}}_{ij} =-y_iA^{\text{ND}}_j +y_j A^{\text{ND}}_i +\epsilon_{ijk}S_k, ~~~~~~F^{\text{ND}}_{i4}=(1+y_4)A^{\text{ND}}_i.  
\ee
The eigenvalues of the $SO(4)$ Landau Hamiltonian (\ref{s04landauhamdr}) are given by  (\ref{eigenvaluespaulischso4}) and the corresponding eigenstates, $i.e.$, the $SO(4)$ monopole harmonics (in the Dirac gauge), are given by \cite{Hasebe-2018, Nair-Daemi-2004}\footnote{(\ref{vectorrepsorep}) constitutes an orthonormal set: 
\be
\langle \bs{\Phi}_{j, m_j;  k, m_k} |\bs{\Phi}_{j', n_j;  k', n_k}\rangle \equiv \int_{S^3} d\Omega_3 ~{\bs{\Phi}_{j,m_j;~ k, m_k}(\Omega_3)}^{\dagger}~\bs{\Phi}_{j',m'_j;~ k', m'_k}(\Omega_3) =\delta_{j,j'}\delta_{k,k'}\delta_{m_j, m'_j}\delta_{m_k, m'_k}. 
\label{normalllso4}
\ee
} 
\be
\bs{\Phi}_{j, m_j; k, m_k}(\chi, \theta, \phi)  =\tilde{g}(\theta, \phi)
\begin{pmatrix}
\Phi_{j, m_j; k, m_k}(\chi, \theta, \phi)_{I/2} \\
\Phi_{j, m_j; k, m_k}(\chi, \theta, \phi)_{I/2 -1}\\
\vdots \\
\Phi_{j, m_j; k, m_k}(\chi, \theta, \phi)_{-{I}/2}
\end{pmatrix} , 
\label{vectorrepsorep}
\ee
where\footnote{$\tilde{g}(\theta, \phi)$ is the gauge function  to relate  the Dirac gauge and the Schwinger gauge in the $SO(4)$ Landau model \cite{Hasebe-2018}.}   
\be
\tilde{g}(\theta, \phi) \equiv   D^{(I/2)}(\phi, \theta, 0)= e^{-i\phi S_z}e^{-i\theta S_y}  , \label{transschdiracso4} 
\ee
and 
\begin{align}
\Phi_{j, m_j; k, m_k}(\Omega_3)_{A} &=\sqrt{\frac{(2j+1)(2k+1)}{2\pi^2 (I+1)}}\nn\\
&\times \sum_{m'_j=-j}^{j}\sum_{m'_k=-k}^{k} \langle {I}/{2}, A|  j, m'_j; k, m'_k\rangle     D^{(l_L)}({\chi}, \theta, \phi)_{m'_j , m_j}  D^{(l_R)}(-\chi, \theta, \phi)_{m'_k , m_k},  
\label{basiswithnormal}
\end{align}
with   the Wigner's $D$-function  
\be
 D^{(l)}(\chi, \theta, \phi) \equiv e^{-i\chi S^{(l)}_z} e^{-i\theta S^{(l)}_y}e^{-i\phi S^{(l)}_z}. 
\ee
With these preparations, we now discuss a relation between the $SO(4)$ Pauli-Schr\"odinger  model and the $SO(4)$ Landau model. 


We have seen that the meron gauge field has the point-like singularity, while the Nair-Daemi's $SU(2)$ monopole has the string-like singularity. 
A similar situation occurs in a lower dimension. 
In 3D  the Wu-Yang $SU(2)$ monopole  \cite{Wu-Yang-1968} has a point-like singularity, while   
the Dirac monopole has the string-like singularity. 
In this sense the meron is  a 4D generalization of the Wu-Yang $SU(2)$ monopole,  while the 
Nair-Daemi $SU(2)$ monopole is a 4D generalization of the Dirac monopole. To find relationship  between the meron and the  Nair-Daemi  monopole gauge field, let us first recall the singular transformation that relates the Wu-Yang monopole and the Dirac monopole configurations \cite{Arafune-Freud-Goebel-1975, Bais-1976}.   
With the $\mathbb{R}^3$ coordinates 
\be
\{z_1, z_2, z_3\}\equiv \{ r \cos\phi\sin\theta, r\sin\phi\sin\theta, r\cos\theta\}, 
\ee
the Wu-Yang monopole  and the Dirac monopole gauge fields  are respectively expressed as 
\be
A^{\text{WY}}=-\frac{1}{r^2}\epsilon_{ijk}z_j S_k dz_i, ~~~~
A^{\text{D}} =-\frac{1}{r(r+z_3)} \epsilon_{ij 3} z_j S_3 dz_i, 
\ee
and are related by the singular transformation
\be
A^{\text{WY}} =g^{\dagger}A^\text{D} g -ig^{\dagger}dg,   
\ee
where\footnote{For $S_i=\frac{1}{2}\sigma_i$, (\ref{su2ggsing}) becomes 
\begin{align} 
g(\theta, \phi)&=e^{-i\frac{\theta}{2} (\hat{z}_2 \sigma_x -\hat{z}_1 \sigma_y)} 
=e^{-i \frac{\phi}{2} \sigma_z} e^{i \frac{\theta}{2} \sigma_y} e^{i \frac{\phi}{2} \sigma_z} 
= 
\begin{pmatrix}
\cos\frac{\theta}{2} & \sin\frac{\theta}{2} e^{-i\phi} \\
-\sin\frac{\theta}{2} e^{i\phi} & \cos\frac{\theta}{2}
\end{pmatrix} 
\nn\\
&=\frac{1}{\sqrt{2(1+z_3)}}(1_2 -iz_1\sigma_x-iz_2\sigma_y +z_3 1_2 ), 
\end{align} 
whose columns are the 1st Hopf spinors (see \cite{Hasebe-2010} for instance). }  
\be 
g(\theta, \phi)
= e^{-i{\theta} (\hat{z}_2 S_x -\hat{z}_1 S_y)}=e^{-i{\phi}S_z} e^{i\theta S_y } e^{i\phi S_z },    \label{su2ggsing}
\ee
with $S^1$-latitude coordinates $\hat{z}_1\equiv \cos\phi,~ \hat{z}_2 \equiv  \sin\phi$. 
A bit of consideration tells that    
the $SU(2)$ monopole field (\ref{nairdaemisu3gaugefield}) and  the meron gauge field (\ref{meronspinmagarbit}) are also related by the following $SU(2)$ singular transformation:  
\be
A^{\text{ND}} =g^{\dagger}A^{\text{AFF}} g -ig^{\dagger}dg,   
\label{andaaffgauge}
\ee
where 
\be
g(\chi, \theta, \phi)=e^{-i{\chi} \sum_{i=1}^3 \hat{y}_i S_i}   ={\tilde{g}}(\theta,\phi) ~e^{-i\chi S_z}~\tilde{g}(\theta, \phi)^{\dagger} .  \label{gwithhigherspinsu2}
\ee
Here $\tilde{g}(\theta, \phi)$ is given by (\ref{transschdiracso4}), and $\hat{y}_i$ are the coordinates on $S^2$-latitude parameterized as  
\be
\hat{y}_{i=1,2,3} =(\sin\theta\cos\phi, \sin\theta\sin\phi,\cos\theta).  
\ee

\begin{figure}[tbph]
\center
\hspace{-0.8cm}
\includegraphics*[width=160mm]{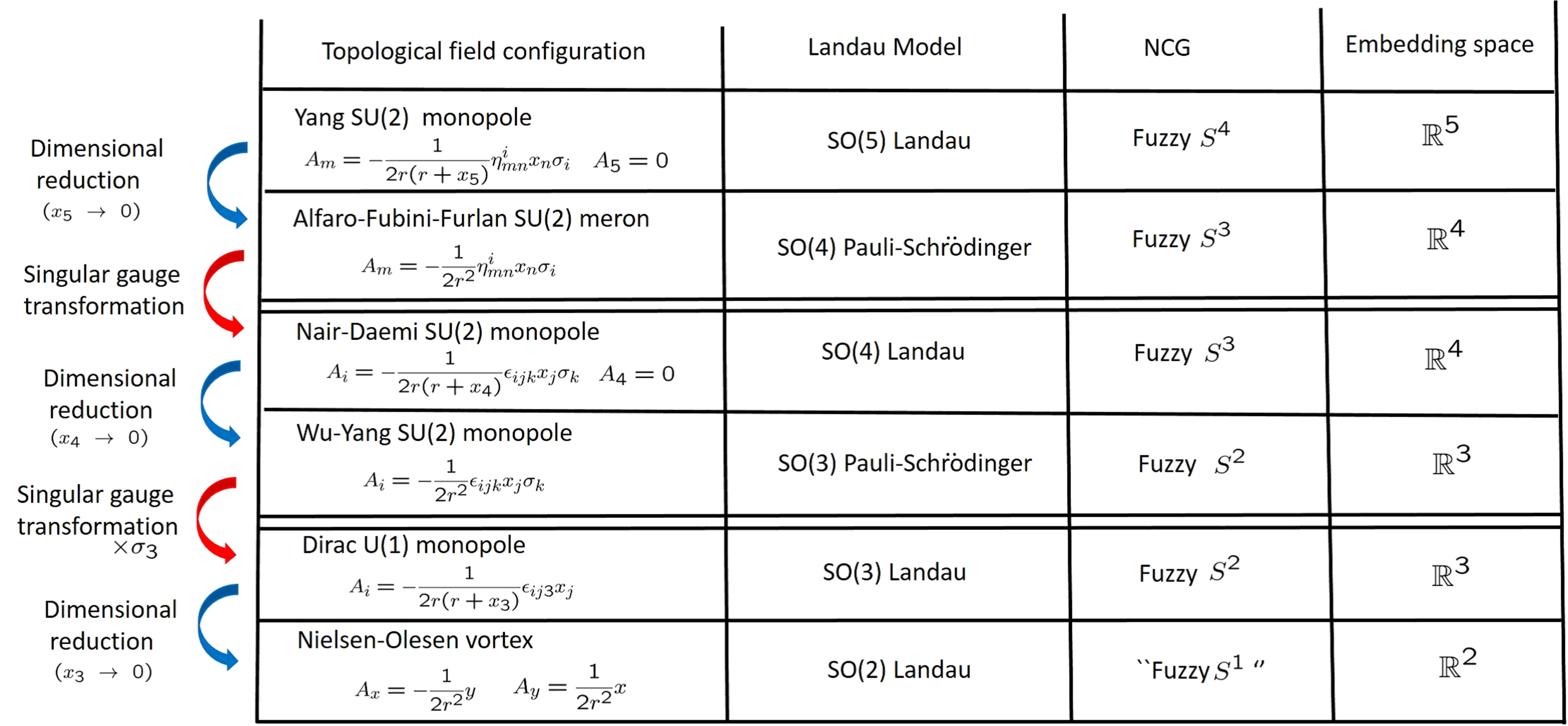}
\caption{Landau models and their background topological field configurations  for the fuzzy sphere geometries. There exist singular gauge transformations and dimensional ladders connecting the Landau models.  }
\label{ladder.fig}
\end{figure}

Therefore, the $SO(4)$ Pauli-Schr\"odinger model is transformed to the $SO(4)$ Landau model by the singular gauge transformation (\ref{andaaffgauge}). 
Indeed,  the $SO(4)$ monopole harmonics (\ref{vectorrepsorep}) and 
the  $SO(4)$ spinor  spherical harmonics (\ref{vectorlikespinsphereharmo}) 
are related as 
\be 
\bs{\Phi}_{j, m_j;~k,m_k}(\chi, \theta, \phi) =(-1)^{\frac{I}{2}+s}~g(\chi, \theta, \phi)^{\dagger} ~\bs{Y}_{j,m_j; k, m_k} (\chi, \theta, \phi). 
\ee 
Consequently, the matrix elements are equivalent:  
\be
\langle \bs{\Phi}_{j', m'_j; k', m'_k}|{y}_{m}| \bs{\Phi}_{j, m_j; k, m_k} \rangle=-\langle \bs{Y}_{j', m'_j; k', m'_k}|{y}_{m}| \bs{Y}_{j, m_j; k, m_k} \rangle, 
\label{relationso4functions}
\ee
where we used $(-1)^{I+s'+s}=-(-1)^{I+2s}=-1$. In Appendix \ref{appendix:so4append},  we rigorously evaluate both sides of  (\ref{relationso4functions})  and explicitly check its validity. 
Therefore, the matrix geometry of the $SO(4)$ Pauli-Schr\"odinger model is exactly equal to the matrix geometry of the $SO(4)$ Landau model, and hence  the $SO(4)$ Landau model describes the  internal fuzzy geometry of the $SO(5)$ Landau model. This demonstrates the idea of the dimensional hierarchy \cite{Hasebe-2017, Hasebe-Kimura-2003} relating the Landau physics in different dimensions.  
In Fig.\ref{ladder.fig}, we summarize the relations among the Landau models  in various dimensions. 
For a better understanding of this section, we also elucidate the case $(j,k) =(1/2, 0)\oplus (0, 1/2)$  in Appendix \ref{appendix;irrpssu2su2}.

\section{Relativistic $SO(5)$ Landau models}\label{sec:relso5landau}

 We explore  relativistic version of the $SO(5)$ Landau model and clarify relationship to the matrix geometry of the Berezin-Toeplitz quantization \cite{Ishiki-Matsumoto-Muraki-2018}.

\subsection{Geometric quantities of $S^4$}

In the parameterization (\ref{polarx1tox5}), the metric of $S^4$ is given by 
\begin{align}
ds^2&={dx_1}^2+ {dx_2}^2+ {dx_3}^2+ {dx_4}^2+ {dx_5}^2 \nn\\
&= d\xi^2+\sin^2\xi d\chi^2+\sin^2\xi\sin^2\chi d\theta^2+\sin^2\xi\sin^2\chi\sin^2\theta d\phi^2. \label{metrics4sngcoord}
\end{align}
We adopt the vierbein in the Schwinger gauge:\footnote{
We choose the numbering of the vierbein as  (\ref{vierbeinchoicerela}) so that the $SO(5)$ Dirac-Landau operator  is reduced to the $SO(4)$ Dirac-Landau operator of \cite{Hasebe-2018} at $\xi=\frac{\pi}{2}$ (see (\ref{reddiracops4tos3})).  
The area of $S^4$ is calculated as 
\be
A(S^4) =\int e^4\wedge e^1\wedge e^2\wedge e^3=\int_0^{\pi} d\xi \sin^3\xi \int_0^{\pi} d\chi \sin^2\chi\int_0^{\pi} d\theta \sin\theta\int_0^{2\pi} d\phi =\frac{8\pi^2}{3}. 
\ee
}
\be
e^4 = d\xi,~~~~e^1=\sin\xi d\chi,~~~~e^2=\sin\xi \sin\chi d\theta,~~~~e^3=\sin\xi\sin\chi\sin\theta d\phi. \label{vierbeinchoicerela}
\ee
With $\omega_{mn}$ satisfying the Cartan structure equation, $de^m+\omega_{mn}e_n=0$,  the spin connections  
\be
\omega_L^i\equiv\frac{1}{2}\eta_{mn}^i\omega^{mn},~~~~\omega_R^i\equiv\frac{1}{2}\bar{\eta}_{mn}^i\omega^{mn}, 
\ee
are given by (see Appendix \ref{append:twogaugestrans} for details)
\begin{align}
&\omega_{\xi}^L=\omega_{\xi}^R=0, ~~~~~\omega_{\chi}^L=\omega_{\chi}^R=-\cos\xi ~i\frac{1}{2}q_x, ~~~~~\omega_{\theta}^L=\omega_{\theta}^R=\cos\xi\sin\chi~i\frac{1}{2}q_y-\cos\chi ~i\frac{1}{2}q_z , \nn\\
&\omega_{\phi}^L=-\omega_{\phi}^R=\cos\xi \sin\chi\sin\theta ~i\frac{1}{2}q_z +\cos\chi\sin\theta ~i\frac{1}{2}q_y +\cos\theta~i\frac{1}{2}q_x. \label{spinconneschwinger}
\end{align}
The $SO(4)$ matrix-valued spin connection is constructed as  
\be
\omega=\begin{pmatrix}
\omega_L & 0 \\
0 & \omega_R
\end{pmatrix} =i\frac{1}{2}\begin{pmatrix}
\omega_L^i ~q_i & 0 \\
0 & \omega_R ^i ~q_i
\end{pmatrix}=i\frac{1}{4}\omega^{mn}\begin{pmatrix}
\eta_{mn}^i ~q_i & 0 \\
0 & \bar{\eta}_{mn}^i ~q_i
\end{pmatrix}, \label{spinconnec}
\ee
which carries   the $SU(2)_L\otimes SU(2)_R$ index:    
\be
(1/2, 0)\oplus (0, 1/2). 
\ee

\subsection{Spinor $SO(5)$  Landau model}

We consider a relativistic spinor particle on $S^4$, which  feels the connection of the base-manifold $S^4$ as well as the external $SU(2)$ monopole gauge field. In other words, the relativistic particle interacts with the synthetic gauge field of the $SO(4)$ connection (\ref{spinconnec}) and the $SU(2)$ monopole field (\ref{zhanghusu2gauge})\footnote{(\ref{spinconneschwinger}) is represented in the Schwinger gauge, while (\ref{zhanghusu2gauge}) is in the Dirac gauge (see Appendix \ref{append:twogaugestrans} for details), so it will be convenient to adopt one gauge in  constructing the synthetic gauge field (\ref{synthegauge}).} 
\be
\mathcal{A}_a =\omega_a \otimes \bs{1}_{I+1} +1_4\otimes A_a. \label{synthegauge}
\ee
For the   $SO(4)\simeq SU(2)_L\otimes SU(2)_R$ gauge group, the synthetic gauge field is irreducibly decomposed as 
\be
((1/2, 0)\oplus (0, 1/2)) \otimes  (0, I/2) = (0, I/2+1/2) \oplus (0, I/2-1/2)\oplus (1/2, I/2), 
\label{so4su2irredecomp}
\ee
and their corresponding dimensions are 
\be
 (\bs{2}\oplus \bs{2}) \otimes(\bs{I+1}) = (\bs{I+2}) \oplus  \bs{I} \oplus   (\bs{2I+2}). 
\ee
The field strength is now constructed as 
\be
\mathcal{F}_{ab} =\partial_a\mathcal{A}_a -\partial_b\mathcal{A}_b +i[\mathcal{A}_a, \mathcal{A}_b]=f_{ab}\otimes \bs{1}_{I+1} +1_4\otimes F_{ab},
\ee
with $f_{ab}=\partial_a\omega_b -\partial_b\omega_a +i[\omega_a,\omega_b]=e_a\wedge e_b$ (Appendix \ref{append:s4curvature}). The $SO(5)$ angular momentum in the synthetic gauge field is 
\be
\mathcal{L}_{ab} = \it{\Lambda}_{ab}+\mathcal{F}_{ab}, \label{spinorso5totalangmo}
\ee
where $\it{\Lambda}_{ab}$ denotes the covariant angular momentum operator 
\be
\it{\Lambda}_{ab} \equiv -ix_a (\partial_b+i\mathcal{A}_b) + ix_b (\partial_a+i\mathcal{A}_a).
\ee
We introduce the spinor $SO(5)$ Landau Hamiltonian as 
\be
H=\frac{1}{2M}\sum_{a<b=1}^5 {\it{\Lambda}_{ab}}^2 = \frac{1}{2M}(\sum_{a<b=1}^5 {\mathcal{L}_{ab}}^2 -\sum_{m<n=1}^4 {\Sigma_{mn}}^2). \label{so5spinorhamlandau}
\ee
The decomposition (\ref{so4su2irredecomp}) implies that, with some appropriate unitary transformation, the spinor $SO(5)$ Landau Hamiltonian 
is  transformed as
\be
H ~\rightarrow~
\begin{pmatrix}
H^{(0, \frac{I}{2}+\frac{1}{2})} & 0 & 0 \\
0 &  H^{(0, \frac{I}{2}-\frac{1}{2})} & 0 \\
0 & 0 &  {H}^{(\frac{1}{2}, \frac{I}{2})} \end{pmatrix}.  \label{decompso5spiham}
\ee
Here $H^{(\frac{I_+}{2}, \frac{I_-}{2})}$ denotes an $SO(5)$ Landau Hamiltonian in the $SO(4)$ monopole background with the $SO(4)$ matrices 
\begin{equation}
\Sigma_{mn}^{(I_+/2,I_-/2)}=\eta_{mn}^i S_i^{(I_+/2)}\otimes \bold{1}_{I_-+1}+\bold{1}_{I_+ +1}\otimes {\bar{\eta}_{mn}}^i S_i^{(I_-/2)}. 
\end{equation}
Notice that $H^{(0, \frac{I}{2}-\frac{1}{2})}$ in (\ref{decompso5spiham}) is just a $\it{non}$-$\it{relativistic}$ Landau Hamiltonian (\ref{origiso5landauhaml}) with the $SU(2)$  (anti-) monopole index $(I-1)/2$.

\subsection{$SO(5)$ Dirac-Landau operator  and zero-modes}

The Dirac-Landau operator on $S^4$, which we call the $SO(5)$ Dirac-Landau operator, is given by 
\begin{align}
-i\fsl{\mathcal{D}}_{S^4} & =-ie_m^{~~\mu}~\gamma^m ~(\partial_{\mu} +i\omega_{\mu}\otimes \bs{1}_{I+1} +i\bs{1}_4\otimes A_{\mu})\nn\\
&=-ie_m^{~~\mu}~\gamma^m ~\otimes ~\tilde{D}_{\mu},  
\label{diraclandauons4}
\end{align}
where $\tilde{D}_{\mu}$ $(\mu=\xi,\chi,\theta,\phi)$ are newly introduced covariant derivatives including the contribution of the spin connection: 
\be
\tilde{D}_{\xi} =D_{\xi} +\frac{3}{2}\cot\xi ~\bs{1}_{I+1}, ~~~\tilde{D}_{\chi} =D_{\chi} +\cot\chi~\bs{1}_{I+1}, ~~~\tilde{D}_{\theta} =D_{\theta} +\frac{1}{2}\cot\theta~\bs{1}_{I+1},~~~\tilde{D}_{\phi} =D_{\phi} 
\label{tildedmudmu}
\ee
with 
\be
D_{\mu}=\partial_{\mu} +iA_{\mu}. \label{dmudefconv}
\ee
The  second terms on the right-hand sides of (\ref{tildedmudmu}) are attributed to the spin connections $\omega_{\mu}$.  We adopt the $SU(2)$ gauge field in (\ref{dmudefconv}) as  
\be
A_{\mu}=\omega_{ R \mu}^1~ S_z^{(I/2)} + \omega_{ R \mu}^2~ S_x^{(I/2)} +\omega_{ R \mu}^3~ S_y^{(I/2)}, 
\ee
or more explicitly,  
\begin{align}
&A_{\xi}=0, ~~~~~A_{\chi}=-\cos\xi ~S^{(I/2)}_z, ~~~~~A_{\theta}=-\cos\xi\sin\chi~S_x^{(I/2)}-\cos\chi S_y^{(I/2)} , \nn\\
&A_{\phi}=-\cos\xi \sin\chi\sin\theta ~S^{(I/2)}_y +\cos\chi\sin\theta ~S^{(I/2)}_x -\cos\theta~S^{(I/2)}_z,   
\label{so5landausu2gauge}
\end{align}
which  denotes  the Yang's (anti-)monopole  in the Schwinger gauge  (see Appendix \ref{append:twogaugestrans}).   
From (\ref{diraclandauons4}), one can find that the Dirac-Landau operator respects the chiral ``symmetry'': 
\be
\{-i\fsl{\mathcal{D}}_{S^4}~, ~\gamma^5\otimes \bs{1}_{I+1}\}=0. \label{chiralsymdiraclandauop}
\ee
Therefore, the positive and negative relativistic Landau levels of the $SO(5)$ Dirac-Landau operator are symmetric with respect to the zero. The $SO(5)$ Dirac-Landau operator does not have any $SO(5)$ indices and  is  invariant under the $SO(5)$ rotations: 
\be
[-i\fsl{\mathcal{D}}_{S^4}, \mathcal{L}_{ab}] =0. 
\label{so5invdiraclandau}
\ee
Here $\mathcal{L}_{ab}$ and $-i\fsl{\mathcal{D}}_{S^4}$ are respectively given by (\ref{spinorso5totalangmo}) and (\ref{diraclandauons4}), and one may in principle verify (\ref{so5invdiraclandau}) by using the explicit forms of the operators. The  Dirac-Landau operator eigenstates are degenerate with respect to the $SO(5)$ rotational symmetry  and can be expanded by the eigenstates of the spinor $SO(5)$ Landau model.  
In particular, the Dirac-Landau operator zero-modes are identical to the lowest Landau level eigenstates of the {non}-{relativistic} Landau Hamiltonian $H^{(0, \frac{I}{2}-\frac{1}{2})}$ of (\ref{decompso5spiham}) as   we shall see in Sec.\ref{sec:zeromodes}.  

\subsubsection{Dimensional reduction to the $SO(4)$ Dirac-Landau operator}

On the equator $\xi=\pi/2$, the $SU(2)$ gauge field (\ref{so5landausu2gauge}) is reduced to the $SU(2)$ gauge field of the $SO(4)$ Landau model \cite{Hasebe-2018}: 
\be
(A_{\chi}, A_{\theta}, A_{\phi}) ~\rightarrow~(0, ~-\cos\chi S_y^{(I/2)}, ~ \cos\chi\sin\theta ~S^{(I/2)}_x -\cos\theta~S^{(I/2)}_z).
\ee
(\ref{diraclandauons4}) can be decomposed as 
\be
-i\fsl{\mathcal{D}}_{S^4} =-i\gamma^4 \otimes \tilde{D}_{\xi} -i\frac{1}{\sin\xi} \biggl(\gamma^1 \otimes \tilde{D}_{\chi} +\frac{1}{\sin\chi}\gamma^2 \otimes \tilde{D}_{\theta} +\frac{1}{\sin\chi \sin\theta} \gamma^3 \otimes \tilde{D}_{\phi}\biggr) ,  \label{diracnewexp}
\ee
where 
\begin{align}
&\tilde{D}_{\xi} \equiv \partial_{\xi}+iA_{\xi} +\frac{3}{2}\cot\xi ~\bs{1}_{I+1}= \partial_{\xi} +\frac{3}{2}\cot\xi ~\bs{1}_{I+1}, \nn\\
&\tilde{D}_{\chi} \equiv \partial_{\chi}+iA_{\chi} +\cot\chi~\bs{1}_{I+1} = \partial_{\chi}-i\cos\xi~S_{z}^{(I/2)} +\cot\chi~\bs{1}_{I+1}, \nn\\
&\tilde{D}_{\theta}\equiv \partial_{\theta}+iA_{\theta} +\frac{1}{2}\cot\theta ~\bs{1}_{I+1}=\partial_{\theta}-i\cos\xi\sin\chi~S_x^{(I/2)}-i\cos\chi~S_y^{(I/2)}+\frac{1}{2}\cot\theta ~\bs{1}_{I+1}, \nn\\
&\tilde{D}_{\phi}\equiv \partial_{\phi}+iA_{\phi}=\partial_{\phi} -i\cos\xi\sin\chi\sin\theta ~S_y^{(I/2)} +i\cos\chi\sin\theta~S_x^{(I/2)}-i\cos\theta~S_z^{(I/2)}. 
\end{align}
With  
 the gamma matrices (\ref{fundgammamatrix})\footnote{The choice (\ref{choicegammarela}) is different from the previous one (\ref{fundgammamatrix}). We adopt  (\ref{choicegammarela}) so that the $SO(4)$ Dirac-Landau operator (\ref{nontildediracs3}) coincides with the expression of \cite{Hasebe-2018}.}  
\be
\gamma^4 =\begin{pmatrix}
0 & 1_2 \\
1_2 & 0 
\end{pmatrix}, ~~\gamma^1 =\begin{pmatrix}
0 & i\sigma_z \\
-i\sigma_z & 0 
\end{pmatrix}, ~~\gamma^2 =\begin{pmatrix}
0 & i\sigma_x \\
-i\sigma_x & 0 
\end{pmatrix}, ~~\gamma^3 =\begin{pmatrix}
0 & i\sigma_y \\
-i\sigma_y & 0 
\end{pmatrix}, \label{choicegammarela}
\ee
we express (\ref{diracnewexp})  as
\footnote{One may readily check that in the absence of the $SU(2)$ monopole gauge field, (\ref{summdiraclandauops4}) is reduced the free $SO(5)$ Dirac operator \cite{CamporesiHiguchi1996}. }   
\begin{align}
-i\fsl{\mathcal{D}}_{S^4} 
&=-i (\partial_{\xi} +\frac{3}{2}\cot\xi )\begin{pmatrix}
0_{2(I+1)}  & 1  \\
 1  & 0_{2(I+1)} 
\end{pmatrix}+\frac{1}{\sin\xi} 
\begin{pmatrix}
 0 & 1 \\
-1 & 0 \end{pmatrix}\otimes 
\tilde{\fsl{\mathcal{D}}}_{S^3} ,    
\label{summdiraclandauops4}
\end{align}
where $-i\tilde{\fsl{\mathcal{D}}}_{S^3}$ is given by 
\be 
-i\tilde{\fsl{\mathcal{D}}}_{S^3} \equiv -i{\fsl{\mathcal{D}}}_{S^3} -\cos\xi~\sum_{i=1}^3\sigma_i \otimes S_i^{(I/2)}.  \label{tildediraclandau}
\ee
$-i\fsl{\mathcal{D}}_{S^3}$ signifies  the $SO(4)$ Dirac-Landau operator on $S^3$ \cite{Hasebe-2018}: 
\begin{align}
-i\fsl{\mathcal{D}}_{S^3} &=-i\sigma^3\otimes ( \partial_{\chi} +\cot\chi~\bs{1}_{I+1}) -i\frac{1}{\sin\chi} \sigma^1\otimes (\partial_{\theta}-i\cos\chi~S_y^{(I/2)}+\frac{1}{2}\cot\theta ~\bs{1}_{I+1})\nn\\
&~~-i\frac{1}{\sin\chi \sin\theta} \sigma^2 \otimes (\partial_{\phi}  +i\cos\chi\sin\theta~S_x^{(I/2)}-i\cos\theta~S_z^{(I/2)}). \label{nontildediracs3}
\end{align}
On the equator of $S^4$ $(\xi=\frac{\pi}{2})$, 
$-i\tilde{\fsl{\mathcal{D}}}_{S^3}$ is reduced to  $-i{\fsl{\mathcal{D}}}_{S^3}$, and so is the $SO(5)$ Dirac-Landau operator:  
\be
-i{\fsl{\mathcal{D}}}_{S^4}|_{\xi=\frac{\pi}{2}} = -i{\fsl{\mathcal{D}}}_{S^3}. 
\label{reddiracops4tos3}
\ee
The relativistic $SO(5)$ Landau model thus embeds the relativistic $SO(4)$ Landau model on the equator  as the non-relativistic $SO(5)$ Landau model does. The fuzzy three-sphere geometry is realized in the $SO(4)$ relativistic Landau model \cite{Hasebe-2018}, and  then the $SO(5)$ relativistic Landau model accommodates a fuzzy three-sphere geometry as its sub-geometry, which suggests that 
  the fuzzy four-sphere  is realized as the whole geometry. 

\subsubsection{Zero-modes and the matrix geometry}\label{sec:zeromodes}

The square of the Dirac-Landau operator  (\ref{diraclandauons4}) and the $SO(5)$   Casimir  (\ref{spinorso5totalangmo}) are related as  \cite{Dolan-2003, Hasebe-2014-1}   
\be
(-i\fsl{\mathcal{D}}_{S^4})^2 = \sum_{a<b =1}^5 {\mathcal{L}_{ab}}^2 - \sum_{a<b =1}^5 {F_{ab}}^2  
+\frac{1}{8}\mathcal{R}_{S^4}= \sum_{a<b =1}^5 {\mathcal{L}_{ab}}^2 -  2\cdot \frac{I}{2}(\frac{I}{2}+1) +\frac{3}{2}. 
\label{dsqlso5cas}
\ee 
Here, we used $\sum_{a<b =1}^5 {F_{ab}}^2=\sum_{m<n}(\bar{\eta}^i_{mn}S_i^{(I/2)})^2 =2~{\bs{S}^{(I/2)}}^2=2\cdot \frac{I}{2}(\frac{I}{2}+1)$ and  
$\mathcal{R}_{S^4}=d(d-1)|_{d=4}=12$ (\ref{riccis4value}).
The square of the Dirac-Landau operator respects the $SO(5)$ rotational symmetry and the chiral symmetry as well:  
\be
[(-i\fsl{\mathcal{D}}_{S^4})^2, \mathcal{L}_{ab}] = [(-i\fsl{\mathcal{D}}_{S^4})^2, \gamma_5\otimes 1_{I+1}] 
=0. 
\ee
Consequently, the eigenvalues of $(-i\fsl{\mathcal{D}}_{S^4})^2$ generally have two kinds of degeneracies   coming from the  $SO(5)$ rotational symmetry and the chiral symmetry. The zero-modes, however,  do not have the degeneracy from the chiral symmetry, and  only have the degeneracy of the $SO(5)$ rotational symmetry.   
 Since the square of the Dirac-Landau operator shares the same $SO(5)$ Casimir  $\sum_{a<b =1}^5 {\mathcal{L}_{ab}}^2$ with  the  spinor Landau Hamiltonian  (\ref{so5spinorhamlandau}),  the eigenvalue problem of (\ref{dsqlso5cas}) is equivalent to that of the  spinor Landau Hamiltonian. 
We then focus on the case 
\be
(p,q)_{SO(5)}=(N+2J, N)\biggr|_{J=\frac{I}{2} - \frac{1}{2}},      
\ee
which corresponds to $H^{(0, \frac{I}{2}-\frac{1}{2})}$ in (\ref{decompso5spiham}).   
In this case,  the $SO(5)$ Casimir eigenvalues are given by   
\be
\sum_{a<b =1}^5  {\mathcal{L}_{ab}}^2 = N^2+N(2J+3) +2J(J+2)\biggr|_{J=\frac{I}{2}-\frac{1}{2}} = N^2+N(I+2) +\frac{1}{2}(I-1)(I+3),  
\ee
and then 
\be
(-i\fsl{\mathcal{D}}_{S^4})^2=
N^2+N(I+2).  
\label{speclandaudiracs4}
\ee
Obviously, the zero eigenvalue is realized at $N=0$.  
Since the eigenvalues of the Dirac-Landau operator take real values,    the zero-modes of  the square of the Dirac-Landau operator are equal to those of the Dirac-Landau operator. 
Notice that $N$ acts as the Landau level index in the {non}-{relativistic} Hamiltonian  $H^{(0,\frac{I}{2}-\frac{1}{2})}$, and so the zero-modes are identical to  the lowest Landau level  eigenstates of $H^{(0,\frac{I}{2}-\frac{1}{2})}$. Then, the zero-mode degeneracy is readily obtained as
\be
D(I-1, N)|_{N=0} =\frac{1}{6}I(I+1)(I+2). 
\label{numberzeromodes} 
\ee
The Atiyah-Singer index theorem  also verifies (\ref{numberzeromodes}),   
$D(I-1, N)|_{N=0}=-c_2(I)=\frac{1}{6}I(I+1)(I+2)$ (\ref{chernnumberantimo}).  
In \cite{Ishiki-Matsumoto-Muraki-2018}, the fuzzy four-sphere geometry was derived in the Berezin-Toeplitz method by taking matrix elements sandwiched by the  zero-modes.\footnote{The authors in \cite{Ishiki-Matsumoto-Muraki-2018} used the stereographic coordinates from $S^4$ in their calculations. 
} Since the zero-modes are identical to the non-relativistic lowest Landau level eigenstates,  the  matrix geometry  obtained in the non-relativistic  analysis (\ref{xalllmat}) exactly coincides with that of the Berezin-Toeplitz quantization.  

\section{Even higher dimensional Landau model and matrix geometry}\label{sec:evenhigherdim}

We extend the discussions  of Sec.\ref{sec:fourspherematgeo}  to even  higher dimensions and investigate the  matrix geometry in the $SO(2k+1)$ Landau model on $S^{2k}$.     

\subsection{Quantum limit}

First we analyze the quantum limit, $I=1$.  
 We introduce the following map analogous to the Hopf maps:  
\be
\Psi~~\rightarrow~~x_a \bs{1}_{2^{k-1}}=\Psi^{\dagger}\gamma_a\Psi.  ~~~(a=1,2,\cdots, 2k+1)
\label{hopfmapgenepsitoxa}
\ee
Here, 
$\gamma_a$ are $SO(2k+1)$ gamma matrices  
\be
\gamma_{i}=\begin{pmatrix}
0 & i\gamma'_i \\
-i\gamma'_i & 0 
\end{pmatrix},~~~\gamma_{2k+1}=\begin{pmatrix}
0 & \bs{1}_{2^{k-1}}  \\
\bs{1}_{2^{k-1}}  & 0 
\end{pmatrix}, ~~~\gamma_{2k+1}=\begin{pmatrix}
-\bs{1}_{2^{k-1}} & 0 \\
0 & \bs{1}_{2^{k-1}} 
\end{pmatrix} ,  
\ee
with $\gamma'_i$ $(i=1,2,\cdots,2k-1)$ being the $SO(2k-1)$ gamma matrices and 
$\Psi$ denotes a $2^k\times 2^{k-1}$ complex spinor  
given by 
\be
\Psi=\frac{1}{\sqrt{2(1+x_{2k+1})}}
\begin{pmatrix}
x_{2k} \bs{1}_{2^{k-1}}+i\sum_{i=1}^{2k-1}\gamma'_i x_i \\
({1}+x_{2k+1}) \bs{1}_{2^{k-1}} 
\end{pmatrix} \cdot g, 
\ee
where $g$ denotes $SO(2k)$ gauge group element, and  $\Psi$ satisfies  
$\Psi^{\dagger}\Psi=\bs{1}_{2^{k-1}}$.   
The corresponding connection is obtained as 
\be
A=-i\Psi^{\dagger}d\Psi =-i\frac{1}{1+x_{2k+1}} \bar{\sigma}_{mn}x_n dx_m, \label{so2kgaugefield}
\ee
with  $SO(2k)$ matrix generators 
\be
\bar{\sigma}_{mn} =-i\frac{1}{4}[\gamma'_m, \gamma'_n].~~~~~~(\gamma'_m\equiv \{\gamma'_m, 1_{2^{k-1}}\})
\ee
(\ref{so2kgaugefield}) signifies the $SO(2k)$ non-Abelian monopole gauge field \cite{Hasebe-Kimura-2003}. We  construct the $SO(2k+1)$ Landau model in a similar manner to Sec.\ref{subsec:so5landaumodel} \cite{Hasebe-Kimura-2003}.   
For instance, 
the $SO(2k+1)$ angular momentum operators are given by 
\be
L_{ab}=-ix_a(\partial_b+iA_b)+ix_b(\partial_a+iA_a) +r^2 F_{ab}. 
\label{so2k+1angularmomopge}
\ee
When we represent $\Psi$ as  
\be
\Psi=\begin{pmatrix}
{\bs{\psi}_1}^{\dagger} \\
{\bs{\psi}_2}^{\dagger} \\
\vdots \\
{\bs{\psi}_{2^{k}}}^{\dagger} 
\end{pmatrix}, 
\ee 
 the  rows  $\bs{\psi}_{\alpha}$ $(\alpha=1,2,\cdots,2^{k})$ denote a set of   $SO(2k+1)$ $2^{k-1}$-component spinors that transform as a multiplet under the $SO(2k+1)$  transformation\footnote{With (\ref{so2k+1angularmomopge}) of $I=1$,  we can explicitly show the $SO(5)$ covariance of $\bs{\psi}_{\alpha}$:  
\be
L_{ab}\bs{\psi}_{\alpha}=(\sigma_{ab})_{\beta\alpha}\bs{\psi}_{\beta}, 
\ee
where 
\be
\sigma_{ab} \equiv -i\frac{1}{4}[\gamma_a, \gamma_b]. 
\ee
}, $i.e.$, the lowest Landau level eigenstates for $I=1$.  
$\Psi$ yields the projection matrix 
\be
P\equiv \Psi \Psi^{\dagger} = \frac{1}{2}(\bs{1}_{2^k}+\sum_{a=1}^{2k+1}x_a \gamma_a), 
\label{projectexplictg}
\ee 
which is a gauge invariant quantity and simply expressed by the $S^{2k}$-coordinates.   
The matrix coordinates are expressed as\footnote{
The coefficient in front of the integration of (\ref{integtodexs}) is added to be accounted for by the normalization of  ${\Psi}$:    
\be
\frac{2}{A(S^{2k})}\int_{S^{2k}} d\Omega_{2k} {\Psi}{\Psi}^{\dagger}=\bs{1}_{2^k}, 
\ee
}  
\be
(X_a)_{\alpha\beta} =\frac{2}{A(S^{2k})}~\int d\Omega_{2k} ~{\bs{\psi}_{\alpha}}^{\dagger}x_a \bs{\psi}_{\beta} , 
\label{integtodexs}
\ee
or 
\be
X_a =\frac{2}{A(S^{2k})}~\int d\Omega_{2k} ~{\Psi} ~x_a ~\Psi^{\dagger}=\frac{2}{A(S^{2k})}~\int d\Omega_{2k} ~P~\gamma_a P  ,  \label{pxpcoodsk+1}
\ee
where $A(S^{2k})=\frac{2^{k+1}\pi^{k}}{(2k-1)!!}$ denotes the area of $S^{2k}$. 
With the formulas 
\be
P\gamma_a P=\frac{1}{2}x_a (1+x_b \gamma_b) ,  ~~~
\int_{S^{2k}} d\Omega_{2k} ~x_a  
=0,~~~\int_{S^{2k}} d\Omega_{2k} ~x_a x_b  =\frac{1}{2k+1}A(S^{2k})~\delta_{ab},    
\ee
we can easily  evaluate (\ref{pxpcoodsk+1}) as 
\be
X_a= \frac{1}{2k+1}\gamma_a.    \label{xahopfmapresultg}
\ee

\subsection{Classical limit}

Next we consider the classical limit $I >\!>1$, in which $L_{ab}$ (\ref{so2k+1angularmomopge}) is reduced to 
\be
L_{ab} ~~\rightarrow~~r^2 F_{ab}. \label{ltofab}
\ee
The coordinates $x_a$ can be extracted from the field strength as \cite{Hasebe-2014-1}
\be
\frac{1}{r^{2k+1}} x_a =\frac{2}{(2k)! c_{k}(I)} \epsilon_{a a_1 a_2 \cdots a_{2k}}\tr(F_{a_1 a_2}F_{a_3 a_4} \cdots F_{a_{2k-1} a_{2k}}), 
\label{xafromfffg}
\ee
where $c_k(I)$ denotes the $k$th Chern number of the $SO(2k)$ gauge field: 
\be
c_k(I)=\frac{1}{(2\pi)^k  k!}\int_{S^{2k}}\tr(F^k). 
\label{chernkform}
\ee
Substituting (\ref{ltofab}) to (\ref{xafromfffg}), we have  
\be
X_a  ~\sim~\frac{2r}{(2k)! c_{k}(I)} \epsilon_{a a_1 a_2 \cdots a_{2k}} L_{a_1 a_2}L_{a_3 a_4} \cdots L_{a_{2k-1} a_{2k}}~\tr (\bs{1}_{\text{internal~space}}).   
\label{xfromlllg}
\ee
Since $L_{ab}$ are the $SO(2k+1)$ operators,  $X_a$ (\ref{xfromlllg}) also become operators. 
 $\tr$ in (\ref{xafromfffg}) (and (\ref{chernkform})) is taken for the ``internal fuzzy space'' $S_{F}^{2k-2}$ with dimension \cite{Hasebe-2014-1,Hasebe-Kimura-2003}
\be
D_{k-1}(I)=\prod_{l=1}^{k-1}\prod_{i=1}^{l}\frac{I+l+i-1}{l+i-1} ,   
\label{expdkig}
\ee
and so 
\be
\tr (\bs{1}_{\text{internal~space}}) =D_{k-1}(I).
\ee
In the lowest Landau level,  the $SO(2k+1)$  operators may be replaced with the $SO(2k+1)$ matrix generators  in the fully symmetric irreducible representation: 
\be
L_{ab}~~\rightarrow~~\Sigma_{ab}, 
\ee
and then (\ref{xfromlllg}) becomes 
\begin{align}
X_a &=\frac{2}{(2k)!}\frac{D_{k-1}(I)}{ c_{k}(I)} \epsilon_{a a_1 a_2 \cdots a_{2k}} \Sigma_{a_1 a_2}\Sigma_{a_3 a_4} \cdots \Sigma_{a_{2k-1} a_{2k}}\nn\\
&= \frac{2}{I} \frac{I}{(2k)!}\frac{D_{k-1}(I)}{D_k(I-1)} \epsilon_{a a_1 a_2 \cdots a_{2k}} \Sigma_{a_1 a_2}\Sigma_{a_3 a_4} \cdots \Sigma_{a_{2k-1} a_{2k}}\nn\\
&=\frac{1}{I} \frac{I!!}{k! (I+2k-2)!!}\epsilon_{a a_1 a_2 \cdots a_{2k}} \Sigma_{a_1 a_2}\Sigma_{a_3 a_4} \cdots \Sigma_{a_{2k-1} a_{2k}}, 
\label{preciseXag}
\end{align}
where   
in the second equation 
 the  Atiyah-Singer index theorem was used \cite{Hasebe-2014-1, Dolan-2003}
\be
c_k(I)=-D_k(I-1). 
\ee
Since the fully symmetric representation $SO(2k+1)$ matrices satisfy  
\be
\epsilon_{a a_1 a_2 \cdots a_{2k}} \Sigma_{a_1 a_2}\Sigma_{a_3 a_4} \cdots \Sigma_{a_{2k-1} a_{2k}} = -\frac{k ! (I+2k-2)!!}{I !!}~\Gamma_a, 
\ee
(\ref{preciseXag})  finally takes a concise form 
\be
X_a =\frac{1}{I}~\Gamma_a. 
\label{xagammaag}
\ee

\subsection{Even higher dimensional matrix geometry}

The results in the two limits, (\ref{xahopfmapresultg}) and (\ref{xagammaag}),  
suggest   the  general form of the matrix coordinates:   
\be
X_a =\frac{1}{I+2k}~\Gamma_a.  \label{generalsk+1xa}
\ee
From the properties of  the $SO(2k+1)$ gamma matrices
\begin{align}
&\sum_{a=1}^{2k+1} \Gamma_a \Gamma_a =  I(I+2k)\bs{1}_{D_k(I)} , \nn \\ 
&[\Gamma_{a_1}, \Gamma_{a_2}, \cdots, \Gamma_{a_{2k}}] = -i^k \frac{(2k)! !~(I+2k-2)!!}{I!!} ~\epsilon_{a_1 a_2 \cdots a_{2k+1}}\Gamma_{a_{2k+1}},   
\label{propgammasggen}
\end{align}
we can find that $X_a$ (\ref{generalsk+1xa}) realize the quantum Nambu geometry of the fuzzy $2k$-sphere \cite{Jabbari2004,DeBellisSS2010}: 
\begin{align}
&\sum_{a=1}^{2k+1} X_a X_a = \frac{I}{I+2k}\bs{1}_{D_k(I)},  
\nn\\
&[X_{a_1}, X_{a_2}, \cdots, X_{a_{2k}}] =-i^k ~C(k, I)~(\frac{2}{I+2k})^{2k-1}~\epsilon_{a_1 a_2 \cdots a_{2k+1}}X_{a_{2k+1}}, \label{nambualgebra2kg}
\end{align}
with 
\be
C(k, I)\equiv \frac{(2k)!!~(I+2k-2)!!}{2^{2k-1} ~I !!}.   
\ee
  The matrix geometry (\ref{generalsk+1xa}) will also be obtained by the Berezin-Toeplitz quantization, since the zero-modes of the Dirac-Landau operator are equal to the lowest Landau level eigenstates  and the  Atiyah-Singer theorem also hold in arbitrary even dimension.

Further, when we take into account 
the low dimensional results including odd dimensions \cite{Hasebe-2015, Hasebe-2018, Hasebe-2017} 
\begin{align}
&\text{$S_F^2$ of $SO(3)$ Landau model}~:~\langle x_i \rangle_{\text{LLL}}=     {\frac{1}{I+2}}  ~2S_i          ,\nn\\
&\text{$S_F^3$ of $SO(4)$ Landau model}~:~ \langle x_{m} \rangle_{\text{LLL}}=      {\frac{1}{I+3}} ~ \Gamma_{m}    , 
\end{align}
 (\ref{generalsk+1xa}) may be naturally generalized to     
\be
X_a =\frac{1}{I+d}~\Gamma_a,  ~~(a=1,2,\cdots, d+1)
\ee
for  the $SO(d+1)$ Landau model. 

\section{Summary}\label{sec:summary}

 In this work, we performed a comprehensive study of the $SO(5)$ Landau models and their matrix geometries.   
With  $SO(5)$ monopole harmonics in a full form, we completely derived the matrix coordinates of four-sphere in any Landau level. In the lowest Landau level,  the matrix geometry is given by the generalized $SO(5)$  gamma matrices realizing the quantum Nambu geometry. 
We showed that the matrix geometry obtained by the Landau level projection  actually interpolates the matrix geometries between the quantum limit and the classical limit.  
 In higher Landau level, the matrix geometry  exhibits a nested fuzzy structure. The  $N$th $SO(5)$ Landau level accommodates $N+1$ inner $SO(4)$ Landau levels each of which realizes quasi-fuzzy four-sphere geometry.  As a whole, there are $N+1$ quasi-fuzzy four-spheres   constituting  
 a  $N+1$ concentric  nested structure with $SO(5)$ symmetry. 
Interestingly, the nested fuzzy structure realizes a pure quantum geometry  having no counterpart in classical geometry.  
We introduced 
an $SO(4)$ Pauli-Schr\"odinger model with meron gauge field background that realizes the inner $SO(4)$ part of the $SO(5)$ Landau model. 
We established a singular gauge transformation between the $SO(4)$ Pauli-Schr\"odinger model and  the $SO(4)$ Landau model and identified the internal fuzzy geometry of the $SO(5)$ Landau model with the $SO(4)$ Landau model matrix geometry.     
Explicit relations among  other low dimensional Landau models  with fuzzy geometries were also exploited. 
 We  analyzed the relativistic $SO(5)$ Landau models and  clarified  relationship to the matrix geometry of the Berezin-Toeplitz quantization. Finally, we  investigated even higher dimensional Landau model and demonstrated a realization of  the associated quantum Nambu geometry in an arbitrary dimension.

Former researches about the non-commutative geometry in  Landau models are  mainly focused on  the lowest Landau level, however in view of the  Landau level projection, there is no  logical reason to exclude higher Landau levels or more generally other energy levels.   
 Indeed, as demonstrated in the present work,  there is a good chance that exotic quantum geometry having not been recognized will be unveiled in  other energy levels.   It may be worthwhile to further proceed in this direction and investigate  quantum geometry by applying the level projection to other physical models.   
The $SO(5)$ Landau model and four-dimensional quantum Hall effect opened  a window to a  research field of  topological phases in higher dimension.  Not  just rendered to be a theoretical issue,     recent technologies of quantum photonics in ultra cold atom have made  experimental explorations 
 possible with the idea of  synthetic dimension  \cite{Zilberberg-Huang-et-al-2018}.  
 The present analysis will be useful not only for theoretical developments of the non-commutative geometry but also for the practical analysis of    higher dimensional topological phases such as quantum Hall effect and Weyl semi-metal \cite{Lian-Zhang-2016,Lian-Zhang-2017}. 

\section*{Acknowledgements}

 I am glad to thank Goro Ishiki for fruitful discussions and inviting me to Tsukuba University for a seminar.  I would like to express my sincere gratitude to late Shou-Cheng Zhang who originally stimulated my interest on this subject.  
 This work was supported by JSPS KAKENHI Grant Number~16K05334 and 16K05138. 

\appendix

\section{Representation theory of the $SO(5)$ group  }\label{sec:represo5theo}

\subsection{$SO(5)$ irreducible representation and the $SO(4)$ decomposition}\label{subsec:so5irrepsandso4decomp}

The $SO(5)$ Casimir operator is constructed as 
\be
\sum_{a<b=1}^5 {L_{ab}}^2 . \label{so4andx5partcasap}
\ee
From the representation theory,   the $SO(5)$ irreducible representation is specified by two integers $(p, q)_{SO(5)}$ with the dimension 
\be
D=\frac{1}{6}(p+2)(q+1)(p+q+3)(p-q+1), \label{simso5pq}
\ee
 and the  $SO(5)$ Casimir eigenvalue of (\ref{so4andx5partcasap}) is given by \footnote{For representation theory of $SO(5)$ (and other classical Lie groups), readers may consult textbooks such as \cite{Wybourne-book} and \cite{Iachello-book}.  In the notation of \cite{Wybourne-book},   
$[[a_1,a_2]] = [[q, p-q]]$ or $[l_1,l_2]=[\frac{1}{2}(p+q), \frac{1}{2}(p-q)]$, and in the notation of  \cite{Iachello-book}, $[\lambda_1, \lambda_2] =[\frac{1}{2}(p+q), \frac{1}{2}(p-q)]$. }
\be
\frac{1}{2}(p^2+q^2)+2p+q . \label{simso5pqeige}
\ee
According to the decomposition $SO(4)\simeq SU(2)_L\otimes SU(2)_R$, we can introduce the $SU(2)_L$ and $SU(2)_R$ angular momentum operators: 
\be
J_i =\frac{1}{4}\eta_{mn}^i L_{mn}, ~~~~K_i=\frac{1}{4}\bar{\eta}_{mn}^i L_{mn}. 
\label{sutwojandkops}
\ee
Here, $\eta_{mn}^i$ and $\bar{\eta}_{mn}^i$ denote the 't Hooft symbols: 
\be
\eta_{mn}^i=\epsilon_{mn i4} +\delta_{m i}\delta_{n 4}-\delta_{m 4}\delta_{n i}, ~~\bar{\eta}_{mn}^i=\epsilon_{mn i4} -\delta_{m i}\delta_{n 4}+\delta_{m 4}\delta_{n i}. 
\ee
$J_i$ and $K_i$ are mutually commutative 
\be
[J_i, K_j]=0, 
\ee
and satisfy 
\be
{J_i}^2=j(j+1),~~~{K_i}^2=k(k+1) ,  
\ee
where each of $j$ and $k$ takes  integer or half-integer values. 
The dimension of the $SO(4)$ irreducible representation  $(j, k)$ is given by 
\be
(2j+1) (2k+1) , 
\ee
and the $SO(4)$ Casimir is expressed as  
\be
{L_{mn}}^2=4(\bs{J}^2+\bs{K}^2)=4(j(j+1)+k(k+1)). 
\ee
The $SO(5)$ irreducible representation is decomposed to the $SO(4)$ irreducible representation as 
\be
(p, q)_{SO(5)} =\sum_{n=0}^q \oplus \biggl(\sum_{s=-\frac{p-q}{2}}^{\frac{p-q}{2}}\oplus ~(j,~k)_{SO(4)}\biggr), 
\ee
where 
\be
j= \frac{n}{2}+\frac{p-q}{4}+\frac{s}{2}, ~~~k=\frac{n}{2}+\frac{p-q}{4}-\frac{s}{2}.
\ee
One may confirm that (\ref{simso5pq}) is given by 
\be
D=\sum_{n=0}^q \sum_{s=-\frac{p-q}{2}}^{\frac{p-q}{2}}(2j+1)(2k+1). 
\ee

\subsection{$SO(5)$ irreducible decomposition of direct products}\label{directproso5}

From (\ref{dcompfavo}), we have 
\begin{align}
&N=0~:~[[1,0]]\otimes [[0, I]] = [[0, I]]\oplus [[1, I ]]\oplus [[ 1, I-2]],  \nn\\
&I=1~:~[[1,0]]\otimes [[N, 1]] = [[N, 1]]\oplus [[N+1, 1 ]]\oplus [[N-1,3]]\oplus [[N-1,1]], \nn\\
&I=0~:~[[1,0]] \otimes [[N, 0]] = [[N+1, 0]]\oplus [[N-1, 2 ]]  \oplus [[N-1, 0]] . 
\end{align} \label{dcompfavo2}
In particular, 
\begin{align}
&[[1,0]]\otimes [[0,1]] =[[1,0]]\oplus [[1,1]] , \nn\\
&[[1,0]]\otimes []1,0]] =[[1,0]]\oplus [[2,0]]   \oplus [[2,-2]]\oplus [[0,2]] \oplus [[0,0]] = [[2,0]]   \oplus [[0,2]] \oplus [[0,0]]   , \nn\\
&[[1,0]]\otimes [[0,2]] =[[0, 2]]\oplus [[1,2]] \oplus [[1, 0]] , 
\end{align}
or 
\begin{align}
&\bs{5}\otimes \bs{4} =\bs{4} \oplus \bs{16}, \label{exmirrepso5deedx} \nn\\
&\bs{5}\otimes \bs{5} =\bs{5}\oplus \bs{14} \oplus (\bs{-5}) \oplus \bs{10}\oplus \bs{1}= \bs{14}  \oplus \bs{10}\oplus \bs{1},  
\nn\\
&\bs{5}\otimes \bs{10} =\bs{10} \oplus \bs{35} \oplus \bs{5}. 
\end{align} 
For other examples of the irreducible decomposition of the tensor product of $SO(5)\simeq USp(4)$, one may consult  \cite{Feger-Kephart-2012} for instance.

\section{Reduction to the $SO(5)$ spherical harmonics}\label{append:so5sphericalharmon}

\subsection{$SO(5)$ free angular momentum Casimir}

In the polar coordinates, the $SO(5)$ free angular momentum operators,  $L_{ab}=-ix_a\frac{\partial}{\partial x_b} +ix_b \frac{\partial}{\partial x_a}$, are expressed as 
\begin{align}
&L_{12}=-i\partial_\phi,~~~L_{13}=i(\cos\phi ~\partial_\theta -\cot\theta\sin\phi ~\partial_{\phi}) , ~~~L_{23} =i(\sin\phi~\partial_{\theta} +\cot\theta\cos\phi~\partial_{\phi}),  \nn\\
&L_{14}=i(\sin\theta\cos\phi\partial_{\chi}+\cot\chi\cos\theta\cos\phi\partial_{\theta}-\cot\chi\frac{1}{\sin\theta}\sin\phi \partial_{\phi}),\nn\\
&L_{24}=i(\sin\theta\sin\phi\partial_{\chi}+\cot\chi\cos\theta\sin\phi\partial_{\theta}+\cot\chi\frac{1}{\sin\theta} \cos\phi\partial_{\phi}),~~L_{34}=i(\cos\theta\partial_{\chi}-\cot\chi \sin\theta\partial_{\theta}),  \nn\\ 
&L_{15}=i(\sin\chi\sin\theta\cos\phi~\partial_{\xi} +\cot\xi\cos\chi\sin\theta\cos\phi ~\partial_\chi +\cot\xi\frac{1}{\sin\chi}\cos\theta\cos\phi~\partial_{\theta}-\cot\xi\frac{1}{\sin\chi}\frac{1}{\sin\theta}\sin\phi ~\partial_{\phi}) ,\nn\\
&L_{25} =i(\sin\chi\sin\theta\sin\phi~\partial_{\xi} +\cot\xi\cos\chi\sin\theta\sin\phi ~\partial_\chi +\cot\xi\frac{1}{\sin\chi}\cos\theta\sin\phi~\partial_{\theta}+\cot\xi\frac{1}{\sin\chi}\frac{1}{\sin\theta}\cos\phi ~\partial_{\phi}), \nn\\
&L_{35} =i(\sin\chi\cos\theta \partial_{\xi}+\cot\xi \cos\chi \cos\theta~\partial_{\chi} -\cot\xi\frac{1}{\sin\chi}\sin\theta~\partial_{\theta}), \nn\\ 
&L_{45} =i(\cos\chi~\partial_{\xi} -\cot\xi\sin\chi~\partial_{\chi}) . 
\label{5dangop}
\end{align}
Notice that $L_{mn}$ $(m,n=1,2,3,4)$ do not depend on $\xi$ and are equal to the polar coordinate expression of the $SO(4)$ free angular momentum operators, $L_{mn} =-iy_m\frac{\partial}{\partial y_n} +iy_n\frac{\partial}{\partial y_m}$.  
The Laplacian on $S^4$ is given by 
\begin{align}
\Delta_{S^4}& =\frac{1}{\sin^3\xi} \partial_{\xi} (\sin^3\xi \partial_{\xi})+\frac{1}{\sin^2\xi}\frac{1}{\sin^2\chi} \partial_{\chi} (\sin^2\chi \partial_{\chi}) +\frac{1}{\sin^2\xi}\frac{1}{\sin^2\chi}  \frac{1}{\sin\theta} \partial_{\theta} (\sin\theta \partial_{\theta}) +\frac{1}{\sin^2\xi}\frac{1}{\sin^2\chi} \frac{1}{\sin^2\theta}{\partial_{\phi}}^2\nn\\
&=-\sum_{a<b=1}^5 {L_{ab}}^2,  
\end{align}
which is related to the Laplacian on $S^3$ as 
\be
\Delta_{S^4} =\frac{1}{\sin^3\xi} \partial_{\xi} (\sin^3\xi \partial_{\xi}) +\frac{1}{\sin^2\xi}\Delta_{S^3}.
\ee
The eigenvalues of the $SO(5)$ free angular momentum  Casimir   is given by 
\be
\sum_{a<b=1}^5 {L_{ab}}^2= N(N+3) , 
\ee
with degeneracy 
\be
D=\frac{1}{6}(N+1)(N+2)(2N+3). 
\ee

\subsection{$SO(5)$ spherical harmonics}

The  $SO(5)$ spherical harmonics are usually given by \cite{Hochstadt-book}
\be 
\Phi_{N n lm} (\Omega_4) =   \Phi_{N n} (\xi)\cdot Y_{nlm}(\Omega_3), \label{so5sphericalharmo1}
\ee
where
\begin{align}
\Phi_{N n} (\xi) &=   \frac{2^{n+1}}{(N+1)!}\sqrt{\frac{(2N+3) ~(N-n)!~(N+n+2)!}{2}}~\cdot~\frac{1}{\sin^{n+2}\xi} P_{N+n+2}^{-(n+1), -(n+1)}(\cos\xi)\nn\\
&=\sqrt{\frac{2N+3}{2}\frac{(N-n)!}{(N+n+2)! }}~\cdot~\frac{1}{\sin\xi}~P_{N+1}^{n+1} (\cos\xi) ,   
\end{align}
with  the associated Legendre polynomials $P^m_n(x)$.\footnote{The associated Legendre polynomials $P_n^m(x)$ are related to the Jacobi polynomials $P_{l}^{(\alpha, \beta)}(x)$ as 
\be
P_n^m(x)=2^m \frac{(n+m)!}{n!}(1-x^2)^{-\frac{m}{2}}\cdot P_{n+m}^{(-m, -m)}(x).  
\label{reljacobilegendre}
\ee
}  
Meanwhile in the present paper,  the  $SO(5)$ monopole harmonics  are given by (\ref{x5so5monopolehamonicsnorm}), which  should be reduced to the $SO(5)$ spherical harmonics at  $I=0$: 
\be
\Psi_{N; \frac{n}{2}, m_L; \frac{n}{2}, m_R} (\Omega_4) = G_{N, \frac{n}{2}, \frac{n}{2}}(\xi) \cdot Y_{\frac{n}{2}, m_L; \frac{n}{2}, m_R}(\Omega_3),~~~~~~(-\frac{n}{2} \le m_L, m_R\le \frac{n}{2}) \label{so5sphericalharmo2}
\ee
where 
\begin{align}
G_{N, \frac{n}{2}, \frac{n}{2}}(\xi) &= \sqrt{N+\frac{3}{2}} ~ \cdot~\frac{1}{\sin\xi}~ d_{N+1, 0, -n-1}(\xi) \nn\\
&=(-1)^{n+1} ~(N+1)!~ \sqrt{\frac{2N+3}{2 ~(N+n+2)!~(N-n)!}}~\cdot~\frac{1}{\sin\xi} 
~\tan^{n+1}(\frac{\xi}{2})~P^{n+1, -(n+1)}_{N+1}(\cos\xi). 
\end{align}
There are superficially distinct  expressions for the $SO(5)$ spherical harmonics, (\ref{so5sphericalharmo1}) and (\ref{so5sphericalharmo2}).  
From the property of the Jacobi polynomials 
\be
(-1)^{n+1} \frac{(N+1)!}{(N-n)!}~ \tan^{n+1}(\frac{\xi}{2})~ P_{N+1}^{n+1, -(n+1)}(\cos\xi) =P_{N+1}^{n+1}(\cos\xi) , 
\ee
we find that the azimuthal parts of the two expressions are identical:   
\be
G_{N, \frac{n}{2}, \frac{n}{2}}(\xi) = \Phi_{N n}(\xi), 
\ee
and the previous $SO(4)$ analysis \cite{Hasebe-2018} tells the relationship between  the $SO(4)$ spherical harmonics parts:  
\be
Y_{nlm}(\Omega_3)=i^l\sum_{m_L, m_R=-n/2}^{n/2} \langle l,m| \frac{n}{2}, m_L; \frac{n}{2}, m_R\rangle  ~Y_{\frac{n}{2}, m; \frac{n}{2}, m'}(\Omega_3). 
\ee
Consequently, 
(\ref{so5sphericalharmo1}) and (\ref{so5sphericalharmo2})  
 are related by the following linear combination  
\be
\Phi_{N n lm} (\Omega_4)  = i^l\sum_{m_L, m_R=-n/2}^{n/2} \langle l,m| \frac{n}{2}, m_L; \frac{n}{2}, m_R\rangle~\Psi_{N; \frac{n}{2}, m_L; \frac{n}{2}, m_R} (\Omega_4)
\ee
or 
\be
 \Psi_{N; \frac{n}{2}, m_L; \frac{n}{2}, m_R} (\Omega_4) =  \sum_{m=-l}^{l} ~(-i)^l \langle \frac{n}{2}, m_L; \frac{n}{2}, m_R | l,m \rangle  ~ \Phi_{N n lm} (\Omega_4), 
\ee
which indicates the equivalence of the two expressions.

\section{The Dirac gauge and the Schwinger gauge  for $S^4$}\label{append:twogaugestrans}

We introduce the Dirac gauge and the Schwinger gauge for $S^4$ and derive a gauge transformation between them.

\subsection{Dirac gauge }

As $S^4$ being a coset space 
\be
S^4 ~\simeq ~SO(5)/SO(4), 
\ee
the non-linear realization is given by \cite{Salam-Strathdee-1982}
\be
\Psi =  e^{i\xi~ y_{m}\sigma_{m 5}} =\frac{1}{\sqrt{2(1+x_5)}} 
\begin{pmatrix}
1+x_5 & x_{m}\bar{q}_{m} \\
-x_{m }q_{m} & 1+x_5
\end{pmatrix} , 
\label{diracmatrix44}
\ee
where $y_m$ are the $S^3$-coordinates (\ref{expliys}), $q_m$ are quaternions (\ref{quaternionsmatrix}),  and $\sigma_{m5}$ are 
\be
\sigma_{m5} =i\frac{1}{2}
\begin{pmatrix}
0 & -\bar{q}_m \\
q_m & 0 
\end{pmatrix}. 
\ee
With the polar coordinates (\ref{polarx1tox5}), (\ref{diracmatrix44}) is expressed as  
\begin{align}
&\Psi=\nn\\
&\begin{pmatrix}
\cos\frac{\xi}{2} & 0 & \sin\frac{\xi}{2} ~(\cos\chi +i\sin\chi\cos\theta) & i\sin\frac{\xi}{2}\sin\chi \sin\theta ~e^{-i\phi} \\
0 & \cos\frac{\xi}{2} & i\sin\frac{\xi}{2}\sin\chi \sin\theta ~e^{i\phi} & \sin\frac{\xi}{2} ~(\cos\chi -i\sin\chi\cos\theta) \\
-\sin\frac{\xi}{2} ~(\cos\chi -i\sin\chi\cos\theta) & i\sin\frac{\xi}{2}\sin\chi \sin\theta ~e^{-i\phi} & \cos\frac{\xi}{2} & 0  \\
i\sin\frac{\xi}{2}\sin\chi \sin\theta ~e^{i\phi} & -\sin\frac{\xi}{2} ~(\cos\chi +i\sin\chi\cos\theta) & 0 & \cos\frac{\xi}{2} 
\end{pmatrix} .  \label{polarcoordiracs4sp}
\end{align}
We decompose $\Psi$ into two $4\times 2$ rectangular matrices: 
\be
\Psi=\begin{pmatrix}
\Psi_L & \Psi_R
\end{pmatrix},  \label{tworecdecomp}
\ee
where  
\be
\Psi_L = \frac{1}{\sqrt{2(1+x_5)}} 
\begin{pmatrix}
1+x_5  \\
-x_{m }q_{m} 
\end{pmatrix},~~~~~\Psi_R = \frac{1}{\sqrt{2(1+x_5)}} 
\begin{pmatrix}
x_{m}\bar{q}_{m} \\
1+x_5 
\end{pmatrix}.
\ee
The 2nd Hopf map (\ref{2ndhopfmap}) can be  expressed as 
\be
x_m 1_2 = -\Psi_L^{\dagger}\gamma_m \Psi_L =\Psi_R^{\dagger}\gamma_m \Psi_R, ~~x_5 1_2 =\Psi_L^{\dagger}\gamma_5 \Psi_L =-\Psi_R^{\dagger}\gamma_5 \Psi_R, 
\ee
and the associated connections are  
\be
\omega_L^{\text{D}}=-i{\Psi_L}^{\dagger}d\Psi_L=-i\frac{1}{2(1+x_5)}\eta_{mn}^i  x_{n} q_i dx_{m}, ~~~~~\omega_R^{\text{D}}=-i{\Psi_R}^{\dagger}d\Psi_R=-i\frac{1}{2(1+x_5)}\bar{\eta}_{mn}^i  x_{n}q_i dx_{m}. 
\ee
Here, $\text{D}$ of $\omega^{\text{D}}$ denotes the Dirac gauge. 
More comprehensively, 
\begin{align}
A^{\text{D}}
\equiv -i{\Psi^{\text{D}}}^{\dagger}d\Psi^{\text{D}} &= \begin{pmatrix}
{-i\Psi_L^{\dagger}d\Psi_L} & -i\Psi_L^{\dagger}d\Psi_R\\
-i\Psi_R^{\dagger}d\Psi_L & {-i\Psi_R^{\dagger}d\Psi_R}
\end{pmatrix} = \begin{pmatrix}
\omega_L^{\text{D}} & -i\Psi_L^{\dagger}d\Psi_R\\
-i\Psi_R^{\dagger}d\Psi_L & \omega^{\text{D}}_R 
\end{pmatrix} \nn\\
&=-i\frac{1}{2(1+x_5)}\begin{pmatrix}
\eta_{mn}^i q_i x_{n}dx_{m} & -x_{m}\bar{q}_{m}dx_5 +(1+x_5)\bar{q}_{m}dx_{m} \\
x_{m}{q}_{m}dx_5 -(1+x_5){q}_{m}dx_{m} & \bar{\eta}_{mn}^i q_i x_{n}dx_{m} 
\end{pmatrix}. \label{diracbiga} 
\end{align}
 $\omega_L^{\text{D}}$ is equal to the $SU(2)$ (anti-)monopole gauge field (\ref{zhanghusu2gauge}) for $I=1$: 
\be
\omega_L^{\text{D}} 
=\frac{1}{4}\omega_{mn}^{\text{D}} \eta_{mn}^i\sigma_i .  
\ee
In the Cartesian coordinates, $\omega_{mn}^{\text{D}}$ are given by 
\be
\omega_{mn}^{\text{D}} = \frac{1}{1+x_5}(x_{m}dx_{n} -x_n dx_m),  
\label{compomegamnd}
\ee
while in the polar coordinates, 
\begin{align}
&\omega_{12}^{\text{D}} =2\sin^2\frac{\xi}{2}\sin^2\chi \sin^2\theta d\phi  , \nn\\
&\omega_{13}^{\text{D}} = -2\sin^2\frac{\xi}{2}\sin^2\chi \cos\phi d\theta +2\sin^2\frac{\xi}{2}\sin^2\chi \sin\theta\cos\theta \sin\phi d\phi , \nn\\
&\omega_{14}^{\text{D}} =  -2\sin^2\frac{\xi}{2}\sin\theta\cos\phi d\chi -2\sin^2\frac{\xi}{2}\sin\chi\cos\chi \cos\theta \cos\phi d\theta +2\sin^2\frac{\xi}{2}\sin\chi\cos\chi \sin\theta \sin\phi d\phi, \nn\\
&\omega_{23}^{\text{D}} =-2\sin^2\frac{\xi}{2}\sin^2\chi \sin\phi d\theta -2\sin^2\frac{\xi}{2}\sin^2\chi \sin\theta\cos\theta\cos \phi d\phi  , \nn\\
&\omega_{24}^{\text{D}} = -2\sin^2\frac{\xi}{2}\sin\theta \sin\phi d\chi -2 \sin^2\frac{\xi}{2}\sin\chi\cos\chi\cos\theta \sin\phi d\theta -2  \sin^2\frac{\xi}{2}\sin\chi\cos\chi\sin\theta \cos\phi d\phi , \nn\\
&\omega_{34}^{\text{D}} = -2\sin^2\frac{\xi}{2}\cos\theta d\chi +2\sin^2\frac{\xi}{2}\sin\chi\cos\chi\sin\theta d\theta . \label{diracgaugespinconne}
\end{align}

\subsection{Schwinger gauge}

$\Psi$ (\ref{polarcoordiracs4sp}) is  factorized as 
\be
\Psi(\xi,\chi, \theta, \phi) =H(\chi,\theta,\phi)^{\dagger}\cdot e^{i\xi \sigma_{45}} \cdot H(\chi,\theta,\phi), \label{factorizationpsidmat}
\ee
where 
\be
e^{i\xi \sigma_{45}}= \begin{pmatrix}
\cos\frac{\xi}{2} ~{1}_2 & \sin\frac{\xi}{2} ~{1}_2 \\
-\sin\frac{\xi}{2} ~{1}_2 & \cos\frac{\xi}{2} ~{1}_2 
\end{pmatrix}   ,    
\ee
and 
\be
H(\chi, \theta, \phi)
 = e^{i\chi \sigma_{43}} ~e^{i\theta \sigma_{31}} ~e^{i\phi\sigma_{12}}=\begin{pmatrix}
H_L(\chi, \theta, \phi) & 0 \\
0 & H_R(\chi, \theta, \phi)
\end{pmatrix}.  \label{44hmat}
\ee
$H_L$ and $H_R$ are  the chiral Hopf spinor matrices \cite{Hasebe-2018} 
\begin{align}
&H_L(\chi, \theta, \phi) =e^{-i\frac{\chi}{2}\sigma_z} e^{i\frac{\theta}{2}\sigma_y}e^{i\frac{\phi}{2}\sigma_z} =
\begin{pmatrix}
\cos(\frac{\theta}{2}) ~e^{-i\frac{1}{2}(\chi-\phi)} & \sin(\frac{\theta}{2}) ~e^{-i\frac{1}{2}(\chi+\phi)} \nn\\
-\sin(\frac{\theta}{2}) ~e^{i\frac{1}{2}(\chi+\phi)} & \cos(\frac{\theta}{2}) ~e^{i\frac{1}{2}(\chi-\phi)}
\end{pmatrix}, \nn\\
&H_R(\chi, \theta, \phi)\equiv H_L(-\chi,\theta,\phi) =e^{i\frac{\chi}{2}\sigma_z} e^{i\frac{\theta}{2}\sigma_y}e^{i\frac{\phi}{2}\sigma_z} =
\begin{pmatrix}
\cos(\frac{\theta}{2}) ~e^{i\frac{1}{2}(\chi+\phi)} & \sin(\frac{\theta}{2}) ~e^{i\frac{1}{2}(\chi-\phi)} \nn\\
-\sin(\frac{\theta}{2}) ~e^{-i\frac{1}{2}(\chi-\phi)} & \cos(\frac{\theta}{2}) ~e^{-i\frac{1}{2}(\chi+\phi)}
\end{pmatrix} . 
\end{align}
Though $H$ is a 4$\times$4 matrix, $H$ carries the $SU(2)$ degrees of freedom parameterized by the three parameters $(\chi, \theta, \phi)$.  
We introduce a new $4\times 4$ matrix $\Phi$ as   
\be
\Psi(\xi,\chi,\theta,\phi ) =\Phi(\xi,\chi,\theta,\phi ) \cdot H (\chi,\theta,\phi), \label{impdiracschwin}
\ee
or 
\be
\Phi(\xi,\chi,\theta,\phi ) =
 H(\chi, \theta, \phi)^{\dagger}\cdot  e^{i\xi \sigma_{45}}.  \label{phihdage45}
\ee
In the polar coordinates, (\ref{phihdage45}) is given by 
\be
\Phi 
=
\begin{pmatrix}
\cos\frac{\xi}{2}~\cos\frac{\theta}{2} ~e^{i\frac{1}{2}(\chi-\phi)}  & -\cos\frac{\xi}{2}~\sin\frac{\theta}{2} ~e^{-i\frac{1}{2}(\chi+\phi)} &  \sin\frac{\xi}{2}~\cos\frac{\theta}{2} ~e^{i\frac{1}{2}(\chi-\phi)}  & -\sin\frac{\xi}{2}~\sin\frac{\theta}{2} ~e^{-i\frac{1}{2}(\chi+\phi)}  \\
\cos\frac{\xi}{2}~\sin\frac{\theta}{2} ~e^{i\frac{1}{2}(\chi+\phi)}  & \cos\frac{\xi}{2}~\cos\frac{\theta}{2} ~e^{-i\frac{1}{2}(\chi-\phi)} &  \sin\frac{\xi}{2}~\sin\frac{\theta}{2} ~e^{i\frac{1}{2}(\chi+\phi)}  & \sin\frac{\xi}{2}~\cos\frac{\theta}{2} ~e^{-i\frac{1}{2}(\chi-\phi)}   \\
-\sin\frac{\xi}{2}~\cos\frac{\theta}{2} ~e^{-i\frac{1}{2}(\chi+\phi)}  & \sin\frac{\xi}{2}~\sin\frac{\theta}{2} ~e^{i\frac{1}{2}(\chi-\phi)} &  \cos\frac{\xi}{2}~\cos\frac{\theta}{2} ~e^{-i\frac{1}{2}(\chi+\phi)}  & -\cos\frac{\xi}{2}~\sin\frac{\theta}{2} ~e^{i\frac{1}{2}(\chi-\phi)} \\
-\sin\frac{\xi}{2}~\sin\frac{\theta}{2} ~e^{-i\frac{1}{2}(\chi-\phi)}  & -\sin\frac{\xi}{2}~\cos\frac{\theta}{2} ~e^{i\frac{1}{2}(\chi+\phi)} &  \cos\frac{\xi}{2}~\sin\frac{\theta}{2} ~e^{-i\frac{1}{2}(\chi-\phi)}  & \cos\frac{\xi}{2}~\cos\frac{\theta}{2} ~e^{i\frac{1}{2}(\chi+\phi)}  \\ 
\end{pmatrix}. 
\label{polarcoorschs4sp}
\ee 
As in the case of $\Psi$ (\ref{tworecdecomp}), we decompose  $\Phi$ as 
\be
\Phi=(\Phi_L~~ \Phi_R) , 
\ee
where 
\be
\Phi_L
= \begin{pmatrix}
\cos\frac{\xi}{2} ~H_L^{\dagger} \\
-\sin\frac{\xi}{2} ~H_R^{\dagger}
\end{pmatrix} , 
~~~~\Phi_R= 
\begin{pmatrix}
\sin\frac{\xi}{2} ~H_L^{\dagger} \\
\cos\frac{\xi}{2} ~H_R^{\dagger}
\end{pmatrix} . 
\ee 
The corresponding connection is derived as 
\be
\omega_L^\text{S}=-i\Phi_L^{\dagger}d\Phi_L =-i\frac{1}{2}(H_L dH_L^{\dagger} +H_R dH_R^{\dagger}) -i\frac{1}{2}\cos\xi ~(H_L dH_L^{\dagger} -H_R dH_R^{\dagger}),  
\label{formgauges}
\ee
where 
\begin{subequations}
\begin{align}
-i\frac{1}{2}(H_L dH_L^{\dagger} +H_R dH_R^{\dagger})&=\cos\chi ~\sin\theta ~d\phi \frac{1}{2}\sigma_x-\cos\chi~ d\theta ~\frac{1}{2}\sigma_y-\cos\theta~ d\phi~ \frac{1}{2}\sigma_z, 
 \label{gaugeons3su2}\\
-i\frac{1}{2}(H_L dH_L^{\dagger} -H_R dH_R^{\dagger})&=\sin\chi  ~d\theta \frac{1}{2}\sigma_x+\sin\chi\sin\theta~ d\phi ~\frac{1}{2}\sigma_y+d\chi~ \frac{1}{2}\sigma_z. 
\end{align}
\end{subequations}
$\text{S}$ of $\omega^{\text{S}}$ stands for the Schwinger gauge.\footnote{(\ref{gaugeons3su2})  denotes the $SU(2)$ gauge field of the $SO(4)$ Landau model in the Schwinger gauge \cite{Hasebe-2018}.}   $\omega_L^\text{S}$   is explicitly given by 
\be
\omega_L^{\text{S}} =i\frac{1}{2}\omega_L^i q_i 
\label{omegaschloneexp}
\ee
where 
\be
\omega_L^x =\cos\chi~\sin\theta~ d\phi +\cos\xi~\sin\chi~d\theta, ~~~\omega_L^y =-\cos\chi ~d\theta +\cos\xi\sin\chi\sin\theta~d\phi, ~~~\omega_L^z=-\cos\theta ~d\phi +\cos\xi ~d\chi. 
\ee 
With 
$\omega_{L}^i \equiv \frac{1}{2}\eta_{mn}^i\omega_{mn}^{\text{S}}$,  
(\ref{omegaschloneexp}) can be rewritten as 
\be
\omega_L^\text{S}
= i\frac{1}{4}\omega^{\text{S}}_{mn}\eta_{mn}^i q_i= i\frac{1}{4} \omega^{\text{S}}_{mn \mu}\eta_{mn}^i q_i 
~dx^{\mu},  ~~~~~(dx^{\mu}= d\theta, d\phi, d\chi, d\xi) \label{defomegschl}
\ee
where 
\begin{align}
&\omega^\text{S}_{12} =-\cos\theta d\phi, ~~~\omega^\text{S}_{13} =\cos\chi d\theta,~~~\omega^\text{S}_{14} =\cos\xi\sin\chi d\theta,  \nn\\
&\omega^\text{S}_{23} =\cos\chi\sin\theta d\phi,  ~~~\omega^\text{S}_{24}=\cos\xi\sin\chi\sin\theta d\phi, ~~~\omega^\text{S}_{34} =\cos\xi d\chi. \label{schwingerspinconncov}
\end{align}
It is straightforward to check that (\ref{schwingerspinconncov}) satisfies  the Cartan structure equation:  
\be
de^{\text{S}}_m  +\omega^{\text{S}}_{mn}e^{\text{S}}_n =0 , 
\ee
with the vierbein in the Schwinger gauge\footnote{The numbering of the vierbein here (\ref{schwingervier}) is different from that of (\ref{vierbeinchoicerela}).  }   
\be
e^{\text{S}}_1 =\sin\xi\sin\chi d\theta, ~~e^{\text{S}}_2 =\sin\xi\sin\chi\sin\theta d\phi, ~~e^{\text{S}}_3  =\sin\xi d\chi, ~~e^{\text{S}}_4 =d\xi. 
\label{schwingervier}
\ee 
Similarly, we have 
\be
\omega_R^{\text{S}}=-i\Phi_R^{\dagger}d\Phi_R =-i\frac{1}{2}(H_L dH_L^{\dagger} +H_R dH_R^{\dagger}) +i\frac{1}{2}\cos\xi ~(H_L dH_L^{\dagger} -H_R dH_R^{\dagger}) 
= i\frac{1}{2}\omega_R^i q_i , 
\ee 
with 
\be
\omega_{R}^i \equiv \frac{1}{2}\bar{\eta}_{mn}^i\omega_{mn}^{\text{S}}, 
\ee
and 
\be
A^{\text{S}} =-i{\Phi}^{\dagger}d\Phi=
\begin{pmatrix}
-i\Phi_L^{\dagger}d\Phi_L & -i\Phi_L^{\dagger}d\Phi_R \\
-i\Phi_R^{\dagger}d\Phi_L & -i\Phi_R^{\dagger}d\Phi_R
\end{pmatrix} = \begin{pmatrix}
\omega_L^S & -i\Phi_L^{\dagger}d\Phi_R \\
-i\Phi_R^{\dagger}d\Phi_L & \omega_R^S
\end{pmatrix}. \label{schwinbiga}
\ee

\subsection{Gauge transformation and vierbein in the Dirac gauge}

From the relation  (\ref{impdiracschwin}), 
we have 
\be
\Psi_L=\Phi_L\cdot H_L, ~~~~~~\Psi_R=\Phi_R\cdot H_R, 
\ee
and so (\ref{diracbiga}) and (\ref{schwinbiga}) are related as  
\be
A^{\text{D}} =H^{\dagger}A^{\text{S}} H-iH^{\dagger}dH.  \label{trangauas}
\ee
(\ref{trangauas}) implies 
\be
\omega_L^{\text{D}} =H_L^{\dagger} ~\omega_L^{\text{S}}~ H_L-iH_L^{\dagger}~dH_L, ~~~~~\omega_R^{\text{D}} =H_R^{\dagger} ~\omega_R^{\text{S}}~ H_R-iH_R^{\dagger}~dH_R,
\ee
or 
\be
\omega_L^{\text{S}} =H_L ~\omega_L^{\text{D}}~ H_L^{\dagger}-iH_LdH_L^{\dagger}, ~~~~~\omega_R^{\text{S}} =H_R ~\omega_R^{\text{D}}~ H_R^{\dagger}-iH_RdH_R^{\dagger}. 
\ee
We then find that the $SO(4)$ matrix-valued spin connections 
\be 
\omega^{\text{D}/\text{S}} \equiv \begin{pmatrix}
\omega_L^{\text{D}/\text{S}} & 0 \\
0 & \omega_R^{\text{D}/\text{S}}
\end{pmatrix} 
= \frac{1}{4}\omega_{mn}^{\text{D}/\text{S}}\begin{pmatrix}
\eta_{mn}^i\sigma_i & 0 \\
0 & \bar{\eta}_{mn}^i\sigma_i
\end{pmatrix}, 
\label{omegamatricesintro}
\ee
are also related by the gauge transformation 
\be
\omega^\text{D} =H^{\dagger} \omega^\text{S} H -iH^{\dagger}dH. 
\ee

Under the $SU(2)$  transformation  $H$ (\ref{44hmat}), the gamma matrices are transformed as 
\be
H^{\dagger} \gamma_m H =\gamma_n O_{nm} \label{su2transfhgam}
\ee
with  
\begin{align}
O &= (e^{i\chi t_{43}} e^{i\theta t_{31} } e^{i\phi t_{12}})^t =e^{-i\phi t_{12}} e^{-i\theta t_{31} }e^{-i\chi t_{43}} \nn\\ 
&=\begin{pmatrix}
\cos\theta \cos\phi & -\sin\phi & \cos\chi\sin\theta\cos\phi & \sin\chi \sin\theta \cos\phi \\
\cos\theta \sin\phi &  \cos\phi & \cos\chi\sin\theta\sin\phi & \sin\chi\sin\theta\sin\phi \\
-\sin\theta  & 0 & \cos\chi\cos\theta & \sin\chi\cos\theta \\
0 & 0 & -\sin\chi & \cos\chi
\end{pmatrix}. 
\end{align}
Here $t_{mn}$ are  the adjoint representation $SO(4)$ generators: 
\be
(t_{mn})_{pq} =-i\delta_{mp}\delta_{nq} +i\delta_{mq}\delta_{np}. 
\ee
Since the vierbein carries local coordinate indices, the vierbein transforms similarly to (\ref{su2transfhgam}).  
Therefore, the vierbein in the Dirac gauge 
can be obtained from the vierbein in the Schwinger gauge:  
\be
e^\text{D}_m =O_{mn} e^\text{S}_n. 
\label{viertrans}
\ee
With the expression of  $e_m^{\text{S}}$ (\ref{schwingervier}), $e^{\text{D}}_m$ are explicitly given by 
\begin{align}
&e^\text{D}_1 =  \sin\chi\sin\theta\cos\phi~d\xi+ \sin\xi \cos\chi\sin\theta\cos\phi~ d\chi +\sin\xi\sin\chi\cos\theta\cos\phi~ d\theta -\sin\xi\sin\chi\sin\theta\sin\phi d\phi  , \nn\\
&e^\text{D}_2 =   \sin\chi\sin\theta\sin\phi~d\xi+ \sin\xi \cos\chi\sin\theta\sin\phi~ d\chi +\sin\xi\sin\chi\cos\theta\sin\phi~ d\theta +\sin\xi\sin\chi\sin\theta\cos\phi d\phi , \nn\\
&e^\text{D}_3=\sin\chi\cos\theta  d\xi+ \sin\xi \cos\chi\cos\theta d\chi -\sin\xi\sin\chi\sin\theta d\theta \nn\\
&e^\text{D}_4=\cos\chi d\xi -\sin\xi\sin\chi d\chi. \label{diracervier}
\end{align}
It is straightforward to show that (\ref{diracgaugespinconne})  and (\ref{diracervier}) satisfy  the Cartan structure equation: 
\be
de^\text{D}_m +\omega^\text{D}_{mn}e^\text{D}_n=0. \label{torstionfreedirac}
\ee
We thus successfully obtained the vierbein in the Dirac gauge from the relation (\ref{viertrans}). 
On the other hand, it will be a formidable task to derive the vierbein in the Dirac gauge  from the Cartan structure equation (\ref{torstionfreedirac}) with the spin connection (\ref{diracgaugespinconne}).  

\subsection{Curvature}\label{append:s4curvature}

The gauge fields $A^{\text{D}}$ (\ref{diracbiga}) and  $A^{\text{S}}$ (\ref{schwinbiga}) are a pure gauge, and indeed $dA^{\text{D}}+i{A^{\text{D}}}^2=dA^{\text{S}}+i{A^{\text{S}}}^2=0$.  Non-zero curvatures are obtained from $\omega^{\text{D/S}}$ (\ref{omegamatricesintro}) as 
\be
f^{\text{D/S}} =d\omega^{\text{D/S}} +i{\omega^{\text{D/S}}}^2 =\frac{1}{2}f^{\text{D/S}}_{mn} \sigma_{mn} 
\ee
with 
\be
f_{mn}^{\text{D/S}} =e^{\text{D/S}}_m \wedge e^{\text{D/S}}_n. 
\ee
$f^{\text{D}}$ and $f^{\text{S}}$ are related by  
\be
f^{\text{D}} =H^{\dagger}f^{\text{S}} ~H . 
\ee
The Riemann curvature can be read off from 
\be
f^{\text{D/S}}_{mn}=\frac{1}{2}R^m_{~~npq}~e^{\text{D/S}}_p\wedge e^{\text{D/S}}_q, 
\ee
as 
\be
R_{1212}=R_{1313}=R_{1414}=R_{2323}=R_{2424}=R_{3434}=1, 
\ee
and the Ricci scalar is obtained as 
\be
R=R^m_{~~nmn}=2\times 6=12. \label{riccis4value}
\ee

\section{Matrix elements for three-sphere coordinates}\label{appendix:so4append} 

Using the integration formula for three $SO(4)$ monopole harmonics [see Sec.6.1 of \cite{Hasebe-2018}]  
\begin{align}
&\int d\Omega_3 
~\bs{\Phi}_{(l_L, m_L; l_R, m_R)}^{\dagger}~ 
{\Phi}_{\frac{1}{2}, \frac{\sigma}{2}; \frac{1}{2},  \frac{\tau}{2}}
 ~\bs{\Phi}_{(l'_L, n_L; l'_R, n_R)}\nn\\
&=\sqrt{\frac{(2l'_L +1)(2l'_R+1) (I+1)2}{\pi^2}} ~\begin{Bmatrix}
l_L & l_R & \frac{I}{2} \\
\frac{1}{2} & \frac{1}{2} & 0 \\
l'_L & l'_R & \frac{I}{2}
\end{Bmatrix} ~C_{\frac{1}{2}, \frac{1}{2}\sigma; ~l'_L, n_L}^{l_L, m_L}~C_{\frac{1}{2}, \frac{1}{2}\tau;~ l'_R, n_R}^{l_R, m_R}\nn\\
&=\sqrt{\frac{(2l'_L +1)(2l'_R+1) (I+1)2}{\pi^2}} ~(-1)^{l_L+l'_L + l_R+l'_R +I+1}~\begin{Bmatrix}
l_L & l_R & \frac{I}{2} \\
l'_L & l'_R & \frac{I}{2} \\
\frac{1}{2} & \frac{1}{2} & 0 
\end{Bmatrix} ~C_{\frac{1}{2}, \frac{1}{2}\sigma; ~l'_L, n_L}^{l_L, m_L}~C_{\frac{1}{2}, \frac{1}{2}\tau;~ l'_R, n_R}^{l_R, m_R}\nn\\
&=\frac{\sqrt{(2l'_L +1)(2l'_R+1)}}{\pi} ~(-1)^{l_L+ 2l'_L + 2l_R+l'_R +\frac{3}{2}(I+1)} ~\begin{Bmatrix}
l_L & l_R & \frac{I}{2} \\
l'_R & l'_L & \frac{1}{2}
\end{Bmatrix} ~C_{\frac{1}{2}, \frac{1}{2}\sigma; ~l'_L, n_L}^{l_L, m_L}~C_{\frac{1}{2}, \frac{1}{2}\tau;~ l'_R, n_R}^{l_R, m_R}, \label{ysmatrixfour}
\end{align}
we will derive the three-sphere matrix coordinates. 

\subsection{About
 the $SO(4)$ spinor spherical harmonics}\label{appendix:so4appendspin}

We evaluate the matrix elements of $y_m$ (\ref{s3coordinatessphericalrel}) sandwiched by the $SO(4)$ spin  spherical harmonics: 
\be
\langle \bs{Y}_{j' m'_j; k' m'_k}| y_{m} |\bs{Y}_{j m_j; k m_k}\rangle =\int  d\Omega_3 
~\bs{Y}_{j' m'_j; k' m'_k}^{\dagger}~ 
y_{m}
 ~\bs{Y}_{j m_j; k m_k}. \label{evasphematel}
\ee
With 
(\ref{vectorlikespinsphereharmo}),  (\ref{evasphematel}) can be rewritten  as 
\begin{align}
\!\!\!\langle \bs{Y}_{j' m'_j; k' m'_k}| y_{m} |\bs{Y}_{j m_j; k m_k}\rangle &=\sum_{A=-\frac{I}{2}}^{\frac{I}{2}}\int  d\Omega_3 
~(\bs{Y}_{j' m'_j; k' m'_k})_A^{*}~ 
y_{m}
 ~(\bs{Y}_{j m_j; k m_k})_A\nn\\
 &=\sum_{A}\sum_{m'_R=-j'}^{j'}\sum_{m_R=-j}^j ~{C_{j' m'_R; \frac{I}{2} A}^{k' m'_k}}~C_{j m_R; \frac{I}{2} A}^{k m_k} ~\int ~d\Omega_3 ~\Phi_{j' m'_j: j' m'_R}^{*} y_{m} \Phi_{j m_j; j m_R}, \label{braketys}
\end{align}
where we used that the Clebsch-Gordan coefficients are real. 
Since the $SO(4)$ spherical harmonics are equal to the monopole harmonics for $I=0$: 
\be
\bs{\Phi}_{j, m_j; k, m_k}|_{j=k=\frac{p}{2}} =\Phi_{j, m_j; j, m_k},  
\ee
(\ref{ysmatrixfour}) gives 
\be 
\int d\Omega_3 ~\Phi_{j' m'_j; j' m'_R}^* ~\Phi_{\frac{1}{2} \frac{\sigma}{2} ; \frac{1}{2} \frac{\tau}{2}} ~\Phi_{j m_j; j, m_R} 
=-\frac{1}{\pi}\sqrt{\frac{2j+1}{2j'+1}}  ~ (-1)^{2(j+j')}~ C_{\frac{1}{2} \frac{\sigma}{2}; j m_j}^{j' m'_j} ~C_{\frac{1}{2} \frac{\tau}{2}; j m_R}^{j' m'_R},  \label{threebodyintsph}
\ee 
where $\begin{Bmatrix} 
j' & j' & 0 \\
j & j & \frac{1}{2} 
\end{Bmatrix} =-i(-1)^{-(j+j')}\frac{1}{\sqrt{(2j+1)(2j'+1)}}$ was used.  Therefore, with respect to  $y_1=-i\frac{\pi}{2}(\Phi_{\frac{1}{2}, \frac{1}{2};\frac{1}{2}, \frac{1}{2} }- \Phi_{\frac{1}{2}, -\frac{1}{2};\frac{1}{2}, -\frac{1}{2} })$,  (\ref{braketys}) becomes  
\begin{align}
&\langle  y_{1} \rangle 
=-i\frac{\pi}{2}  \sum_{A}\sum_{m'_R=-j'}^{j'}\sum_{m_R=-j}^j ~{C_{j' m'_R; \frac{I}{2} A }^{k' m'_k}}~C_{j m_R; \frac{I}{2} A}^{k m_k} ~\int ~d\Omega_3 ~\Phi_{j' m'_j: j' m'_R}^{*} (\Phi_{\frac{1}{2}, \frac{1}{2}; \frac{1}{2}, \frac{1}{2}}-  \Phi_{\frac{1}{2}, -\frac{1}{2}; \frac{1}{2}, -\frac{1}{2}})\Phi_{j m_j; j m_R} \nn\\
&=i(-1)^{2(j+j')}\frac{1}{2}\sqrt{\frac{2j+1}{2j'+1}}\sum_{A}\sum_{m'_R=-j'}^{j'}\sum_{m_R=-j}^j ~{C_{j' m'_R ; \frac{I}{2} A }^{k' m'_k}}~C_{j m_R; \frac{I}{2} A}^{k m_k}   ~(C_{\frac{1}{2} \frac{1}{2}; j m_j}^{j' m'_j} ~C_{\frac{1}{2} \frac{1}{2}; j m_R}^{j' m'_R}- C_{\frac{1}{2} -\frac{1}{2}; j m_j}^{j' m'_j} ~C_{\frac{1}{2} -\frac{1}{2}; j m_R}^{j' m'_R} )\nn\\
&=i(-1)^{2(j+j')}\frac{1}{2}\sqrt{\frac{2j+1}{2j'+1}} \nn\\
&~~~~\times \biggl(C_{\frac{1}{2} \frac{1}{2}; j m_j}^{j' m'_j} \overbrace{\sum_{A, m'_R, m_R}~C_{j m_R; \frac{I}{2} A}^{k m_k}~C_{j' m'_R ; \frac{I}{2} A }^{k' m'_k}    ~C_{\frac{1}{2} \frac{1}{2}; j m_R}^{j' m'_R}}-     C_{\frac{1}{2} -\frac{1}{2}; j m_j}^{j' m'_j} \overbrace{\sum_{A, m'_R, m_R} ~C_{j m_R; \frac{I}{2} A}^{k m_k} ~C_{j' m'_R ; \frac{I}{2} A }^{k' m'_k}~C_{\frac{1}{2} -\frac{1}{2}; j m_R}^{j' m'_R}} \biggr). \label{halfwaycalcu}
\end{align}
To proceed further,  we need to calculate 
\begin{align}
\overbrace{\sum_{A, m'_R, m_R}~C_{j m_R; \frac{I}{2} A}^{k m_k}   ~C_{j' m'_R ; \frac{I}{2} A }^{k' m'_k} ~C_{\frac{1}{2} \frac{\tau}{2}; j m_R}^{j' m'_R}}
&=(-1)^{\frac{1}{2}+j-j'}
~\sum_{A, m'_R, m_R} ~C_{j m_R; \frac{I}{2} A}^{k m_k} ~C_{j' m'_R; \frac{I}{2} A}^{k' m'_k} ~C_{j m_R; \frac{1}{2} \frac{\tau}{2}}^{j' m'_R}
\nn\\
&=-(-1)^{j+k+\frac{I}{2}} ~\sqrt{(2k+1)(2j'+1)} ~C_{k m_k; \frac{1}{2} \frac{\tau}{2}}^{k' m'_k}
\begin{Bmatrix}
j  & k & \frac{I}{2} \\
k'  & j' & \frac{1}{2}
\end{Bmatrix} \nn\\
&=-i(-1)^{j+2k+\frac{I}{2}-k'} ~\sqrt{(2k+1)(2j'+1)} ~C_{\frac{1}{2} \frac{\tau}{2}; ~k m_k}^{k' m'_k}
\begin{Bmatrix}
j  & k & \frac{I}{2} \\
k'  & j' & \frac{1}{2}
\end{Bmatrix} , 
\end{align}
where  in the first and last equations we used 
\be 
 C_{j m; j' m'}^{k n} =(-1)^{j+j'-k}C_{j' m'; j m}^{k n} , 
\ee
and in the second equation,  (Eq.(12) in p.260 of \cite{Varshalovich-et-al-Book}) 
\be
\sum_{\alpha, \beta, \delta}~C_{a \alpha; b \beta}^{c \gamma}~C_{d \delta; b \beta}^{e \epsilon}    ~C_{ a \alpha; f \varphi}^{d \delta}  =(-1)^{b+c+d+f} ~\sqrt{(2c+1)(2d+1)}~C_{c \gamma ; f \varphi}^{e \epsilon}
\begin{Bmatrix}
a & c & b \\
e & d & f
\end{Bmatrix}. 
\ee
Consequently, 
\begin{align}
 &\langle \bs{Y}_{j' m'_j; k' m'_k}| y_{1} |\bs{Y}_{j m_j; k m_k}\rangle  \nn\\
&=\frac{\sqrt{(2j+1)(2k+1)}}{2}(-1)^{-j+2j'+2k-k'+\frac{I}{2}}  ~\begin{Bmatrix}
 j' & k' & \frac{I}{2} \\
 k & j & \frac{1}{2}\end{Bmatrix}
 ~(C_{\frac{1}{2} \frac{1}{2}; j m_j}^{j' m_j'}C_{\frac{1}{2} \frac{1}{2}; k m_k}^{k' m_k'} - C_{\frac{1}{2} -\frac{1}{2}; j m_j}^{j' m_j'} C_{\frac{1}{2} -\frac{1}{2}; k m_k}^{k' m_k'}    ). 
 \label{matricaly1el}
\end{align}
We used the fact that $j$ takes a half-integer or integer value and so $(-1)^{4j}=1$, and the property of the 6j symbol, $\begin{Bmatrix}
 a & b & c \\
 d & e & f \end{Bmatrix} = \begin{Bmatrix}
 e & d & c \\
 b & a & f \end{Bmatrix}$. 

When $j+k=j'+k'=n+\frac{I}{2}$, the signature part of (\ref{matricaly1el}) is simplified as 
\be
(-1)^{-j+2j'+2k-k'+\frac{I}{2}} = (-1)^{n+I-k+k'}, 
\ee
and  the matrix elements of $y_1$ and other coordinates become 
\begin{align}
&\langle \bs{Y}_{j' m'_j; k' m'_k}| y_{m=1,2} |\bs{Y}_{j m_j; k m_k}\rangle  \nn\\
&~~~~~~~~~~~=(-i)^{m-1}\frac{\sqrt{(2j+1)(2k+1)} }{2}(-1)^{n+I-k+k'} ~\begin{Bmatrix}
 j' & k' & \frac{I}{2} \\
 k & j & \frac{1}{2}\end{Bmatrix}
 ~(C_{\frac{1}{2} \frac{1}{2}; j m_j}^{j' m_j'}C_{\frac{1}{2} \frac{1}{2}; k m_k}^{k' m_k'} +(-1)^m C_{\frac{1}{2} -\frac{1}{2}; j m_j}^{j' m_j'} C_{\frac{1}{2} \frac{1}{2}; k m_k}^{k' m_k'}    ), \nn\\
&\langle \bs{Y}_{j' m'_j; k' m'_k}| y_{m=3,4} |\bs{Y}_{j m_j; k m_k}\rangle  \nn\\
&~~~~~~~~~~~=(-i)^{m-1}\frac{\sqrt{(2j+1)(2k+1)} }{2}(-1)^{n+I-k+k'} ~\begin{Bmatrix}
 j' & k' & \frac{I}{2} \\
 k & j & \frac{1}{2}\end{Bmatrix}
 ~(C_{\frac{1}{2} \frac{1}{2}; j m_j}^{j' m_j'}C_{\frac{1}{2} -\frac{1}{2}; k m_k}^{k' m_k'} -(-1)^{m} C_{\frac{1}{2} -\frac{1}{2}; j m_j}^{j' m_j'} C_{\frac{1}{2} \frac{1}{2}; k m_k}^{k' m_k'}    ). 
 \label{summatrielespinohanew}
\end{align}
With the explicit form of Clebsh-Gordan coefficients
\be
C_{\frac{1}{2}  \pm \frac{1}{2};~j, m}^{j', m'} = \delta_{j',j+\frac{1}{2}} ~ \delta_{m',m\pm \frac{1}{2}}  ~\sqrt{\frac{j\pm m+1}{2j+1}} ~\pm \delta_{j',j-\frac{1}{2}}  ~\delta_{m',m\pm \frac{1}{2}}~  \sqrt{\frac{j\mp m}{2j+1}}, \label{clebsh1/2form}
\ee 
(\ref{summatrielespinohanew}) yields (\ref{explfcmu1}) and (\ref{explfcmu2}).

\subsection{About the $SO(4)$ monopole harmonics}\label{appendix:so4appendmono}

Next, we evaluate the matrix elements of $y_m$ sandwiched by the $SO(4)$ monopole harmonics. From (\ref{ysmatrixfour}) and (\ref{s3coordinatessphericalrel}), we immediately have 
\bse
\begin{align}
\langle \bs{\Phi}_{j', m'_j;  k', m'_k}|{y}_{m=1,2} |\bs{\Phi}_{j, m_j;  k, m_k}\rangle &=(-i)^{m} \frac{\sqrt{(2j +1)(2k+1)}}{2} ~(-1)^{j'+ 2j + 2k'+k +\frac{3}{2}(I+1)} ~\begin{Bmatrix}
j' & k' & \frac{I}{2} \\
k & j & \frac{1}{2}
\end{Bmatrix} ~\nn\\
&~\times (C_{\frac{1}{2}, \frac{1}{2}; ~j, m_j}^{j', m'_j}~C_{\frac{1}{2}, \frac{1}{2};~ k, m_k}^{k', m'_k}+(-1)^{m}C_{\frac{1}{2}, -\frac{1}{2}; ~j, m'_j}^{j', m'_j}~C_{\frac{1}{2}, -\frac{1}{2};~ k, m_k}^{k', m'_k} ), \\
\langle \bs{\Phi}_{j', m'_j;  k', m'_k}|{y}_{m=3,4} |\bs{\Phi}_{j, m_j;  k, m_k}\rangle &=(-i)^{m} \frac{\sqrt{(2j +1)(2k+1)}}{2} ~(-1)^{j'+ 2j + 2k'+k +\frac{3}{2}(I+1)} ~\begin{Bmatrix}
j' & k' & \frac{I}{2} \\
k & j & \frac{1}{2}
\end{Bmatrix} ~\nn\\
&~\times (C_{\frac{1}{2}, \frac{1}{2}; ~j, m_j}^{j', m'_j}~C_{\frac{1}{2}, -\frac{1}{2};~ k, m_k}^{k', m'_k} -(-1)^{m}C_{\frac{1}{2}, -\frac{1}{2}; ~j, m_j}^{j', m'_j}~C_{\frac{1}{2}, \frac{1}{2};~ k, m_k}^{k', m'_k} ). 
\end{align}  \label{monopoleharmatrixso4}
\ese
When 
$j+k=j'+k'=n+\frac{I}{2}$,  
the sign-part is simplified as 
\be
(-1)^{j'+2j+2k'+k+\frac{3}{2}(I+1)} =-i(-1)^{(j'+k'+j+k) + k'+j+\frac{3}{2}I}=-i (-1)^{2(n+I)+ j+k'+\frac{I}{2} } =-i (-1)^{j+k'+\frac{I}{2}}, 
\ee
and with (\ref{clebsh1/2form}) 
we have 
\be
\langle \bs{\Phi}_{j', m'_j; k', m'_k}|{y}_{m}| \bs{\Phi}_{j, m_j; k, m_k} \rangle = \sum_{\sigma=+,-} \mathcal{Y}_{m}^{(\sigma, -\sigma)}(j, ~k)_{(m_j',m'_k; ~m_j, m_k)}~\cdot~ \delta_{j',j+\frac{\sigma}{2}}~\delta_{k',k-\frac{\sigma}{2}}, 
\ee
where 
$\mathcal{Y}_{m}^{(\sigma, -\sigma)}(j,k)_{(m_j',m'_k;~m_j, m_k)} \equiv \langle \bs{\Phi}_{j+\frac{\sigma}{2}, m'_j; k-\frac{\sigma}{2}, m'_k}|{y}_{m}| \bs{\Phi}_{j, m_j; k, m_k}\rangle$ 
are given by 
\begin{align}
\mathcal{Y}_{m=1,2}^{(+-)}(j,k) &=-(-i)^m \frac{1}{2} (-1)^{n+I}  \begin{Bmatrix}  
j+\frac{1}{2} & k-\frac{1}{2} & \frac{I}{2} \\
k & j & \frac{1}{2}
\end{Bmatrix}\nn\\
&\times  (\delta_{m_j', m_j+\frac{1}{2}}\delta_{m_k', m_k+\frac{1}{2}} \sqrt{(j+m_j+1)(k-m_k)} -(-1)^{m} \delta_{m_j', m_j-\frac{1}{2}}\delta_{m_k', m_k-\frac{1}{2}} \sqrt{(j-m_j+1)(k+m_k)} ), \nn\\%
\mathcal{Y}_{m=3,4}^{(+-)}(j,k) &=(-i)^m \frac{1}{2} (-1)^{n+I}  \begin{Bmatrix}  
j+\frac{1}{2} & k-\frac{1}{2} & \frac{I}{2} \\
k & j & \frac{1}{2}
\end{Bmatrix}\nn\\
&\times  (\delta_{m_j', m_j+\frac{1}{2}}\delta_{m_k', m_k-\frac{1}{2}} \sqrt{(j+m_j+1)(k+m_k)} +(-1)^{m} \delta_{m_j', m_j-\frac{1}{2}}\delta_{m_k', m_k+\frac{1}{2}} \sqrt{(j-m_j+1)(k-m_k)} )\nn\\
\mathcal{Y}_{m=1,2}^{(-+)}(j,k) &=(-i)^m \frac{1}{2} (-1)^{n+I}  \begin{Bmatrix}  
j-\frac{1}{2} & k+\frac{1}{2} & \frac{I}{2} \\
k & j & \frac{1}{2}
\end{Bmatrix}\nn\\
&\times  (\delta_{m_j', m_j+\frac{1}{2}}\delta_{m_k', m_k+\frac{1}{2}} \sqrt{(j-m_j)(k+m_k+1)} -(-1)^{m} \delta_{m_j', m_j-\frac{1}{2}}\delta_{m_k', m_k-\frac{1}{2}} \sqrt{(j+m_j)(k-m_k+1)} ), \nn\\
\mathcal{Y}_{m=3,4}^{(-+)}(j,k) &=(-i)^m \frac{1}{2} (-1)^{n+I}  \begin{Bmatrix}  
j-\frac{1}{2} & k+\frac{1}{2} & \frac{I}{2} \\
k & j & \frac{1}{2}
\end{Bmatrix}\nn\\
&\times  (\delta_{m_j', m_j+\frac{1}{2}}\delta_{m_k', m_k-\frac{1}{2}} \sqrt{(j-m_j)(k-m_k+1)} +(-1)^{m} \delta_{m_j', m_j-\frac{1}{2}}\delta_{m_k', m_k+\frac{1}{2}} \sqrt{(j+m_j)(k+m_k+1)} ). 
\label{pexplfcmu2}
\end{align}
On may find that (\ref{pexplfcmu2}) is  simply related to  (\ref{explfcmu1}) and (\ref{explfcmu2}) as 
\be
\mathcal{Y}_{m}^{(\sigma,-\sigma)}(j,k)=-{Y}_{m=1,2}^{(\sigma,-\sigma)}(j,k). 
\ee

\subsection{Special case  $(j, k)=(1/2, 0)\oplus (0,1/2)$}\label{appendix;irrpssu2su2}

With a $4\times 2$ matrix 
\be
\Phi^{\text{AFF}}(\Omega_3) \equiv \frac{1}{\sqrt{2}}\begin{pmatrix}
y_m\bar{q}_m \\
1_2 
\end{pmatrix}, 
\ee
we can obtain the meron gauge field configuration (\ref{merongaugefielda}): 
\be
A^{\text{AFF}}=-i{\Phi^{\text{AFF}}}^{\dagger}d\Phi^{\text{AFF}} =-\frac{1}{2}\bar{\eta}_{mn}^i y_{n}\sigma_i ~dy_{m}, 
\ee
where  $\bar{q}_m q_n =\delta_{mn}-\bar{\eta}_{mn}^iq_i$ was used.  
$\Phi^{\text{AFF}}(\Omega_3)$ can be represented by the $SO(4)$ spinor spherical harmonics $\bs{Y}_{j, m_j;~k, m_k}$ (\ref{vectorlikespinsphereharmo}) as 
\be
\Phi^{\text{AFF}}(\Omega_3)^{\dagger} =  \pi
\begin{pmatrix}
|\bs{\Phi}^{\text{AFF}}_{1}\rangle & |\bs{\Phi}^{\text{AFF}}_{2}\rangle & |\bs{\Phi}^{\text{AFF}}_{3}\rangle & |\bs{\Phi}^{\text{AFF}}_{4}\rangle
\end{pmatrix} =
\pi \begin{pmatrix}
-|\bs{Y}_{1}\rangle  & -|\bs{Y}_{2}\rangle  & |\bs{Y}_{3}\rangle & |\bs{Y}_{4}\rangle
\end{pmatrix} 
\ee
where 
\begin{subequations}
\begin{align}
&
~~~~~~~~~~~~|\bs{Y}_{1}\rangle \equiv \bs{Y}_{1/2, 1/2;~ 0,0} =  \frac{1}{\sqrt{2}\pi} 
\begin{pmatrix}
-y_4+iy_3 \\
iy_1 -y_2 
\end{pmatrix}
 =  \frac{1}{\sqrt{2}\pi} \begin{pmatrix}
-\cos\chi +i\sin\chi\cos\theta \\
i\sin\chi \sin\theta e^{i\phi}
\end{pmatrix},\nn\\
&~~~~~~~~~~~~|\bs{Y}_{2}\rangle \equiv \bs{Y}_{1/2, -1/2;~ 0,0} =\frac{1}{\sqrt{2}\pi} 
\begin{pmatrix}
iy_1 +y_2  \\
-y_4-iy_3 
\end{pmatrix}
   =\frac{1}{\sqrt{2}\pi} \begin{pmatrix}
i\sin\chi \sin\theta e^{-i\phi} \\
-\cos\chi -i\sin\chi\cos\theta
\end{pmatrix}, \\
&
~~~~~~~~~~~~|\bs{Y}_{3}\rangle\equiv  \bs{Y}_{0,0; ~1/2, 1/2} =\frac{1}{\sqrt{2}\pi} \begin{pmatrix}
1\\
0 
\end{pmatrix},\nn\\
&~~~~~~~~~~~~|\bs{Y}_{4}\rangle\equiv\bs{Y}_{0, 0; ~1/2, -1/2} =\frac{1}{\sqrt{2}\pi} \begin{pmatrix}
0\\
1
\end{pmatrix}. 
\end{align}  \label{spinorso4spheminim}
\end{subequations}
It is not difficult to  derive the $4\times 4$ matrix elements of $y_{m}$  by performing the integration $\langle \bs{\Phi}^{\text{AFF}}_\alpha|y_m|\bs{\Phi}^{\text{AFF}}_\beta\rangle$ such as $\langle\bs{\phi}_1^{\text{AFF}}|y_1|\bs{\phi}_4^{\text{AFF}}\rangle =\int d\Omega_3(-\bs{Y}_{1/2,1/2;~ 0,0})^{\dagger}y_1 \bs{Y}_{0,0;~1/2,-1/2}=\frac{1}{4}i$: 
\be
\langle y_{m}\rangle_{\Phi^{\text{AFF}}} =\frac{1}{4}\gamma_{m},  
\label{affmatym}
\ee
and similarly 
\be
\langle y_{m}\rangle_{Y} =-\frac{1}{4}\gamma_{m}. 
\label{spinym}
\ee
We introduce another $4\times 2$ matrix 
\be
\Phi= \Phi^{\text{AFF}} \cdot g =\frac{1}{2\sqrt{1+y_4}}
\begin{pmatrix}
1_2 +y_{m}q_{m} \\
1_2 +y_{m}\bar{q}_{m}
\end{pmatrix},  \label{defpshindfromaff}
\ee
where $g$ (\ref{gwithhigherspinsu2}) is given by 
\begin{align}
g(\Omega_3)&=e^{-i\frac{\chi}{2} \sum_{i=1}^3 \hat{y}_i \sigma_i} =\tilde{g}(\theta, \phi)~e^{-i\frac{\chi}{2} \sigma_z} ~\tilde{g}(\theta, \phi)^{\dagger}
=  
\begin{pmatrix}
\cos\frac{\chi}{2} -i\sin\frac{\chi}{2}\cos\theta & -i\sin\frac{\chi}{2}\sin\theta  e^{-i\phi} \\
-i\sin\frac{\chi}{2}\sin\theta e^{i\phi} & \cos\frac{\chi}{2} +i\sin\frac{\chi}{2}\cos\theta 
\end{pmatrix} 
\nn\\
&=\frac{1}{\sqrt{2(1+{y}_4)}} (1_2 +{y}_{m}{q}_{m}),  \label{genemerontonaridaemi}
\end{align} 
with 
\be
\tilde{g}(\theta, \phi)=e^{-i\frac{\phi}{2}\sigma_z} e^{-i\frac{\theta}{2}\sigma_y} =
\begin{pmatrix}
\cos\frac{\theta}{2}e^{-i\frac{\phi}{2}} & -\sin\frac{\theta}{2}e^{-i\frac{\phi}{2}} \\
\sin\frac{\theta}{2}e^{i\frac{\phi}{2}}  & \cos\frac{\theta}{2}e^{i\frac{\phi}{2}}
\end{pmatrix} . \label{I1/2tildegmat}
\ee
$\Phi$ (\ref{defpshindfromaff}) is related to the $SO(4)$ monopole harmonics (in the Dirac gauge) $\bs{\Phi}_{j,m_j;~k, m_k}$ (\ref{vectorrepsorep}) as 
\be
{\Phi}(\Omega_3)^{\dagger}\equiv \frac{1}{2\sqrt{1+y_4}} \begin{pmatrix}
1_2 +y_mq_m & 1_2 +y_m\bar{q}_m 
 \end{pmatrix}
= \pi \begin{pmatrix}
|\bs{\Phi}_1 \rangle & 
|\bs{\Phi}_2 \rangle& 
|\bs{\Phi}_3 \rangle& 
|\bs{\Phi}_4\rangle 
\end{pmatrix}  
\ee
where\footnote{In the Schwinger gauge, the $SO(4)$ monopole harmonics are represented as 
\begin{subequations}
\begin{align}
&
\bs{\Phi}^S_{1/2, 1/2;~ 0,0} =\frac{1}{\sqrt{2}\pi} \begin{pmatrix}
\cos\frac{\theta}{2} e^{-i\frac{1}{2}(\chi-\phi)} \\
-\sin\frac{\theta}{2} e^{i\frac{1}{2}(\chi+\phi)}
\end{pmatrix},~~~\bs{\Phi}^S_{1/2, -1/2;~ 0,0} =\frac{1}{\sqrt{2}\pi} \begin{pmatrix}
\sin\frac{\theta}{2} e^{-i\frac{1}{2}(\chi+\phi)} \\
\cos\frac{\theta}{2} e^{i\frac{1}{2}(\chi-\phi)}
\end{pmatrix}, \\
&
\bs{\Phi}^S_{0,0; ~1/2, 1/2} =\frac{1}{\sqrt{2}\pi} \begin{pmatrix}
\cos\frac{\theta}{2} e^{i\frac{1}{2}(\chi+\phi)} \\
-\sin\frac{\theta}{2} e^{-i\frac{1}{2}(\chi-\phi)}
\end{pmatrix},~~~\bs{\Phi}^S_{0,0;~1/2, -1/2} =\frac{1}{\sqrt{2}\pi} \begin{pmatrix}
\sin\frac{\theta}{2} e^{i\frac{1}{2}(\chi-\phi)} \\
\cos\frac{\theta}{2} e^{-i\frac{1}{2}(\chi+\phi)}
\end{pmatrix}. 
\end{align} \label{schwingergamonosimpl}
\end{subequations}
They are related to the Dirac gauge (\ref{diracgamonosimpl}) as 
\be
\bs{\Phi}_{j,m_j;~k,m_k} =\tilde{g}~ \bs{\Phi}^S_{j,m_j;~k,m_k}
\ee
with $\tilde{g}$ (\ref{I1/2tildegmat}). 
} 
\begin{subequations}
\begin{align}
&|\bs{\Phi}_1\rangle \equiv \bs{\Phi}_{1/2, 1/2;~ 0,0} =\frac{1}{\sqrt{2}\pi} \begin{pmatrix}
\cos\frac{\chi}{2} -i\sin\frac{\chi}{2}\cos\theta \\
-i\sin\frac{\chi}{2} \sin\theta ~e^{i\phi}
\end{pmatrix},~~|\bs{\Phi}_2\rangle \equiv\bs{\Phi}_{1/2, -1/2;~ 0,0} =\frac{1}{\sqrt{2}\pi} \begin{pmatrix}
-i\sin\frac{\chi}{2} \sin\theta e^{-i\phi} \\
\cos\frac{\chi}{2} +i\sin\frac{\chi}{2}\cos\theta
\end{pmatrix}, \\
&|\bs{\Phi}_3\rangle \equiv \bs{\Phi}_{0,0; ~1/2, 1/2} =\frac{1}{\sqrt{2}\pi}\begin{pmatrix}
\cos\frac{\chi}{2} +i\sin\frac{\chi}{2}\cos\theta \\
i\sin\frac{\chi}{2} \sin\theta ~e^{i\phi}
\end{pmatrix},~~~|\bs{\Phi}_4\rangle \equiv \bs{\Phi}_{0,0;~1/2, -1/2} =\frac{1}{\sqrt{2}\pi} \begin{pmatrix}
i\sin\frac{\chi}{2} \sin\theta e^{-i\phi} \\
\cos\frac{\chi}{2} -i\sin\frac{\chi}{2}\cos\theta
\end{pmatrix}. 
\end{align} \label{diracgamonosimpl}
\end{subequations}
The  corresponding matrix coordinates are 
\be
\langle y_{m}\rangle_{\Phi}=\frac{1}{4}\gamma_m. 
\label{mateleymfund}
\ee
(\ref{defpshindfromaff}) implies  that 
\be
|\bs{\Phi}_{\alpha}\rangle 
 = g^{\dagger} |\bs{\Phi}_{\alpha}^{\text{AFF}}\rangle, 
\ee
and consequently  
\be
\langle y_{m}\rangle_{\Phi^{\text{AFF}}}=\langle y_{m}\rangle_{\Phi}, 
\ee
which can also be confirmed by comparing (\ref{affmatym}) with (\ref{mateleymfund}).  
Similarly, we have
\be
|\bs{\Phi}_{\alpha}\rangle = -g^{\dagger} |\bs{Y}_{\alpha}\rangle~~(\alpha=1,2), ~~~~~~ |\bs{\Phi}_{\alpha}\rangle = g^{\dagger} |\bs{Y}_{\alpha}\rangle~~(\alpha=3,4),   \label{transpsidspinsph}
\ee
which implies 
\be
\langle \bs{\Phi}_{\alpha}|y_m|\bs{\Phi}_{\beta}\rangle = - \langle \bs{Y}_{\alpha}|y_m|\bs{Y}_{\beta}\rangle. 
\label{relmatricespsidys}
\ee
This relation is also obtained from (\ref{mateleymfund}) and (\ref{spinym}). 
(\ref{relmatricespsidys}) realizes the simplest version of  (\ref{relationso4functions}). 



\end{document}